\documentclass[12pt]{article}
\usepackage[dvipdfmx]{graphicx}
\usepackage{hyperref,amssymb,amsmath}
\usepackage{bbold}

\usepackage{mathtools}
\usepackage{bm}
\usepackage{url}
\usepackage{color}

\newcommand{\LQ}{\Lambda_{\rm QCD}}
\newcommand{\alfs}{\alpha_{s}}
\newcommand{\msbar}{\overline{\rm MS}}
\newcommand{\mbar}{\overline{m}}
\newcommand{\LMS}{\Lambda_{\rm \overline{MS}}}
\newcommand{\EQ}[1]{\begin{equation}#1\end{equation}}

\newcommand \bra[1]{\left< {#1} \,\right\vert}
\newcommand \ket[1]{\left\vert\, {#1} \, \right>}

\newcommand{\bea}{\begin{eqnarray}}
\newcommand{\eea}{\end{eqnarray}}
\newcommand{\simgt}{\hbox{ \raise3pt\hbox to 0pt{$>$}\raise-3pt\hbox{$\sim$} }}
\newcommand{\simlt}{\hbox{ \raise3pt\hbox to 0pt{$<$}\raise-3pt\hbox{$\sim$} }}

\newcommand{\be}{\begin{equation}}
\newcommand{\ee}{\end{equation}}

\newcommand{\lt}{\left}
\newcommand{\rt}{\right}

\newcommand{\non}{\nonumber \\}

\begin{document}

\begin{titlepage}

    \begin{flushright}
      \normalsize TU--1120\\
      \normalsize KEK--TH--2326\\
      \today
    \end{flushright}

\vskip2.5cm
\begin{center}
\Large
Renormalon subtraction in OPE using Fourier transform: \\
Formulation and application to various observables
\end{center}

\vspace*{0.8cm}
\begin{center}
{\sc Yuuki Hayashi}$^{a}$ ,
{\sc Yukinari Sumino}$^{a}$ and
{\sc Hiromasa Takaura}$^{b}$\\[5mm]
  {\small$^a$\it Department of Physics, Tohoku University}\\[0.1cm]
  {\small\it Sendai, 980-8578 Japan} \\
  \vspace{0.5cm}
  {\small$^b$\it Institute of Particle and Nuclear studies, KEK}\\[0.1cm]
  {\small\it Tsukuba, 305-0801, Japan}
\end{center}

\vspace*{2.8cm}
\begin{abstract}
\small
Properly separating and subtracting renormalons 
in the framework of the operator product expansion (OPE) is
a way to realize high precision computation of QCD effects
in high energy physics.
We propose a new method (FTRS method), 
which enables
to subtract multiple renormalons
simultaneously from a general observable.
It utilizes a property of Fourier transform, and
the leading Wilson coefficient is written in a one-parameter integral
form whose integrand has suppressed (or vanishing) renormalons.
The renormalon subtraction scheme coincides with 
the usual principal-value prescription at large orders.
We perform test analyses and subtract 
the ${\cal O}(\LQ^4)$ renormalon from the Adler function, 
the ${\cal O}(\LQ^2)$ renormalon from the $B\to X_ul\bar{\nu}$ decay width, 
and the ${\cal O}(\LQ)$ and ${\cal O}(\LQ^2)$ renormalons from 
the $B,\,D$ meson masses. 
The analyses show good consistency with theoretical expectations,
such as improved convergence and scale dependence.
In particular
we obtain
$
\bar{\Lambda}_{\rm FTRS}=0.495\pm0.053~\text{GeV} 
$ and
$
(\mu_\pi^2)_{\rm FTRS}=-0.12\pm 0.23~\text{GeV}^2
$
for the non-perturbative parameters of HQET.
We explain the formulation and analyses in detail.

\vspace*{0.8cm}
\noindent

\end{abstract}


\vfil
\end{titlepage}

\newpage
\tableofcontents
\clearpage

\section{Introduction}

After the discovery of the Higgs boson, particle physics 
entered the era of high precision physics through  frontier experiments, such as the
experiments
at the LHC, the BELLE II experiment, etc.
Under these circumstances, 
it has become indispensable to achieve high precision calculations
of QCD effects.
Those calculations 
play important roles, in an essential way, in precision studies
of the Higgs boson properties, in direct new physics searches, in probing
beyond the SM physics through precision flavor physics, and so on.

Thanks to recent developments of perturbative QCD and lattice QCD
calculations, the accuracy of theoretical calculations of QCD effects
has  improved greatly.
In particular, the asymptotic freedom of QCD allows us 
to predict an observable in the high energy region by perturbative calculations, 
and recent developments of computational technologies 
enable access to higher-order corrections for many observables.
With such progress, 
it has become visible that ambiguities arise
in higher-order perturbative QCD calculations,
which are called renormalon ambiguities \cite{Beneke:1998ui}.
Renormalons are the source of a rapid growth of perturbative coefficients,
which lead to bad convergence of perturbative series.
Renormalons limit achievable accuracies of theoretical
predictions and can obstruct 
high precision theoretical calculations in  particle physics.

The use of the operator-product expansion (OPE) in combination
with renormalon subtraction is a way to achieve
high precision calculation of QCD effects.
The OPE framework is especially suited to solve the renormalon problem
since it treats the perturbative and non-perturbative counterparts together.
In the OPE of an observable, ultraviolet (UV)
and infrared (IR) contributions are factorized  into the Wilson coefficients
and non-perturbative matrix elements, respectively.
The former are calculated in perturbative QCD, while the
latter are determined by non-perturbative methods.
It has been recognized that (IR) renormalons
are contained in the respective parts.
Since a physical observable as a whole should not contain renormalon uncertainties, 
what we need is to separate the renormalons
from the respective parts and cancel them.

To separate and cancel renormalons, 
many technologies have been developed over the past decades.
Historically cancellation of renormalons made a strong impact 
in heavy quarkonium physics.
The perturbative series for the quarkonium energy levels turned out to be
poorly convergent when the quark pole mass was used to express
the levels, following the convention of QED bound state calculations.
When they were re-expressed by a short-distance quark mass,
the convergence of the perturbative series improved dramatically.
This was understood as due to the cancellation of the ${\cal O}(\LQ)$
renormalons between the pole mass and binding energy \cite{Pineda:id, Hoang:1998nz, Beneke:1998rk}.
A similar cancellation was observed in the 
$B$ meson partial
widths in the semileptonic decay modes \cite{Bigi:1994em,Neubert:1994wq,Ball:1995wa}.
These features were applied successfully in accurate determinations of fundamental
physical constants such as 
the heavy quark masses \cite{Penin:2014zaa,Kiyo:2015ufa,Beneke:2016oox,Bazavov:2018omf,Peset:2018ria}, some of the 
Cabbibo-Kobayashi-Maskawa matrix elements \cite{Hoang:1998hm,Alberti:2014yda},
and the strong coupling constant $\alfs$ \cite{Bazavov:2012ka}.
It is also notable that, recently in some OPEs 
which implement the ${\cal O}(\LQ)$
renormalon cancellation,
the validity of the OPE
has been checked using a multitude of observables;
see e.g., \cite{Alberti:2014yda,Bazavov:2018omf}.


Analyses including cancellation of renormalons beyond the ${\cal O}(\LQ)$
renormalon
of the pole mass have started only recently.
The ${\cal O}(\LQ)$ and ${\cal O}(\LQ^3)$ renormalons
of the static QCD potential $V_{\rm QCD}(r)$ were subtracted
 and combined with
the corresponding non-perturbative matrix elements 
\cite{Sumino:2005cq, Takaura:2018lpw, Takaura:2018vcy}.
Then the potential in the OPE was compared
with a lattice result.
By the renormalon subtraction,
perturbative uncertainty of the Wilson coefficient
reduced considerably and the matching range with the lattice
data became significantly wider. 
By the matching
$\alfs$ was determined with a
good accuracy, which agreed with other measurements.
Also the ${\cal O}(\LQ^4)$ renormalon contained in the lattice
plaquette action was subtracted and absorbed into
the local gluon condensate \cite{Ayala:2020pxq}.

In this paper, we generalize the method which was previously established
only for $V_{\rm QCD}(r)$ \cite{Sumino:2005cq}.
The method for $V_{\rm QCD}(r)$ is based on the fact that 
the renormalons 
are eliminated or highly suppressed
in $\tilde{V}_{\rm QCD}(q)$ 
(Fourier transform of $V_{\rm QCD}(r)$)
\cite{Hoang:1998nz, Beneke:1998rk, Brambilla:1999qa},
and that
the renormalons can be represented by contour integrals of $q$.
Recently, it has been clarified that the
properties of the Fourier transform
can explain the renormalon suppression \cite{Sumino:2020mxk}.
We can extend this mechanism to a general single-scale observable, 
such that the dominant renormalons are suppressed 
in an artificial momentum ($\tau$) space using Fourier transform.
Then the multiple renormalons of a general observable 
are separated simultaneously
in the form of simple $\tau$ integrals, 
consistently with the OPE formulation.
We present the formulation of the new method
and the necessary formulas for practical use.
The formulation is simple and easy to compute, once
the standard perturbative series is given for each
observable.

As a first theoretical test, 
we apply the method to the following observables to subtract the renormalons: 
(1) the ${\cal O}(\LQ^4)$ renormalon from the Adler function,
(2) the ${\cal O}(\LQ^2)$ renormalon from the $B\to X_ul\bar{\nu}$
decay width,
(3) the ${\cal O}(\LQ)$ and ${\cal O}(\LQ^2)$ renormalons simultaneously from 
the $B$ or $D$ meson mass.
The ${\cal O}(\LQ^2)$ renormalons in the $B$ decay width and $B$, $D$ meson masses
are subtracted for the first time in this study. 
All the analyses show good consistency with theoretical expectations,
e.g., good convergence and consistent behavior with the OPE. 

We have already reported part of the results in the letter 
article \cite{Hayashi:2020ylq}.
In this paper, we present the details of the study, adding various examinations.
Furthermore, we update the analysis of $\Gamma(B\to X_ul\bar{\nu})$
incorporating the recently calculated ${\cal O}(\alfs^3)$ corrections \cite{Fael:2020tow}.

The paper is organized as follows.
In Sec.~2, we explain the theoretical formulation:
first, we review the renormalon structure and 
renormalon subtraction for a general observable in the framework of the OPE (Sec.~2.1), 
and then we
construct a new method for renormalon subtraction using Fourier transform
(Sec.~2.2).
In addition, we apply our renormalon subtraction method to two simplified models (Sec.~2.3).
Through Secs.~3--5, test analyses of our method are presented.
In Sec.~3, we examine the Adler function and the local gluon condensate.
In Sec.~4, we examine the $B\to X_ul\bar{\nu}$ decay width.
In Sec.~5, we examine 
the $B$ and $D$ meson masses and extract the non-perturbative parameters
$\bar{\Lambda}$, $\mu_\pi^2$. 
Summary  and conclusions are given in Sec.~6.
Details are presented in Appendices.
In App.~\ref{App:PertCoeffs}, we collect
the perturbative coefficients used in our analyses.
In App.~\ref{appB}, we discuss the inclusion of logarithmic corrections to the renormalons
into our method.
In App.~\ref{AppC}, we derive the relation between the Borel integral and Fourier 
transform.
In App.~\ref{AppD}, we derive the formula for the renormalon-subtracted
Wilson coefficient in a different way to discuss relation with the Wilsonian picture.
In App.~\ref{AppE}, we explain how to resum the 
artificial UV renormalons 
which emerge from Fourier transform.
In App.~\ref{App:PhenoModel}, we explain
the phenomenological model of the
$R$-ratio used in Sec.~3.
In App.~\ref{App:cond(iii)}, we investigate the case where unsuppressed
renormalons (or singularities in the Borel plane) remain in the momentum space
in the FTRS method and estimate the effects.

\section{Theoretical formulation}
\subsection{OPE framework and renormalon subtraction}
In this section, we review the renormalon structure of a general observable 
in the framework of the operator product expansion (OPE) and 
the renormalization group equation (RGE) \cite{Beneke:1998ui}.
This is followed by an explanation of 
the principle of renormalon subtraction in the same framework.

Let us consider a dimensionless observable $A(Q)$ with a hard scale $Q\,(\gg\LQ)$.
In the OPE formulation, we separate
UV contributions (from energy scale$\,\simgt Q$) and 
IR contributions (from energy scale$\,\ll Q$) 
and express $A(Q)$ as 
\be
A(Q)=\sum_iX^A_{{\cal O}_i}(Q;\mu,\alpha_s)
\frac{\langle{{\cal O}_i}\rangle(\mu)}{Q^{d_i}}.
\label{OPE}
\ee
Throughout this paper $\alfs\equiv\alpha_s(\mu)$ denotes the strong coupling
constant in the $\overline{{\rm MS}}$ scheme,
renormalized at the renormalization scale $\mu$.
(Whenever $\mu$ is set to a special value, the argument of $\alfs$ will be shown
explicitly.)
The right-hand side of eq.~\eqref{OPE}
gives an expansion of $A(Q)$ by the inverse power of $Q$, 
and $d_i \ge 0$ represents the canonical dimension of 
a low-energy local operator ${\cal O}_i$. 
In each term, 
the information of the UV and IR contributions, respectively, is encoded into 
the Wilson coefficient $X_{{\cal O}_i}^A$ and the matrix element 
$\langle{\cal O}_i\rangle$.
Since the natural size of $\langle{{\cal O}_i}\rangle$ is
order $\LQ^{\,d_i}$, contributions from 
higher-dimensional operators are more suppressed.

Often the leading order contribution 
turns out to be the Wilson coefficient of the identity operator,\footnote{
Even in the case that the leading operator is not the 
identity operator, 
often a similar argument holds.
For example, in the Heavy Quark Effective Theory (HQET) calculation of the 
$B$ meson  semileptonic 
decay width $\Gamma(B\to X l\overline{\nu})$, 
the leading operator is the heavy quark number operator 
$\bar{b}_vb_v$.
From the heavy quark number conservation, it follows that
$\bra{B}\bar{b}_vb_v\ket{B}=1$.
More details will be discussed in later sections.}
\be
X(Q)\equiv X^A_{\bf 1}(Q;\mu,\alpha_s) \langle{\bf 1}\rangle
\, .
\ee
Since ${\langle{\bf 1}\rangle}=1$, 
$X$ coincides with the naive perturbative QCD calculation of $A(Q)$,
\be
X(Q)=A(Q)_{PT}=\sum_{n=0}^\infty c_n(\mu/Q)\alpha_s^{n+1}.
\label{inf}
\ee
The next-to-leading term of eq.~(\ref{OPE}) is denoted as
$X^A_{{\cal O}_1}\langle{\cal O}_1\rangle/Q^{d_1}$, with
the Wilson coefficient $X^A_{{\cal O}_1}$
and the nonperturbative matrix element $\langle{\cal O}_1\rangle$. 
(For simplicity we assume that only one operator contributes at order
$1/Q^{d_1}$.)
The former can be calculated by perturbative QCD,
and the latter should be determined in a nonperturbative way, 
e.g., from experimental data or by lattice QCD calculation. 
Since the 
next-to-leading term is RG invariant,
we can determine the $\mu/Q$ dependence as \cite{Beneke:1998ui}
\be
X^A_{{\cal O}_1}(Q;\mu,\alpha_s)\frac{\langle{\cal O}_1\rangle(\mu)}{Q^{d_1}}=
K\,
e^{-\frac{d_1}{2b_0\alpha_s}}
\biggl(\frac{\mu^2}{Q^2}\biggr)^{\!\frac{d_1}{2}}
\alpha_s^{\frac{\gamma_0}{b_0}-\frac{d_1b_1}{2b_0^2}}\sum_{n=0}^\infty s_n(\mu/Q)\alpha_s^{n},
~~~
s_0=1\,.
\label{NPterm}
\ee
Apart from the overall normalization $K$,
the parameters $b_0$, $b_1$, $\gamma_0$
and the expansion coefficients $s_n\,(n\ge 1)$ can be calculated
in perturbative QCD.
$b_i$ denotes the  coefficients of the QCD beta function 
\bea
\mu^2 \frac{d \alpha_s}{d \mu^2}=\beta(\alpha_s)=-\sum_{i=0}^\infty b_i \, \alpha_s^{i+2}
\label{MSbarRGeq-alfs}
\,.
\eea
The first two coefficients are given explicitly by
\be
b_0=\frac{1}{4 \pi} \lt(11-\frac{2}{3} n_f \rt),
~~~~~
b_1=\frac{1}{(4 \pi)^2} \lt(102-\frac{38}{3} n_f \rt) \, ,
\ee
where $n_f$ is the number of quark flavors.
$\gamma_0$ denotes the 
one-loop coefficient of the anomalous dimension of ${\cal O}_1$,
\be
\Big[\mu^2\frac{d}{d\mu^2}-\gamma(\alpha_s)\Big]X^A_{{\cal O}_1}=0,\, 
~~~
\gamma(\alpha_s)=\sum_{n=0}^\infty \gamma_n \alfs^{n+1}.
\ee
$s_n$ is constructed from the coefficients of the perturbative
expansion of $X^A_{{\cal O}_1}$, 
the QCD beta function, and the anomalous dimension $\gamma(\alpha_s)$.

It is conjectured that the perturbative series of $X(Q)$
[eq.~(\ref{inf})] diverges
as $c_n \sim n!$ at high orders.
To quantify the ambiguity induced  by such behavior, its Borel transform is considered:
\be
 B_X(u)=\sum_n \frac{c_n(\mu/Q)}{n!}\Bigl(\frac{u}{b_0}\Bigr)^n \,.
\label{BorelTrX}
\ee
Formally we can reconstruct $X$ from
its Borel transform $B_X(u)$ by the inverse Borel transform given by 
the integral
\be
X``=" \frac{1}{b_0}\int_0^{\infty} du \,B_X(u) e^{-u/(b_0\alfs)} \, . \label{invBorelTr}
\ee
However, if there are singularities on the positive
real axis, the integral is ill defined.
It is conjectured that such singularities (branch points) do exist,
which are known as (IR) renormalons,
from analyses of a certain class of diagrams.
The ambiguity induced by renormalons is defined
from the discontinuities of the corresponding singularities in the Borel plane:
\be
\delta X = \frac{1}{2ib_0}\int_{C_+ - C_-} \!\!\!\!\!\!\!\!\!
du \, e^{-u/(b_0\alfs)}  B_X(u)  \,.
\label{ren-by-BI}
\ee
In the complex Borel ($u$) plane
the integral contours $C_\pm(u)$ connect $0\pm i\epsilon$ and $+\infty\pm i\epsilon$
infinitesimally above/below the positive real axis on which the discontinuities are located.
(We always assume that the discontinuities extend toward $+\infty$.)

The location of a renormalon singularity $u_*$ and the form of 
$\delta X$ due to this singularity
(apart from its overall normalization $N_{u_*}$) can be
determined from the OPE and RGE as follows.
It is assumed that the leading renormalon of $X$, induced by the
singularity closest to the origin, can be absorbed by the next-to-leading term
of eq.~(\ref{NPterm}).
Then the singularity structure of $B_{X}(u)$ in the vicinity of $u=u_*$ is 
determined to be
\be
B_{X}(u)=\left(\frac{\mu^2}{Q^2}\right)^{u_*}\frac{N_{u_*}}{(1-u/u_*)^{1+\nu_{u_*}}}\sum_{n=0}^\infty s'_n(\mu/Q)(1-u/u_*)^n+({\rm regular\,part}),\label{Bsing}
\ee
\be
\nu_{u_*}= \frac{b_1}{b_0^2} u_*-\frac{\gamma_0}{b_0},\,~~ u_*=\frac{d_1}{2}.
\ee
Except for the normalization parameter $N_{u_*}$, 
all the parameter dependence in eq.~(\ref{Bsing}) is specified by the OPE and RGE. 
In fact, the ambiguity generated by this renormalon takes the form
\be
\delta X_{u_*}=\frac{\pi \, N_{u_*}\, {u_*^{1+\nu_{u_*}}}}{b_0\,\Gamma(1+\nu_{u_*})}\left(\frac{\Lambda_{\rm \overline{MS}}^2}{Q^2}\right)^{\! u_*}\!\!\!
(b_0\alpha_s(Q))^{\gamma_0/b_0}\sum_{n=0}^\infty {s}_n(\mu/Q=1)\alpha_s(Q)^{n}~~~;~~~
{s}_0=1.
\label{renu}
\ee
Here, we have chosen $\mu=Q$ and written the argument of $\alfs$ explicitly.
(From eqs.~\eqref{NPterm} and \eqref{renu} one can read off the relation between
$N_{u_*}$ and ${\rm Im}\, K$.)
$\LMS$ denotes the dynamical scale defined in the $\overline{{\rm MS}}$ scheme, 
which is given by
\be
\frac{\Lambda_{\overline{\rm MS}}^2}{\mu^2}=\exp \left[-\left\{\frac{1}{b_{0} \alpha_{s}}+\frac{b_{1}}{b_{0}^{2}} \log \left(b_{0} \alpha_{s}\right)+\int_{0}^{\alpha_{s}} d x\left(\frac{1}{\beta(x)}+\frac{1}{b_{0} x^{2}}-\frac{b_{1}}{b_{0}^{2} x}\right)\right\}\right].
\label{LAMBDA}
\ee
Higher order ambiguities generated by the singularities further from the origin
can be determined,
besides their normalization, in some cases in the similar manner. 

We stress again that,
since renormalon
ambiguities are assumed to have the same $Q$ dependence as 
the non-perturbative terms in the OPE,
they can be absorbed into the non-perturbative matrix elements in the OPE framework.
This is because a physical observable should be free of renormalon ambiguities,
and if the OPE is a legitimate  framework to treat observables beyond perturbation
theory, the renormalon ambiguities of Wilson coefficients
should be canceled as above.\footnote{
This is similar to the concept that the leading renormalon included in $V_{\rm QCD}(r)$
of order $\LQ$ can be absorbed into the twice of the quark pole  mass.
An important point is to separate the order $\LQ$ renormalons in $V_{\rm QCD}(r)$
and $2 m_{\rm pole}$ explicitly and cancel them.
}
Since renormalons arise from IR dynamics
in the calculation of Wilson coefficients, to
absorb them into matrix elements is in agreement with the concept of the OPE.

There is a scheme dependence in how to absorb renormalons into the
matrix elements.
A conventional prescription,
the ``principal value (PV) prescription," is 
to define a renormalon-subtracted Wilson coefficient by 
\be
[X(Q)]_{\rm PV}= \frac{1}{b_0}\int_{0,\rm PV}^{\infty}
du \, e^{-u/(b_0\alfs)}  B_X(u)\,.
\label{PV-Bl}
\ee
It takes the PV of the Borel resummation integral, that is,
it takes the average over the contours $C_\pm(u)$.
By definition, the renormalon ambiguities are minimally subtracted from $X$.
At the same time, the matrix elements in the PV prescription are also defined free of
renormalon ambiguities, after absorbing the renormalons into
the matrix elements.
Below we will advocate a renormalon subtraction scheme, which coincides with 
the usual principal-value prescription at large orders.

\subsection{Renormalon subtraction using Fourier transform}
\label{sec:FTRS}

In practice, it is a non-trivial task to evaluate the PV integral eq.~(\ref{PV-Bl})
from known perturbative series (up to ${\rm N}^k{\rm LO}$).
In this section, we explain a new method to obtain the
PV integral in a systematic approximation using a different integral representation.
The method conforms well with the OPE (expansion in $1/Q$) and RGE.

\subsubsection*{Relation between PV prescription and Fourier transform}
%

To evaluate $[X(Q)]_{\rm PV}$,
we extend the formulation developed in ref.~\cite{Sumino:2005cq},
which works for $V_{\rm QCD}(r)$.
Renormalons of $V_{\rm QCD}(r)$ are located
at $u_*=1/2$, $3/2$, $\dots$.
It is known \cite{Hoang:1998nz, Beneke:1998rk, Brambilla:1999qa, Sumino:2020mxk} 
that 
for the momentum-space potential
[i.e., the Fourier transform of 
the coordinate-space potential $V_{\rm QCD}(r)$]
these renormalons are absent (or highly suppressed) and 
the perturbative series exhibits a good convergence.
Conversely, $V_{\rm QCD}(r)$ is given by the inverse Fourier integral
of the momentum-space potential which is (largely) free from the renormalons.
This indicates that the renormalons of $V_{\rm QCD}(r)$
arise from (IR region of) the inverse Fourier integral.
Using the formulation of ref.~\cite{Sumino:2005cq}, 
one can avoid the renormalon uncertainties reviving from the inverse Fourier transform
and  give a renormalon-subtracted prediction for $V_{\rm QCD}(r)$.
This is realized by a proper deformation of the integration contour of the inverse Fourier transform.
In this method, one minimally subtracts the renormalons using Fourier transform, 
and the renormalon-subtracted prediction is actually equivalent to the 
PV integral eq.~(\ref{PV-Bl})
as long as the momentum-space potential is free of renormalons.
We propose a generalized method using an analogous mechanism. 
The key to achieving this goal is to find a proper Fourier transform
which suppresses the renormalons included in the original observable.

We define the Wilson coefficient in the ``momentum space" by the Fourier transform 
as
\be
\tilde{X}(\tau)=\int d^3\vec{x}\,e^{-i\vec{\tau}\cdot\vec{x}}r^{2au'}X(r^{-a}),
\label{tildeX}
\ee
where we define $r=|\vec{x}|=Q^{-1/a}$.
This Fourier transform includes the parameters $a,u'$, 
which are important for suppressing the renormalons in momentum space. 
We consider the case where the effects of the anomalous dimension
are subdominant and 
the ambiguity induced by the $u=u_*$ renormalon can be
approximated by [cf.\  eq.~\eqref{renu}]
\be
\delta X(Q)\approx\frac{\pi}{b_0}\frac{N_{u_*}}{\Gamma(1+\nu_{u_*})}\, u_*^{1+\nu_{u_*}}\left(\frac{\Lambda_{\rm \overline{MS}}^2}{Q^2}\right)^{u_*}.
\label{defu*renormalon}
\ee
(We discuss how to include the corrections to this approximation in App.~\ref{appB}.
On the other hand,
if $\gamma_0=0$ and $r_n=0$ for $n\ge 1$, namely if there are no
logarithmic corrections, the following procedure eliminates the corresponding
renormalon exactly.)
Since the Fourier transform and the Borel resummation mutually commute, 
we obtain
\bea
&&\delta \tilde{X}(\tau)=\int d^3\vec{x}\,e^{-i\vec{\tau}\cdot\vec{x}}r^{2au'}\delta X(r^{-a})\non
&&~~~
= -\frac{\pi}{b_0}\frac{4\pi}{\tau^{3+2au'}}\frac{N_{u_*}}{\Gamma(1+\nu_{u_*})}\, 
u_*^{1+\nu_{u_*}}\left(\frac{\LMS^2}{\tau^{2a}}\right)^{u_*}\!\!\!\!\sin\left(\pi a(u_*+u')\right)\Gamma\left(2a(u_*+u')+2\right),
\nonumber\\
&&
\label{FTdelX}
\eea
where we have assumed eq.~\eqref{defu*renormalon}
and used analytical continuation of the result for 
$(0< \nolinebreak )3/2+a(u_*+u')<1$.
The ambiguities
at $u_*=-u'+ n/{a}$ for 
$n= 0, 1, 2,\cdots$ are eliminated 
due to the multiple roots of the sine factor.
We can adjust the parameters $a,u'$ so that the ambiguities are eliminated 
or suppressed for 
desired $u_*$'s.\footnote{
In addition we can vary the dimension of the Fourier transform
to $d^n\vec{x}$. For simplicity we set $n=3$.
}
In the case of the static QCD potential $X(Q=1/r)=rV_{\rm QCD}(r)$,
we choose $(a,u')=(1,-1/2)$, 
which suppresses the renormalons at $u=1/2,\,3/2,\,5/2,\,\cdots$. 
$\tilde{X}(\tau=q)$ reduces to the standard momentum-space potential,
and the renormalon at $u=1/2$ is eliminated, while that at $u=3/2$ is
highly suppressed.
(In particular, in the large-$\beta_0$ approximation IR renormalons are totally
absent in the momentum-space potential.)
 In the case of a general observable $X(Q)$, we can
adjust the parameters $a,u'$
to cancel or suppress the dominant renormalons of $\tilde{X}$.
The
level of suppression depends on the observable, but at least the first two renormalons closest
to the origin can always be suppressed.

Naively we can reconstruct $X(Q)$ by the inverse Fourier transform.
After angular integration, it can be expressed by the one-parameter integral as
\bea
&&
X(Q)=r^{-2 a u'} \int \frac{d^3\vec{\tau}}{(2\pi)^3}\,
e^{i\vec{\tau}\cdot\vec{x}}\,\tilde{X}(\tau)
=\frac{r^{-2 a u'-1}}{2 \pi^2} \int_0^{\infty} d \tau \, \tau \sin(\tau r) \tilde{X}(\tau) \, .
\label{InvFT-X}
\eea
$X(Q)$ has renormalons, while the dominant renormalons are
suppressed in $\tilde{X}$ in the integrand.
The dominant renormalons are generated by the $\tau$-integral
of logarithms $\log^n(\mu^2/\tau^{2 a})$ at small $\tau$ in the perturbative series for $\tilde{X}(\tau)$.
When we consider resummation of the logarithms by RG alternatively 
(as we will do in practice),
they stem from the singularity of the running coupling constant $\alfs(\tau)$
in  $\tilde{X}$ located on the positive $\tau$ axis.
$\delta X$ is generated by the integral surrounding the
discontinuity of this singularity.
The expected power dependence on $\LMS$ is obtained once we expand
$\sin(\tau r)$ in $\tau$.

We propose to compute the renormalon-subtracted
$X(Q)$ in the PV prescription, $[X(Q)]_{\rm PV}$, in the following way.
We take the PV of the above integral, that is, take
the average over the contours $C_\pm(\tau)$:
\be
[X(Q)]_{\rm FTRS}=
\frac{r^{-2 a u'-1}}{2 \pi^2} \int_{0,\rm PV}^{\infty} d \tau \, \tau \sin(\tau r) \tilde{X}(\tau) \, .
\label{XFTRS}
\ee
(``FTRS" stands for the renormalon subtraction using Fourier transform.)
Here, $\tilde{X}(\tau)$ is evaluated by RG-improvement up to
a certain order (see below) and
has a singularity (Landau singularity) on the positive $\tau$-axis.
In the case of $V_{\rm QCD}(r)$,
this quantity coincides with the renormalon-subtracted leading
Wilson coefficient of $V_{\rm QCD}(r)$ used in the analyses \cite{Sumino:2005cq, Takaura:2018lpw}.
Since renormalons of $\tilde{X}(\tau)$ are suppressed, the only source of
renormalons in eq.~(\ref{InvFT-X}) is 
that from the integral of the singularity of $\tilde{X}(\tau)$.
Then the PV prescription in eq.~\eqref{XFTRS}, which minimally regulates the singularity of the integrand 
(or more specifically, that of the running coupling
constant), corresponds to the minimally renormalon-subtracted quantity
eq.~(\ref{PV-Bl}).
An argument is given for the
equivalence of eqs.~(\ref{PV-Bl}) and (\ref{XFTRS})
in the appendix of ref.~\cite{Sumino:2020mxk}, while ample numerical
evidence to support this relation
for the static potential is presented in the main body of that paper. 
We present a refined argument which is applicable up to 
N$^4$LL approximation in App.~\ref{AppC}.
We note that the equivalence holds only when $\tilde{X}(\tau)$
does not have renormalons.
If renormalons remain in $\tilde{X}(\tau)$,
renormalons cannot be removed from the $Q$-space quantity merely by the PV integral in eq.~\eqref{XFTRS}, 
which only regulates the Landau singularity of the running coupling constant.
Hence, the renormalon suppression for the $\tau$-space quantity [cf. eq.~\eqref{FTdelX}]
is crucial for renormalon subtraction.

\subsubsection*{How to compute: Contour deformation and expansion in $Q^{-2/a}$}
In practice,
$\tilde{X}(\tau)$ has to be estimated approximately from the
known perturbative series of $X(Q)$ up to N$^k$LO.
We calculate $\tilde{X}(\tau)$ in the N$^k$LL approximation in the
following manner.
From the coefficients of the series up to 
$k$-th order perturbation 
\bea
X(Q)=\sum_{n=0}^{k}c_n\alfs(Q)^{n+1}\,,
\label{PTXcn}
\eea 
($c_n$ is the perturbative coefficient at the renormalization scale $\mu=Q$,)
$\tilde{X}$ is given by
\be
\tilde{X}(\tau)\to\tilde{X}^{(k)}(\tau)=\frac{4\pi}{\tau^{3+2au'}}
\sum_{n=0}^k \tilde{c}_n(0)\,\alfs(\tau^a)^{n+1}
\, ,
\label{FTXpert}
\ee
where $\tilde{c}_n(L_\tau)$ is defined by the following relation
\be
F(\hat{H},L_\tau)\sum_{n=0}^\infty c_n\alfs^{n+1}=\sum_{n=0}^\infty \tilde{c}_n(L_\tau)\alfs^{n+1}\,,
\label{relcandtildec}
\ee
\be
F(u,L_\tau)=-\sin\left(\pi a(u+u')\right)\Gamma\left(2a(u+u')+2\right)e^{L_\tau u}\,.
\label{Fu}
\ee
Here, $L_\tau=\log(\mu^2/\tau^{2a})\,,\,
\hat{H}=-\beta(\alfs)\frac{\partial}{\partial \alfs}\,$.
Comparing both sides of eq.~(\ref{relcandtildec}) at each order of the series expansion in $\alfs$, 
$\tilde{c}_n(L_\tau)$ is expressed by the coefficients of the original series 
$c_0\,,\,c_1\,,\,\cdots\,,\, c_n$ as
\bea
&&
\tilde{c}_0(L_\tau)=F(0,0)c_0\,,\,
\\ &&
\tilde{c}_1(L_\tau)=F(0,0)c_1+\partial_u F(0,L_\tau)b_0c_0\,,\,
\\ &&
\tilde{c}_2(L_\tau)=F(0,0)c_2+\partial_u F(0,L_\tau)b_1c_0+2\partial_uF(0,L_\tau)b_0c_1+\partial_u^2F(0,L_\tau)b_0^2c_0\,,
\\ &&
~~~~~~~~~\vdots~~\,.
\nonumber
\eea
The relations (\ref{relcandtildec}),(\ref{Fu}) follow from
the Fourier transform of $X=(\mu^2r^{2a})^{\hat{H}}\sum_{n=0}^\infty c_n\alfs^{n+1}$,
cf.\  eqs.~(\ref{defu*renormalon}) and (\ref{FTdelX}).
Since the renormalons in $\tilde{X}(\tau)$ are suppressed
and its perturbative series has a good convergence, 
it is natural to perform RG improvement in the $\tau$ space, and
$\tilde{X}^{(k)}(\tau)$ for a larger $k$ should be
a more accurate approximation of $\tilde{X}(\tau)$.
Accuracy tests by including higher orders will be given in the
toy model analysis and the test analyses below. 

%
\begin{figure}[t]
\centering
\includegraphics[width=5cm]{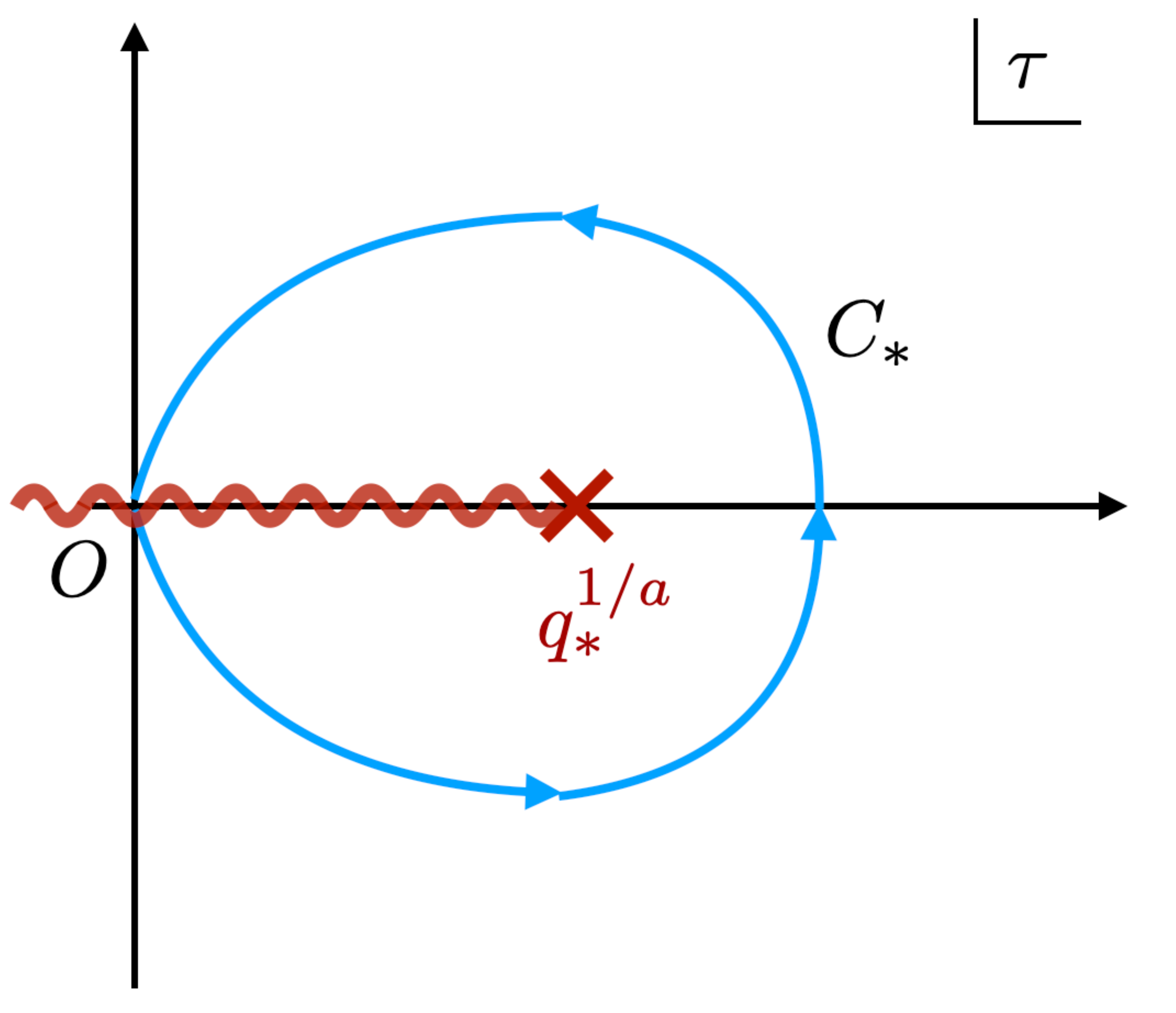}
\caption{Integration contour $C_*$.
The branch cut, shown by the wavy line, stems from the Landau 
singularity 
of the running coupling constant.
[$\alfs(q)$ diverges at $q=q_*$.]}
\label{fig:contour}
\end{figure}

In the numerical evaluation of eq.~(\ref{XFTRS}) it is useful to
decompose 
$[X(Q)]_{\rm FTRS}$ into two parts
by deforming the integral contour in the complex plane as follows. 
Hereafter, $\tilde{X}(\tau)$ stands for $\tilde{X}^{(k)}(\tau)$.
We can take the branch cut of $\alfs (q)$ such that 
$[\alpha_s(q+i\epsilon)]^*=\alpha_s(q-i\epsilon)$ for $q\in \mathbf{R}_{>0}$.
Hence, $\tilde{X}(\tau)$ has no imaginary part 
when its average is taken over $C_\pm(\tau)$, and we have
\bea
[X(Q)]_{\rm FTRS}&=&
\frac{r^{-2 a u'-1}}{2 \pi^2} \int_{0,\rm PV}^{\infty} d \tau \, \tau\, {\rm Im}\,[\exp(i \tau r)] \tilde{X}(\tau) \non
&=&\frac{r^{-2 a u'-1}}{2 \pi^2}{\rm Im}\, \int_{0,\rm PV}^{\infty} d \tau \, \tau\, \exp(i \tau r) \tilde{X}(\tau)\,.
\eea
The integral path $C_+$ can be deformed to 
the imaginary axis in the upper half $\tau$ plane\footnote{
This deformation is justified if $au'\ge -1$.
Here, we note that
$|\tilde{X}(\tau)| \sim 1/(\tau^{3+2au'} \log |\tau|)$ for $|\tau| \gg 1$.
This condition is satisfied in all the examples studied here, where $au'=-1$.
}
($\tau=it$), 
\be
{\rm Im}\,\int_{C_+} d \tau \, \tau\, \exp(i \tau r) \tilde{X}(\tau)
=-\int_{0}^{\infty} d t \, t\, \exp(-t r) {\rm Im}\,[\tilde{X}(it)]\,,
\ee
while the integral along the path $C_-$ requires
an additional contribution from the discontinuity,
\be
{\rm Im}\,\int_{C_-} d \tau \, \tau\, \exp(i \tau r) \tilde{X}(\tau)
=-\int_{0}^{\infty} d t \, t\, \exp(-t r) {\rm Im}\,[\tilde{X}(it)]
+{\rm Im}\,\int_{C_*} d \tau \, \tau\, \exp(i \tau r) \tilde{X}(\tau)\,.
\ee
The integration contour $C_*$ is shown in Fig.~\ref{fig:contour}. 
The integral along this contour can be rewritten as follows: 
\bea
{\rm Im}\,\int_{C_*} d \tau \, \tau\, \exp(i \tau r) \tilde{X}(\tau)
&=&{\rm Im}\,\int_{C_*} d \tau \, \tau\, [\cos(\tau r)+i\sin(\tau r)] \tilde{X}(\tau)\non
&=&{\rm Im}\,\int_{C_*} d \tau \, \tau\, \cos(\tau r)\tilde{X}(\tau)+{\rm Re}\,\int_{C_*} d \tau \, \tau\,\sin(\tau r) \tilde{X}(\tau)\non
&=&\frac{1}{i}\int_{C_*} d \tau \, \tau\, \cos(\tau r)\tilde{X}(\tau).
\eea
In the last equality, we used the fact that
each integral along $C_*$ gives a pure imaginary value
since the integrand $g(\tau)$ satisfies $[g(\tau)]^*=g(\tau^*)$.
Thus, $[X(Q)]_{\rm FTRS}$ is given by 
\be
[X(Q)]_{\rm FTRS}=\bar{X}_0(Q)+\bar{X}_{\rm pow}(Q)\,,\,
\label{XFTRS-decomp}
\ee
\be
\bar{X}_0(Q)=\frac{-r^{-2 a u'-1}}{2\pi^2} \int_0^{\infty} dt \, t\, e^{-tr}\,{\rm Im}\,[\tilde{X}(it)]  \,,
\label{X0}
\ee
\be
\bar{X}_{\rm pow}(Q)=\frac{r^{-2 a u'-1}}{4\pi^2 i}\int_{C_*} d \tau \, \tau  \cos(\tau r) \tilde{X}(\tau)\,.
\label{Xpow}
\ee
In the case $au'=-1/2$, it is convenient to combine the first term 
of $\bar{X}_{\rm pow}$ in expansion in $\tau r$
with $\bar{X}_0$ and decompose $[X(Q)]_{\rm FTRS}$ as
\be
[X(Q)]_{\rm FTRS}={X}_0(Q)+{X}_{\rm pow}(Q)\,,\,
\label{XFTRS-decomp-redefine}
\ee
\be
{X}_0(Q)=\frac{-r^{-2 a u'-1}}{2\pi^2} \int_0^{\infty} dt \, t\, e^{-tr}\,{\rm Im}\,[\tilde{X}(it)] 
+\frac{r^{-2 a u'-1}}{4\pi^2 i}\int_{C_*} d \tau \, \tau \tilde{X}(\tau)
 \,,
\label{X0-redefine}
\ee
\be
{X}_{\rm pow}(Q)=\frac{r^{-2 a u'-1}}{4\pi^2 i}\int_{C_*} d \tau \, \tau  [\cos(\tau r)-1] \tilde{X}(\tau)\,.
\label{Xpow-redefine}
\ee

${X}_0$ can be identified with the leading contribution in
expansion in $Q^{-2/a}$. 
In the first term of eq.~(\ref{X0-redefine}),
$e^{-i\tau r}$ is replaced by a damping factor $e^{-tr}$ and the integral
converges swiftly at large $t$.
The asymptotic behavior of $X_0$ for large $Q$ (small $r$) coincides with
that of $X(Q)$ and is determined by the 
RGE.
(The $r=Q^{-1}$ dependence of $X_0$ in the case of the static potential has been
analyzed in detail in \cite{Sumino:2003yp,Sumino:2005cq}.
$X_0/r$ behaves as a Coulombic potential $\sim \text{const.}/r$ 
with
mild logarithmic corrections in the
entire range of $r$.)

$X_{\rm pow}$ can be expanded by $r=Q^{-1/a}$ 
once $[\cos(\tau r)-1]$ is expanded in $\tau$,
\bea
X_{\rm pow}(Q)&=&\frac{r^{-2au'-1}}{4\pi^2 i}\int_{C_*} d \tau \, \tau  
\left[-(\tau r)^2/2!+(\tau r)^4/4!+\cdots\right] \tilde{X}(\tau)\,
\non
&=&
\frac{P_1}{Q^{d_1+1/a}}+\frac{P_2}{Q^{d_1+3/a}}+\cdots,~~~(u'=-d_1/2)
\eea
and the coefficients of this power series $P_1,P_2,\,...$ are real.
($d_1$ is the canonical dimension of the operator ${\cal O}_1$.)
It should be expanded at least to the order of the eliminated renormalon.
Then eq.~(\ref{XFTRS-decomp-redefine}) gives the renormalon-subtracted prediction 
of a general observable $X(Q)$
with an appropriate power accuracy of $1/Q$.\footnote{
The results scarcely change by varying the truncation order
beyond the minimum necessary order in expansion in $Q^{-1/a}$, 
in the tested range of $Q$ in the examples below.
}

For large $k$,
$[X(Q)]_{\rm FTRS}$ converges to the 
renormalon-subtracted Wilson coefficient in the PV prescription,
provided that the following assumptions which we made are satisfied:
(i)Renormalons cancel between the Wilson
coefficient and the corresponding operator matrix elements
in the OPE; 
(ii)The QCD
beta function beyond five loops does not alter the analytic structure
of the roots of the beta function
which holds up to N$^4$LL; 
(iii)There are no singularities except 
those which we suppress by the sine factor (on the 
positive real axis) in the Borel plane.
See App.~\ref{AppC} for the relevant argument.

On the other hand, the subtracted renormalons are
given as follows.
If we take the integration contour as $C_{\pm}$ instead of the PV integral in eq.~\eqref{XFTRS},
we also have power dependence with imaginary coefficients whose sign depends on which contour is chosen.
The power series with imaginary coefficients is identified with the renormalon uncertainties.
Thus,
\be
\left[ \delta X_{u_*=-u'+n/a} \right]_{\rm FTRS}=
\frac{(-1)^n}{2(2n+1)! \pi^2} 
\frac{1}{Q^{2(-u'+n/a)}} \int_{C_*} d \tau \, \tau^{2(n+1)}\tilde{X}(\tau) \, .
\label{delXFTRS}
\ee
They can be absorbed into the non-perturbative matrix elements, while
they do not appear in eq.~\eqref{XFTRS}
or (\ref{XFTRS-decomp}), where the average over $C_{\pm}$ is taken.

Let us briefly discuss what happens if the condition (iii) is not
satisfied and unsuppressed singularities exist in the right-half Borel plane.
(Further details on this point are discussed in App.~\ref{App:cond(iii)}.)
Suppose that 
there is an unsuppressed singularity at $u=R>0$ in the Borel plane. 
Then,
the series
of $[X(Q)]_{\rm FTRS}$, eq.~\eqref{PTXcn}, 
is apparently converging at $n < n_*(R,Q)$ but starts to diverge
at $n \geq n_*(R,Q)$,
where $n_*(R,Q)\approx R/[b_0 \alfs(Q)]$ for a large $R$.
[See eq.~\eqref{accurate-n*} for a more accurate expression of $n_*(R,Q)$.]
The size of the minimal term $|c_{n_*}\alfs(Q)^{n_*+1}|$ is
order $(\LQ^2/Q^2)^R$.
This behavior is similar to the 
behavior of the usual perturbative series with a singularity at $u=R$
in the Borel plane.
See App.~\ref{App:cond(iii)} for the effect of singularities not located on the
real axis.
The effect is similar and determined by the distances of the
singularities from the origin.
If there is more than one unsuppressed singularities in the Borel plane,
the effect of the one closest to the origin tends to dominate
(depending on the residues of the singularities).
Thus, the FTRS method 
subtracts only the divergent behaviors corresponding to the
singularities that are suppressed by the sine factor, and
others remain.
This feature of the 
FTRS method is in contrast to the standard PV prescription eq.~\eqref{PV-Bl},
which in principle always gives a well-defined value.
Namely, if the condition (iii) is violated,
the series of $[X(Q)]_{\rm FTRS}$ does not converge to the
value of the PV prescription 
but has (typically) an order $(\LQ^2/Q^2)^R$
uncertainty, where $R$ is the distance to the unsuppressed singularity closest to the
origin.

There is another method \cite{Sumino:2004ht,Sumino:2014qpa}
to obtain the FTRS formula eq.~(\ref{XFTRS-decomp})
in closer connection with the Wilsonian picture,
which separates UV and IR contributions by  introducing a cut off
(factorization scale).
It may give some insight into the physical picture of the
renormalon subtraction in the FTRS method.
We present its derivation in App.~\ref{AppD}.

The advantages of our method can be stated as follows.
First, our formulation to subtract renormalons works without knowing 
normalization constants $N_{u_*}$ of the renormalons to be subtracted,
following the above calculation procedure.
In other methods \cite{Lee:2002sn,Ayala:2019uaw,Ayala:2019hkn,Ayala:2020odx,Takaura:2020byt},
in order to subtract renormalons 
one needs normalization constants $N_{u_*}$ of the corresponding renormalons.
Normalization constants of renormalons far from the origin are generally difficult to estimate.
Although we certainly need to know large order perturbative series to improve the accuracy of renormalon-subtracted results,
the above feature of our method practically facilitates subtracting multiple renormalons even with a
small number of known perturbative coefficients. 
Secondly, we can give predictions free from the unphysical singularity 
around $Q \sim \LMS$ caused by the running of the coupling, 
in the same way as the previous study of $V_{\rm QCD}(r)$ \cite{Sumino:2005cq}.
Since renormalons and the unphysical singularity are the main sources destabilizing
perturbative results at IR regions, the removal of these factors 
is a marked feature of our method.

As seen in 
eq.~(\ref{FTdelX}), the Fourier transform 
generates artificial UV renormalons in $\tilde{X}$
at $u=-u'-(2\ell+3)/(2a)$ for $\ell =0, 1, 2,\cdots$.\footnote{
We note that the static QCD potential is actually a special case
where UV renormalons are not generated by Fourier transform. 
This is because the sine factor in eq. (18) actually suppresses UV renormalons as well
for the special parameter choice $(a, u')=(1,-1/2)$, corresponding to the momentum-space potential.
In this case, we do not need a resummation of the momentum-space series.}
They are Borel summable, and we perform the Borel summation 
whenever the induced UV renormalons are located closer to the
origin than the IR renormalons of our interest.
The formula for resummation of these artificial UV renormalons is given in App.~\ref{AppE}.\footnote{
A further study has shown that the formula for the resummation of 
the artificial UV renormalons can be extended to
the  resummation including the UV renormalons in the original perturbative series.
We are now preparing a new paper on the extension of the formula.
}
At the end of App.~\ref{AppE}, 
the formula to resum all of the artificial UV renormalons is also given.

\subsection{Application of FTRS method to simplified models}
\label{Sec2.3}
To demonstrate how the FTRS method works, in this section
we apply renormalon subtraction to two 
simplified series using the FTRS method. 
In this demonstration, we use the beta function at one loop. 
We investigate the result of the FTRS method by truncation of the series 
in the momentum space at a finite order.

First, we study the running coupling constant $\alfs(Q)$, whose perturbative expansion
in $\alpha_s(\mu)$ is given by a geometric series,
to investigate what happens if the FTRS method is applied to a convergent series.
In this approximation, $\alfs(Q)$ is given by
\be
\alfs(Q)=\frac{1}{b_0\log\big(Q^2/\LQ^2\big)}.
\label{as-exact}
\ee
Since the perturbative series of $\alfs(Q)$ in $\alpha_s(\mu)$ is convergent,
the result of the PV prescription is exactly equal to eq.~\eqref{as-exact} 
[$[\alfs(Q)]_{\rm PV}=\alpha_s(Q)$].

When the FTRS method is applied to $\alfs(Q)$, 
the Fourier transform of $\alpha_s(Q)$ is obtained:
\bea
&&~~~~\tilde{\alpha_s}(\tau)=\frac{4\pi}{\tau}\Bigg[\sin(\pi u)\Gamma(2u)\Bigg]_{u=\hat{H}}
\!\!\!\!\alfs(\mu=\tau)\non
&&\approx\frac{4\pi}{b_0\tau}
\Big[1.571 x_\tau - 1.813x_\tau^2 + 
 7.261x_\tau^3 - 50.53x_\tau^4 + 
 377.4x_\tau^5 - 3849x_\tau^6+\cdots\Big]\,,
 \label{tildeas}
\eea
where the parameters $a$ and $u'$ are adjusted to $(a,u')=(1,-1)$ and $x_\tau=b_0\alpha_s(\tau)$.
After the Fourier transform, the UV renormalons are generated at $u=-1/2,\,-3/2,\,\cdots$,
which cause the sign-alternating divergent behavior of $\tilde{\alpha_s}(\tau)$.
We then separate the contribution from these artificial UV renormalons from eq.~\eqref{tildeas}
using the method explained in App.~\ref{AppE}.
The subtracted series is given by
\bea
&&~~~~\tilde{\alpha_s}^{\rm subt.}(\tau)
=\tilde{\alpha_s}(\tau)-(\rm contribution\, from\,  UV\, renormalons)\non
&&\approx\frac{4\pi}{b_0\tau}\Big[0.6247x_\tau + 0.1502x_\tau^2 - 0.6900x_\tau^3 - 2.623x_\tau^4 - 6.308x_\tau^5 - 9.766x_\tau^6+\cdots\Big]\,.
\label{tildeas-subt}
\eea
The subtracted series shows a stable behavior
since there are no UV (or IR) renormalons in eq.~\eqref{tildeas-subt}.\footnote{
After inverse Fourier transform the series becomes even more 
convergent. 
This is because
(i)the integral in eq.~\eqref{X0} generates a power-like suppression 
(we can set an upper bound proportional to $\pi^{-n}$
on the absolute value of the
$n$-th term of the series of $\bar{X}_0$), and
(ii)${\cal O}(1/n!)$ factor stems from the closed path integral in eq.~\eqref{Xpow}.
}

\begin{figure}[t]
\centering
\includegraphics[width=16cm]{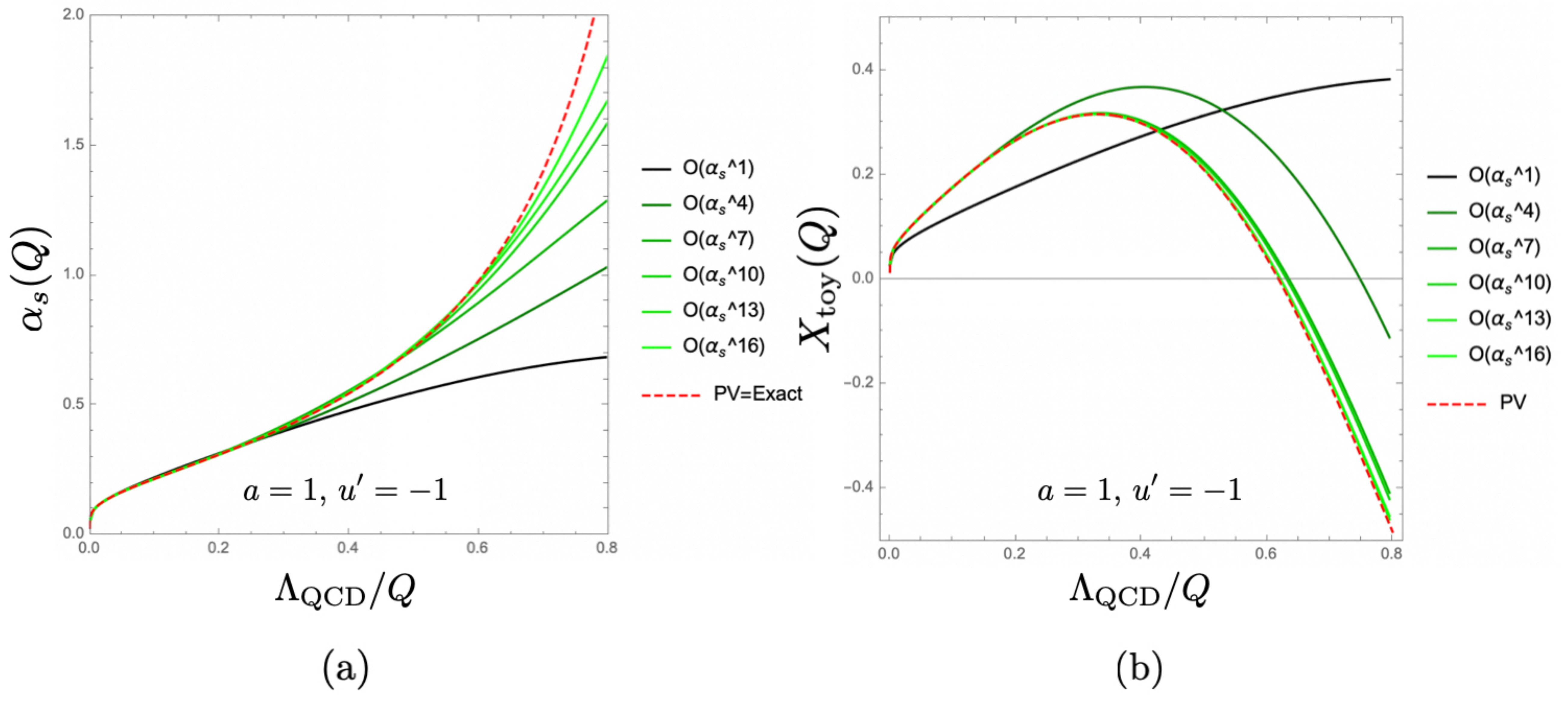}
\caption{\small Comparison of PV prescription (dashed red line) 
and FTRS method (green lines with gradation) for 
(a) $\alfs(Q)$ and 
(b) $X_{\rm toy}(Q)$.
We set $b_0=1$.
The parameters of the Fourier transform are adjusted to $a=1$ and $u'=-1$.
Eqs.~\eqref{tildeas-subt} and \eqref{tildeXtoy-subt} are truncated
at order $\alpha_s(\tau)^k$ ($k=1,\cdots,16$).
}
\label{fig:Xtoy}
\end{figure}

We compare the exact form of $\alfs(Q)$ [eq.~\eqref{as-exact}] (obtained in the PV prescription) 
to that obtained in the FTRS method.
The latter is evaluated as the sum of
the PV-integral of eq.~\eqref{tildeas-subt} by truncating the series
and the resummed UV renormalons.
Fig.~\ref{fig:Xtoy}~(a) shows the result.
The dashed red line is evaluated by the former, 
and the green lines with gradation show the latter.
It can be seen that, as the truncation order increases, 
the result of the FTRS method gradually approaches the result of the PV prescription.

The second toy model is defined by the Borel transform of a divergent series.
The model $X_{\rm toy}(Q)$ is RG invariant 
and its Borel function $B_{X_{\rm toy}}(u)$ is given by,
\be
B_{X_{\rm toy}}(u)=\frac{1}{(1-u)(2-u)}\bigg(\frac{\mu^2}{Q^2}\bigg)^u\,,
\label{Xtoy-Bl}
\ee
which has the IR renormalons at $u=1$ and $u=2$.
The perturbative series of $X_{\rm toy}(Q)$ is given by
\bea
&&[X_{\rm toy}(Q)]_{\rm PT}
=\sum_{n=0}^\infty\frac{2^{n+1}-1}{2^{n+1}}n!b_0^n\alpha_s(Q)^{n+1}\non
&\quad&\approx\frac{1}{b_0}\Big[0.5x_Q+ 0.75x_Q^2 + 1.75x_Q^3 + 5.625x_Q^4 + 23.25x_Q^5 + 118.1x_Q^6 + \cdots\Big]\,,
\label{Xtoy-PT}
\eea
where $x_Q=b_0\alpha_s(Q)$.
The PV prescription [eq.~\eqref{PV-Bl}] defines a finite contribution of $X_{\rm toy}(Q)$,
\bea
[X_{\rm toy}(Q)]_{\rm PV}
&=&\frac{1}{b_0}\int_{0,\rm PV}^\infty du\,e^{-\frac{u}{b_0\alfs(\mu)}}B_{X_{\rm toy}}(u)\non
&=&\frac{e^{-\frac{1}{b_0\alfs(Q)}}}{b_0}{\rm Ei}\Big(\frac{1}{b_0\alfs(Q)}\Big)-\frac{e^{-\frac{2}{b_0\alfs(Q)}}}{b_0}{\rm Ei}\Big(\frac{2}{b_0\alfs(Q)}\Big),
\label{Xtoy-BlPV}
\eea
where ${\rm Ei}(z)=\int_{-z,\rm PV}^\infty dx\,\frac{e^{-x}}{x}$ is
the exponential integral.

We evaluate $X_{\rm toy}(Q)$ using the FTRS method 
and compare it with the above result given by the PV prescription.
To eliminate the renormalon poles at $u=1,\,2$, 
we take the parameters $a=1,\,u'=-1$.
The momentum-space series is given by\footnote{
In separating the artificial UV renormalons from eq.~\eqref{tildeXtoy},
the subtracted part [eq.~\eqref{tildeXtoy-subt}] does not contain the IR renormalons
[see eqs.~\eqref{sep-UVren}--\eqref{elim-IRren} in App.~\ref{AppE}].
}
\bea
&&\tilde{X}_{\rm toy}(\tau)
=\frac{4\pi}{\tau}\Bigg[\sin\big(\pi u\big)\Gamma\big(2u\big)\Bigg]_{u=\hat{H}}\sum_{n=0}^\infty\frac{2^{n+1}-1}{2^{n+1}}n!b_0^n\alpha_s(\tau)^{n+1}\non
&\quad&\approx\frac{4\pi}{b_0\tau}\Big[0.7854x_\tau + 0.2714x_\tau^2 + 3.659 x_\tau^3 - 9.609 x_\tau^4 + 109.1 x_\tau^5 - 1010 x_\tau^6+\cdots\Big]\,,\quad
\label{tildeXtoy}
\eea
where $x_\tau=b_0\alpha_s(\tau)$.
This series does not include IR renormalons but 
it shows a sign-alternating divergent behavior. 
This is because the UV renormalons emerge at $u=-1/2,\,-3/2,\,\cdots$.
We separate the contribution from the UV renormalons from eq.~\eqref{tildeXtoy}
in the same way as in the first example.
The subtracted series is given by
\bea
&&\tilde{X}_{\rm toy}^{\rm subt.}(\tau)
=\tilde{X}_{\rm toy}(\tau)-(\rm contribution\, from\,  UV\, renormalons)\non
&\quad&\approx\frac{4\pi}{b_0\tau}\Big[0.5250x_\tau + 0.8006x_\tau^2 + 1.532x_\tau^3 + 3.179x_\tau^4 + 6.732x_\tau^5 + 13.82x_\tau^6+\cdots\Big]\,.\quad
\label{tildeXtoy-subt}
\eea
Indeed, the series shows a better convergence than eq.~\eqref{Xtoy-PT}
in the momentum space,
owing to the removal of the IR renormalons and the separation of the UV renormalons.
The subtracted UV renormalons should be resummed according to App.~\ref{AppE}.

We compare the $X_{\rm toy}(Q)$ obtained in the PV prescription [eq.~\eqref{Xtoy-BlPV}]
to that obtained in the FTRS method.
The latter is evaluated as the sum of 
the PV-integral of eq.~\eqref{tildeXtoy-subt} by truncating the series
and the resummed UV renormalons.
Fig.~\ref{fig:Xtoy}~(b) shows the result.
The dashed red line is evaluated by the former, 
and the green lines with gradation correspond to the latter.
It can be seen that, as the truncation order increases, 
the result of the FTRS method gradually approaches the result of the PV prescription.

\begin{figure}[t]
\centering
\includegraphics[width=9cm]{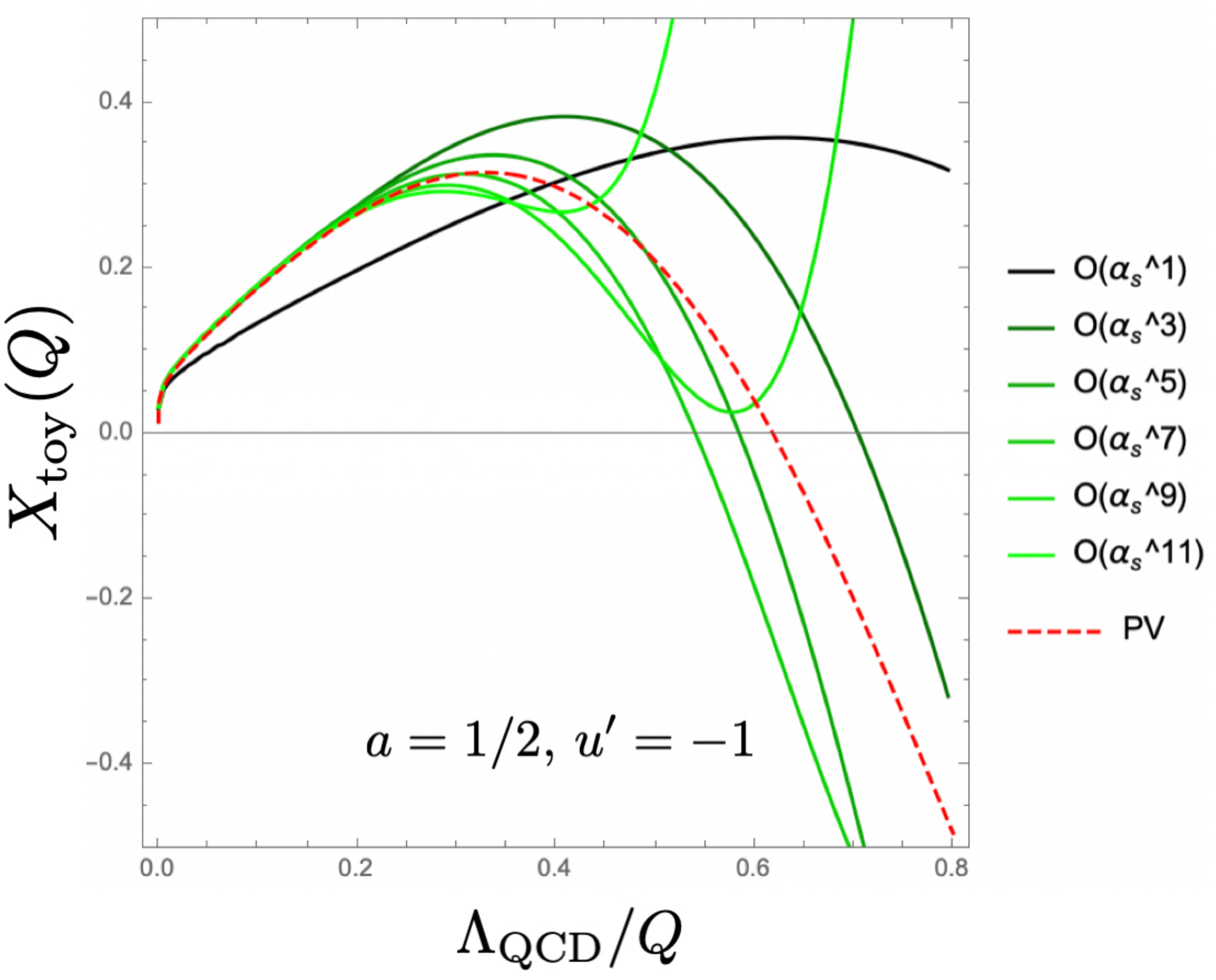}
\caption{\small Comparison of PV prescription (dashed red line) 
and FTRS method (green lines with gradation) for $X_{\rm toy}(Q)$.
We set $b_0=1$.
The parameters of the Fourier transform are adjusted to $a=1/2$ and $u'=-1$
such that the $u=2$ renormalon remains unsuppressed.
The momentum-space series is truncated
at order $\alpha_s(\tau)^k$ ($k=1,\cdots,11$).
}
\label{fig:Xtoymis}
\end{figure}

We emphasize two important points here.
The first is to choose the parameters $a$ and $u'$ correctly.
Fig.~\ref{fig:Xtoymis} shows the result of the FTRS method for $X_{\rm toy}$ 
when the parameters are adjusted to $(a,u')=(1/2,-1)$ so as to
suppress in the momentum space the leading ($u=1$) renormalon
but not the subleading ($u=2$) renormalon.
If the parameters are chosen incorrectly such that unsuppressed renormalons remain
in the momentum-space series,
the result of the FTRS method starts to diverge from a certain order,
consistently with the estimate discussed below eq.~\eqref{delXFTRS}
and in App.~\ref{App:cond(iii)}.\footnote{
For the quantity whose perturbative series is convergent such as $\alpha_s(Q)$,
any adjustment of $(a,u')$ does not ruin the convergence of FTRS method
since no IR renormalon remains in the momentum space.
We confirm the convergence of the FTRS-method result for $\alpha_s(Q)$
with the parameters adjusted to $(a,\,u')=(1/2,-1)$.
}
This situation can be avoided by a
proper adjustment of the parameters $a$ and $u'$.

The other is the resummation of the artificial UV renormalons.
When UV renormalons are present, sign-alternating divergent behavior appears as in eq.~\eqref{tildeXtoy}.
The UV renomalons can be resummed thanks to the full knowledge of its residue.
If the contribution of the UV renormalons is not resummed, 
a truncation analysis as in Fig.~\ref{fig:Xtoy} does not converge to the result of the PV prescription.
\vspace{5mm}

From Sec.~3 to Sec.~5,
we will test the validity of the above method 
(FTRS method) by applying it to the following observables:
the ${\cal O}(\LQ^4)$ renormalon of the Adler function; 
the ${\cal O}(\LQ^2)$ renormalon of the $B\to X_ul\bar{\nu}$; 
the ${\cal O}(\LQ)$ and ${\cal O}(\LQ^2)$ renormalons simultaneously of the $B$ or $D$ meson mass. 
The renormalons of ${\cal O}(\LQ^2)$ in the $B$ decay and the $B$, $D$ meson masses 
are subtracted for the first time in this paper. 
We show that our results meet theoretical expectations, 
e.g., good convergence and consistent behaviors with the OPE.

\section{Observable I : Adler function}

We perform theoretical tests of the FTRS method using different observables.
(The formulas for the perturbative coefficients necessary for the analyses
are collected in App.~\ref{App:PertCoeffs}.)
The first observable is the Adler function.

\subsection{OPE and renormalon}

Consider the hadronic contribution to
the photon vacuum polarization function, $\Pi_h(Q^2)$, in the Euclidean region:
\be
(q_\mu q_\nu - g_{\mu\nu} q^2)
\Pi_h(Q^2)=-i\int d^4x\,e^{-iq\cdot x}\langle J_\mu(x)J_\nu(0)\rangle
~~~;~~~
Q^2=-q^2>0 \,,
\ee
where
\be
J_\mu(x)=\sum_f Q_f\overline{\psi}_f(x)\gamma_\mu\psi_f(x)=\frac{2}{3}\overline{u}(x)\gamma_\mu u(x)-\frac{1}{3}\overline{d}(x)\gamma_\mu d(x)+\cdots
\ee
denotes the electromagnetic current of quarks.
The Adler function is defined by  
\be
D(Q^2)=12\pi^2Q^2\frac{d \Pi_h(Q^2)}{dQ^2}.
\ee
Its OPE for $Q^2 \gg \LQ^2 $ is given by
\be
D_{\rm OPE}(Q^2)=X^D(Q^2)+2\pi^2\Big(\sum_{f}Q_f^2\Big)X^D_{GG}(Q^2)\frac{\langle G^2\rangle}{Q^4}+\cdots.
\ee
Throughout the analysis of the Adler function, 
we set the number of active quark flavors to $n_f=2$ 
and the quark masses to zero, $m_u=m_d=0$.
In this case
we can ignore the effect of the quark condensate 
$m_q\langle \bar{q}q\rangle/Q^4$ in the OPE of the Adler function.
The leading Wilson coefficient can be computed in perturbative QCD as
\bea
&&
X^D(Q^2)\equiv X_{\bf 1}^D(Q^2)\non
&&=3\biggl(\sum_f Q_f^2\biggr)\left(1+\sum_{n=0}^N a^{NS}_n\alpha_s(Q^2)^{n+1}\right)
+3\biggl(\sum_f Q_f\biggr)^2\left(\sum_{n=2}^N a^{SI}_n\alpha_s(Q^2)^{n+1}\right).
\eea
The coefficients $a^{NS}_n,\,a^{SI}_n$ are known 
up to $N=3$ (up to ${\cal O}(\alfs^4)$) \cite{Baikov:2010je,Baikov:2012zn}.
The non-perturbative matrix element in
the second term in the OPE is known as the local gluon condensate
\be
\langle G^2\rangle\equiv-\langle0|\frac{\beta(\alpha_s)}{\pi b_0\alfs}G_{\mu\nu}G^{\mu\nu}|0\rangle
=\langle0|\frac{\alpha_s}{\pi }G_{\mu\nu}G^{\mu\nu}|0\rangle\Big(1+{\cal O}(\alpha_s)\Big).
\ee
(We define it in the same manner as, for instance, in Ref.~\cite{Ayala:2020pxq}.)
The Wilson coefficient of $ G^2$ is given by
\be
2\pi^2\Big(\sum_{f}Q_f^2\Big)X^D_{GG}(Q^2)=\frac{10\pi^2}{9}\Big(1+{\cal O}(\alfs(Q))\Big),
\ee
which is known up to order $\alfs$ \cite{Chetyrkin:1985kn}.

The Adler function is also related to 
the $R$-ratio,
\be
R(s)=\frac{\sigma(e^+e^-\to\rm hadrons)}{4\pi\alpha_{EM}(s)/(3s)}\,,
\ee
by
\be
D(Q^2)=Q^2\int_0^\infty ds\,\frac{R(s)}{(s+Q^2)^2}\,,
\ee
which follows from the dispersion relation.
Since the $R$-ratio is an experimentally measurable quantity,
a non-perturbative determination of the Adler function is
possible through this relation.

In principle, we can determine the value of
the local gluon condensate $\langle G^2\rangle$ 
by comparing such a non-perturbative determination with the OPE.
Since, however, the perturbative prediction 
of the
leading Wilson coefficient $X^D(Q^2)$
contains a renormalon
of order $\LQ^4$, it generates
an ambiguity of the same order of magnitude as $\langle G^2\rangle$ itself.
This effect prevents any accurate
determination of $\langle G^2\rangle$ and calls for the subtraction of
the corresponding renormalon.

In this analysis, 
in order to determine a well-defined $\langle G^2\rangle$, 
we separate the order ${\cal O}(\LQ^4)$ renormalon 
(corresponding to $u=2$)
from $X^D(Q^2)$ by the FTRS method,
\be
D_{\rm OPE}(Q^2)=X^D_{\rm FTRS}(Q^2)+\frac{10\pi^2}{9}\frac{\langle G^2\rangle_{\rm FTRS}}{Q^4}.
\ee
We use the LO result $X^D_{GG}(Q^2)=1$ for simplicity.
In this manner the local gluon condensate $\langle G^2\rangle_{\rm FTRS}$
(which should coincide with the usual PV prescription at high orders) is 
defined free from the dominant renormalon.
Below we estimate the value of $\langle G^2\rangle_{\rm FTRS}$
using a phenomenological model of the $R$-ratio as an input.


\subsection{\boldmath
FTRS formula and phenomenological model}

We construct
the renormalon-subtracted leading Wilson coefficient 
in the FTRS method as follows.
The perturbative coefficients of the Adler function are given by
\be
X^D(Q^2)=3\sum_f Q_f^2+\sum_{n=0}^3 a_n\alpha_s(Q)^{n+1},
\ee
\bea
a_n&=&3\sum_f Q_f^2a_n^{NS}+3\bigg(\sum_fQ_f\bigg)^2a_n^{SI}=\frac{5}{3}a_n^{NS}+\frac{1}{3}a_n^{SI}\quad{\rm for}\,\,\,n_f=2.
\eea
The formulas for the necessary perturbative coefficients are collected
in App.~\ref{App:PertCoeffs}.
We choose the parameters $(a,u')=(1/2,-2)$ in the Fourier transform
such that the
renormalons in the artificial momentum space ($\tau$ space)
are suppressed at $u=2,4,6,\cdots$.
According to eqs.~\eqref{XFTRS-decomp}--\eqref{Xpow},
the FTRS Wilson coefficient can be decomposed as
\be
X^D_{\rm FTRS}(Q^2)=\frac{5}{3}+\bar{X}_{0}^D(Q^2)+\bar{X}_{\rm pow}^D(Q^2),
\label{decomp-X-Adler}
\ee
with
\be
\bar{X}^D_{0}(Q^2)=\frac{-1}{2\pi^2Q^2}\int_0^\infty dt \,t\, e^{-t/Q^2}{\rm Im}\,[\tilde{X}^D(\tau=it)],
\ee
\be
\bar{X}^D_{\rm pow}(Q^2)=\frac{1}{4\pi^2 iQ^2}\int_{C_{*}} d\tau\,\tau\, \tilde{X}^D(\tau).
\ee
$\bar{X}^D_{\rm pow}(Q^2)$ is given as a power series 
$\LMS^2/Q^2,\,\LMS^6/Q^6\,,
\LMS^{10}/Q^{10},\cdots$, and
we truncate it at ${\cal O}(\LMS^2/Q^2)$. 
The Fourier transform generates artificial UV renormalons in $\tilde{X}^D(\tau)$, 
which can be resummed by the formula in App.~\ref{AppE}.
With our setup, the
UV renormalons at $u=-1,-3,-5,\cdots$ are generated.
After resummation of the artificial $u=-1$ renormalon,
$\tilde{X}^D(\tau)$ in the $\rm N^3LL$ approximation is calculated as\footnote{
To resum UV renormalons we separate the series into two parts.
Although the sum of the two parts is free of IR renormalons, they appear
in each part.
}

\bea
\tilde{X}^D(\tau)\biggr|_{\rm N^3LL}
&=&\frac{4\pi}{\tau}\Biggl[
\biggl[\sin\Big(\frac{\pi}{2}u_*\Big)\Gamma(u_*)-\frac{1}{u_*+1}
\biggr]_{u_*\to\hat{H}}\sum_{n=0}^\infty a_n\alpha_s(\sqrt{\tau})^{n+1}
\Biggr]_\text{up to ${\cal O}(\alfs(\sqrt{\tau})^4)$}
\nonumber\\
&&+\frac{8\pi}{\tau}\int_0^1{dv}\, v\sum_{n=0}^3a_n\alpha_s(\sqrt{\tau}/v)^{n+1}\nonumber\\
&\approx&\frac{4\pi}{\tau}\Bigg[0.3028  a_\tau + 0.2073 a_\tau^2 + 0.3177  a_\tau^3 + 0.6159 a_\tau^4 \nonumber\\
&\quad\quad&+\int_0^1 dv\, 2v\Big(0.5305 a_{\tau, v} + 0.2964 a_{\tau, v}^2 +0.5415a_{\tau, v}^3 + 1.253 a_{\tau, v}^4\Big)\Bigg]\non
&\equiv&\frac{4\pi}{\tau}
\Bigg[\sum_{n=0}^3\tilde{a}_n^{\rm subt}a_\tau^{n+1}+\int_0^1dv\,2v\sum_{n=0}^3{a}_na_{\tau,v}^{n+1}\Bigg],
\label{FTRSformula-AdlerFn}
\eea
where $a_\tau=\alfs(\sqrt{\tau})$, $a_{\tau, v}=\alfs(\sqrt{\tau}/v)$.
In the numerical analysis below, we vary the renormalization scale $\mu$ of eq.~\eqref{FTRSformula-AdlerFn} 
between $[\sqrt{\tau}/2,2\sqrt{\tau}]$ to
investigate the scale dependence.
Since eq.~\eqref{FTRSformula-AdlerFn} is RG invariant,
$\tilde{X}^D(\tau)|_{\rm N^3LL}$ for $\mu=2\sqrt{\tau}$ and $\mu=\sqrt{\tau}/2$ are given by\footnote{
In the numerical analyses in Secs. 4 and 5,
we investigate the scale dependence in the same manner.
In the N$^k$LL approximation, 
terms up to ${\cal O}(\alfs^{k+1})$ are used.
}
\bea
\tilde{X}^D(\tau)\biggr|_{\rm N^3LL}^{\mu=2\sqrt{\tau}}\!\!
&=&\frac{4\pi}{\tau}
\Bigg[\bigg(\frac{\mu^2}{\tau}\bigg)^{\hat{H}}\sum_{n=0}^3\tilde{a}_n^{\rm subt}\alfs(\mu)^{n+1}\Bigg]^{\mu=2\sqrt{\tau}}_\text{up to ${\cal O}(\alfs(2\sqrt{\tau})^4)$}\non
&\quad\quad&+\int_0^1dv\,2v\Bigg[\bigg(\frac{\mu^2}{\tau/v^2}\bigg)^{\hat{H}}\sum_{n=0}^3{a}_n\alfs(\mu)^{n+1}\Bigg]^{\mu=2\sqrt{\tau}/v}_\text{up to ${\cal O}(\alfs(2\sqrt{\tau}/v)^4)$},
\label{FTRSformula-AdlerFn-twice}
\eea
\bea
\tilde{X}^D(\tau)\biggr|_{\rm N^3LL}^{\mu=\sqrt{\tau}/2}\!\!
&=&\frac{4\pi}{\tau}
\Bigg[\bigg(\frac{\mu^2}{\tau}\bigg)^{\hat{H}}\sum_{n=0}^3\tilde{a}_n^{\rm subt}\alfs(\mu)^{n+1}\Bigg]^{\mu=\sqrt{\tau}/2}_\text{up to ${\cal O}(\alfs(\sqrt{\tau}/2)^4)$}\non
&\quad\quad&+\int_0^1dv\,2v\Bigg[\bigg(\frac{\mu^2}{\tau/v^2}\bigg)^{\hat{H}}\sum_{n=0}^3{a}_n\alfs(\mu)^{n+1}\Bigg]^{\mu=\sqrt{\tau}/(2v)}_\text{up to ${\cal O}(\alfs(\sqrt{\tau}/(2v))^4)$}.
\label{FTRSformula-AdlerFn-half}
\eea

Next, we explain a phenomenological model of the Adler function.
In ref.~\cite{Bernecker:2011gh}
a phenomenological model for the $R$-ratio,
$R_{\rm model}(s)$, is constructed from experimental data. 
The formula for $R_{\rm model}(s)$ is summarized in App.~\ref{App:PhenoModel}.\footnote{
We adjust the original model, which was constructed
for $n_f=3$, to the $n_f=2$ case.
}
We define the Adler function constructed from $R_{\rm model}(s)$ as 
\be
D_{\rm pheno}(Q^2)=Q^2\int_0^\infty ds\,\frac{R_{\rm model}(s)}{(s+Q^2)^2}.
\ee

\subsection{Consistency checks and
estimate of $\langle G^2\rangle_{\rm FTRS}$}

In this section, we present consistency checks of the OPE in the
FTRS method and estimate
the local gluon condensate $\langle G^2\rangle_{\rm FTRS}$.
We use $D_{\rm pheno}(Q^2)$ as a reference.
Throughout the test analyses (also in Secs.~4, 5),
we evaluate the running coupling constant 
by solving the RGE numerically with the 5-loop beta function \cite{Baikov:2017ayn}.

The OPE prediction is given by
\be
D_{\rm OPE}(Q^2;\LMS,\langle G^2\rangle)=X^D_{\rm FTRS}(Q^2)+\frac{10\pi^2}{9}\frac{\langle G^2\rangle}{Q^4},
\ee
where $\langle G^2\rangle$ and $\LMS$ are taken as the fitting parameters.\footnote{
In ref.~\cite{Hayashi:2020ylq}, 
we chose $\LMS$ and $A= 2\pi^2 (\sum_f Q_f^2) X^D_{GG} \langle G^2 \rangle/\Lambda_{\overline{\rm MS}}^4 $ as the fitting parameters 
while we choose $\Lambda_{\overline{\rm MS}}$ and $\langle G^2\rangle$ 
as the fitting parameters here. 
We obtain $A=-25(11)$~\cite{Hayashi:2020ylq} and 
$\langle G^2\rangle=-0.00123(10)$. 
Although the central values are mutually consistent 
(by converting one into the other assuming the LO Wilson coefficient $X^D_{GG}$), 
the error sizes are largely different; the relative error sizes are 40\% and 15\% 
for $A$ and $\langle G^2\rangle$, respectively. 
We estimate that the large error for $A$ is introduced due to the large uncertainty of 
$\Lambda_{\overline{\rm MS}}$, 
which is contained in $A$ as $\Lambda_{\overline{\rm MS}}^{-4}$. 
We also estimate the impact of the one-loop correction of 
$X^D_{GG}$~\cite{Chetyrkin:1985kn} on the gluon condensate 
$\langle G^2\rangle$.  We find that it can shift the value by about 10 \%.
}
$X^D_{\rm FTRS}(Q^2)$ depends on $\LMS$ through the
running coupling constant.
Here,
\be
\LMS=\LMS^{n_f=2},~~~
\langle G^2\rangle=\langle G^2\rangle_{\rm FTRS}^{n_f=2}.
\ee
We include $\LMS^{n_f=2}$ in the fitting parameters.
We match $D_{\rm OPE}(Q^2)$ and $D_{\rm pheno}(Q^2)$
in the range
between $Q^2=0.6\,\rm GeV^2$ and $Q^2=2\,\rm GeV^2$ 
($0.8\,{\rm GeV}\lesssim Q\lesssim1.4\,{\rm GeV}$).
The result of the fit is given by
\bea
\LMS^{n_f=2}=0.271(39)\,{\rm GeV},~~~
\langle G^2\rangle_{\rm FTRS}^{n_f=2}=-0.0123(10)~{\rm GeV}^4
.
\label{AdlerFitRes}
\eea
The error is estimated only from the scale dependence of the FTRS method, 
where the scale $\mu$ is varied between $\mu=2\sqrt{\tau}$ and $\mu=\sqrt{\tau}/2$. 

\begin{figure}[t]
 \begin{minipage}[t]{0.49\hsize}
  \begin{center}
   \includegraphics[width=81mm]{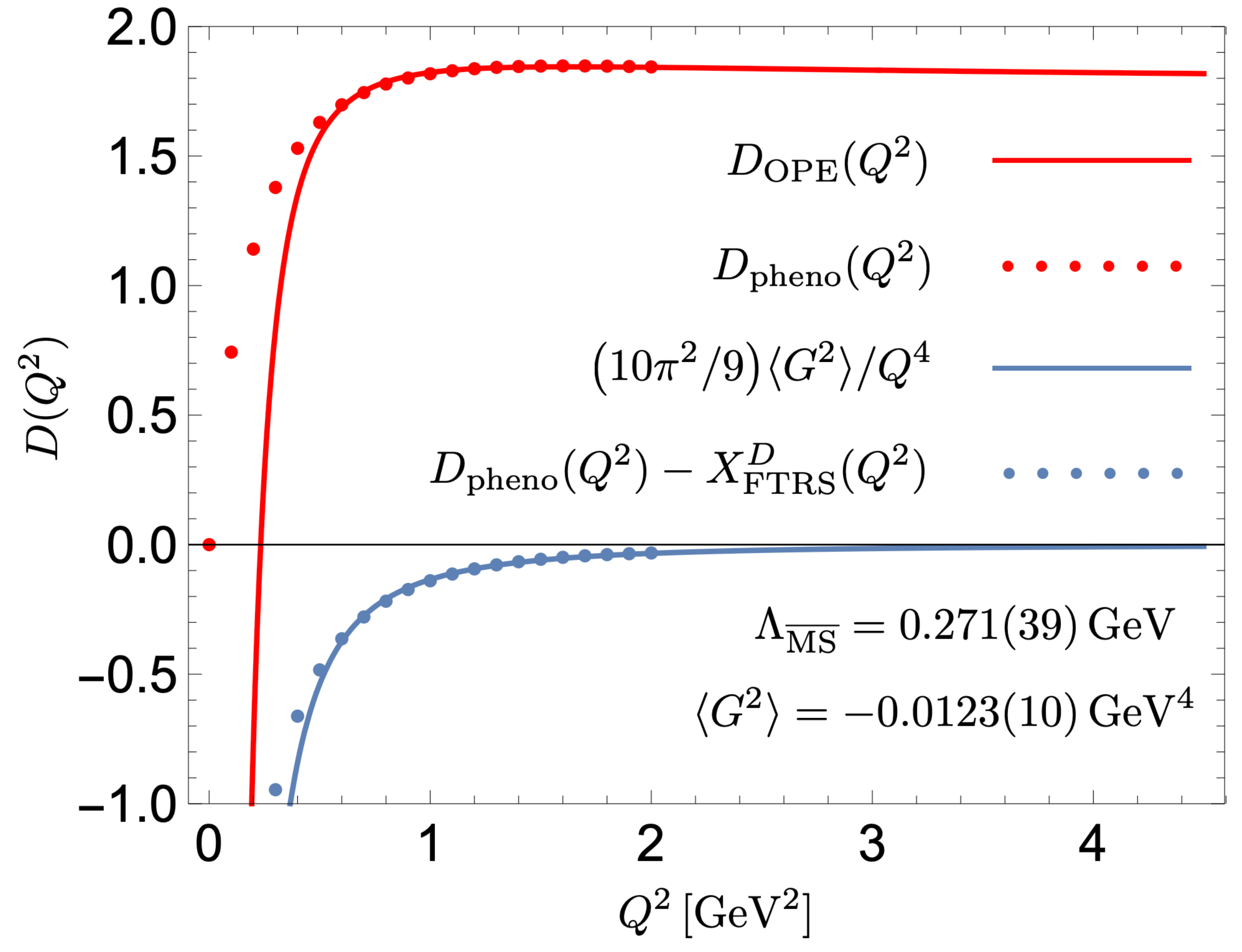}
  \end{center}
  \caption{\small Comparison of $D_{\rm OPE}(Q^2)$ (red solid) and 
  $D_{\rm pheno}(Q^2)$ (red dots). 
  The former is determined by fitting to the latter
  in  the range $0.6 \leq Q^2\leq 2.0\,{\rm GeV^2}$. 
  Also $(10\pi^2/9)\langle G^2\rangle/{Q^4}$  (blue solid)
  and the difference $D_{\rm pheno}(Q^2)-X^D_{\rm FTRS}(Q^2)$
 (blue dots) are compared.
  }
  \label{phenoOPE}
 \end{minipage}
 \hfill
 \begin{minipage}[t]{0.49 \hsize}
  \begin{center}
   \includegraphics[width=81mm]{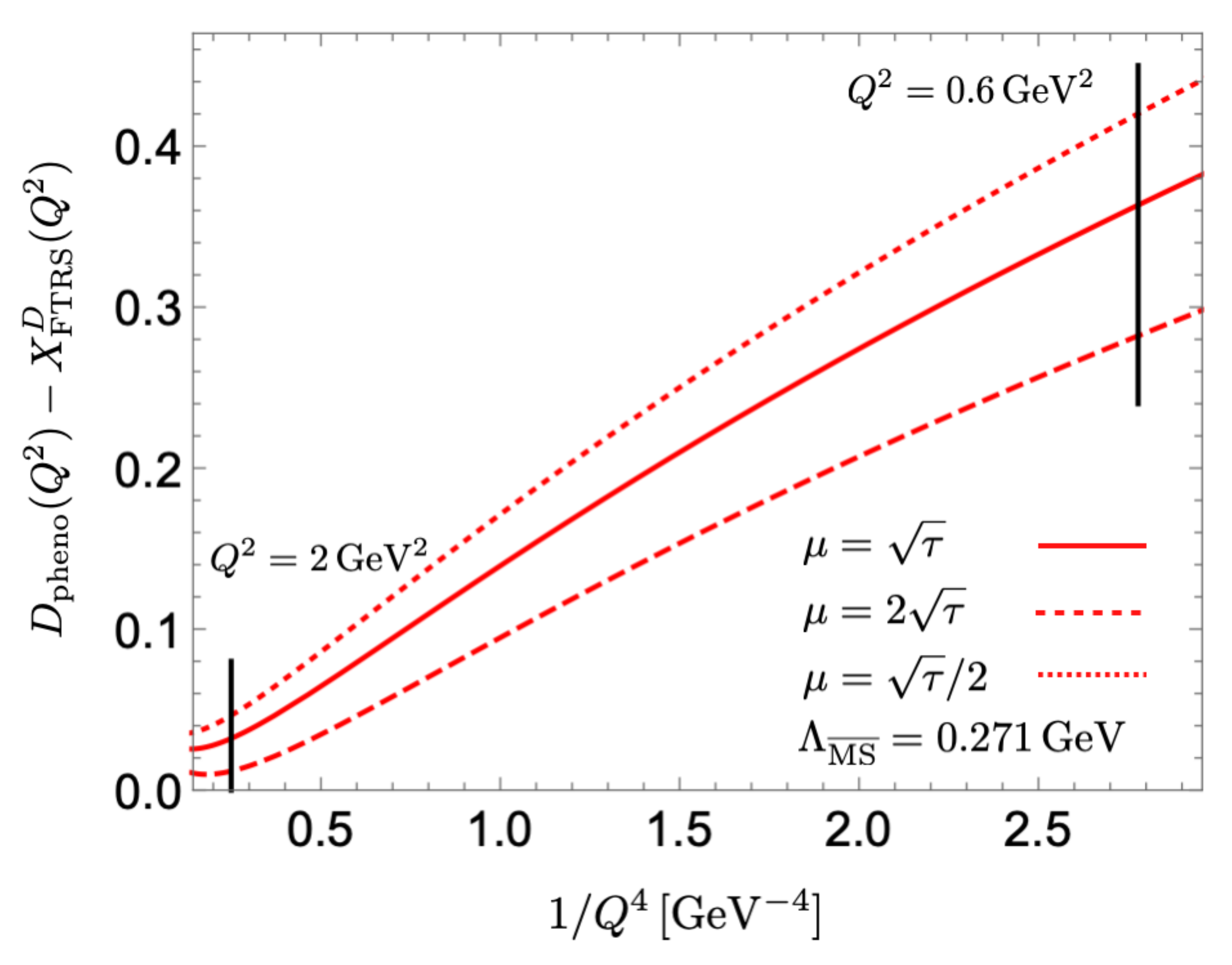}
  \end{center}
  \caption{\small 
  $D_{\rm pheno}(Q^2)-X^D_{\rm FTRS}(Q^2)$ vs.\ $1/Q^4$ 
  with $\LMS=0.271\,{\rm GeV}$.
  The solid, dashed, and dotted lines correspond to 
  $\mu=\sqrt{\tau},2\sqrt{\tau},\sqrt{\tau}/2$, respectively.
  Vertical short lines show the boundaries of the range used for the fit.
  In this range, the behavior is almost proportional to $1/Q^4$, which is consistent with OPE.}
  \label{glcon}
 \end{minipage}
\end{figure}
\begin{figure}[h]
        \centering
        \includegraphics[width=16cm]{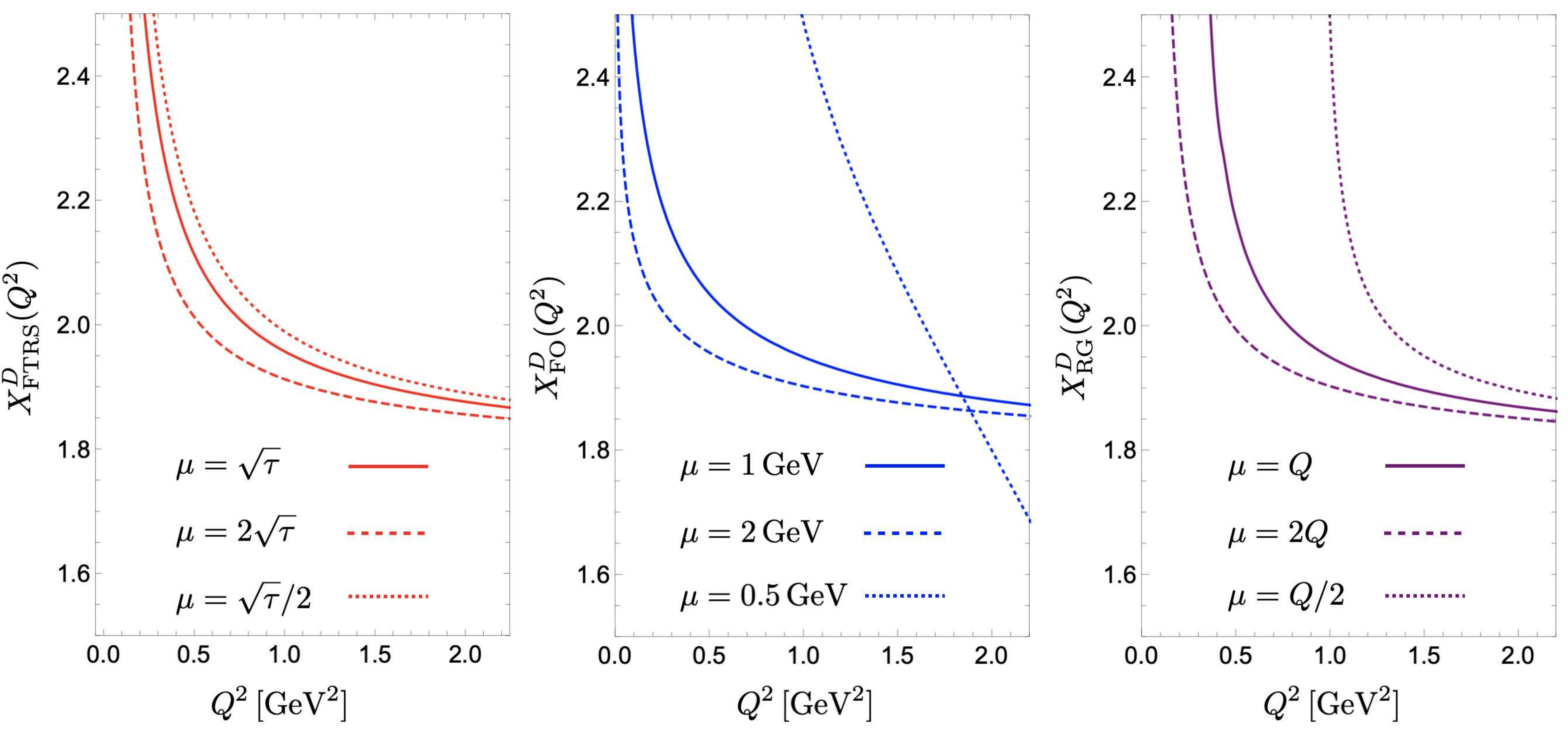}
        \caption{\small Comparison of $X^D_{\rm FTRS}(Q^2)$, $X^D_{\rm FO}(Q^2)$ and $X^D_{\rm RG}(Q^2)$ for $\LMS=0.271 \,{\rm GeV}$.
        The dashed and dotted lines correspond to the variation of the scale
        by a factor 2 and 1/2, respectively, from its standard value (solid).
}
        \label{hikaku}
\end{figure}

Figs.~\ref{phenoOPE} and \ref{glcon} show the results of the fit. 
Within the range of matching, 
we observe an overall consistency of the OPE $D_{\rm OPE}(Q^2)$ and
the reference $D_{\rm pheno}(Q^2)$.
More in detail, we see that
$D_{\rm pheno}(Q^2)-D_{\rm FTRS}(Q^2)$ is almost
proportional to $1/Q^4$ 
in this range, consistently with the OPE.

In Fig.~\ref{hikaku} we compare the leading Wilson coefficient in
the FTRS method, $X^D_{\rm FTRS}(Q^2)$, with the
fixed-order (FO) calculation $X^D_{\rm FO}(Q^2)$ and 
RG-improved calculation $X^D_{\rm RG}(Q^2)$,
where we take $\LMS=0.271 \,{\rm GeV}$.
The IR renormalons are not subtracted from the latter two quantities.
Explicitly, we define
\be
X^D_{\rm FO}(Q^2)=\frac{5}{3}+\sum_{n=0}^3 a_n(\mu_0/Q)\alpha_s(\mu_0)^{n+1},
~~X^D_{\rm RG}(Q^2)=\frac{5}{3}+\sum_{n=0}^3 a_n(1)\alpha_s(Q)^{n+1}.
\ee 
In the FO calculation, we choose the renormalization scale $\mu_0=1\,\rm GeV$.
In each figure we vary the scale by a factor 2 and 1/2 to test the stability of the
prediction.
The scale dependence of $X^D_{\rm FTRS}(Q^2)$ is considerably smaller than 
the other two in the range 
$0.6\,{\rm GeV^2}\leq Q^2\leq2\,{\rm GeV^2}$.
This is consistent with the expectation that
convergence of the perturbative series improves 
by subtraction of the $u=2$ IR renormalon.
The removal of the unphysical singularity of $\alfs(Q)$ also plays
a significant role.

The value of $\LMS^{n_f=2}$ in eq.~\eqref{AdlerFitRes} 
can be compared with the determination by 
lattice QCD simulations from various observables 
\cite{Aoki:2019cca} ($\LMS^{n_f=2}=330^{+21}_{-63}$~MeV).
Although
our first test 
analysis is fairly crude, with unknown uncertainties contained in the model
cross section
(including our naive adaptation to the $n_f=2$ case), etc., 
it is interesting that we observe a rough consistency
with today's world-average value.

As seen in Fig.~\ref{hikaku} the LO Wilson coefficient
increases as $Q^2$ is reduced
(in all of the FTRS, FO and RG-improved calculations).
Oppositely, $D_{\rm pheno}(Q^2)$ decreases as $Q^2$ is reduced.
(See Fig.~\ref{phenoOPE}.)
The latter behavior is a natural consequence of the resonance
shape of the $R$-ratio $R(s)$ in the low $s$ region.
In the above matching of the OPE of the Adler function
with $D_{\rm pheno}(Q^2)$, this behavior is 
caused by the $\langle G^2\rangle/Q^4$ term.
Hence, the sign of $\langle G^2\rangle$ is determined to be
negative by the fit.


\begin{table}[t]
\begin{center}
  \begin{tabular}{c|c|c}
  \hline
    Range of $Q^2$ [GeV$^2$] & $\LMS$ [GeV] & $\langle G^2\rangle$  \\
   \hline
    $[0.6,\,2.0]$ & $0.271(39)$ & $-0.0123(10)$  \\
    $[0.6,\,1.0]$ & $0.247(38)$ & $-0.0106(7)$  \\
    $[1.0,\,2.0]$ & $0.285(38)$ & $-0.0141(17)$  \\
    $[0.9,\,1.5]$ & $0.289(42)$ & $-0.0146(14)$  \\
   \hline
  \end{tabular}
\caption{\small
Results of the fit as we vary the range of the matching of 
$D_{\rm OPE}(Q^2)$ and $D_{\rm pheno}(Q^2)$.
The errors correspond to the change of scale by a factor of 2 or 1/2
from $\mu=\sqrt{\tau}$.
}
\label{tab:VaryFitRangeAdler}
  \end{center}
\end{table}

We test if the fit result is sensitive to the range we adopt for the matching.
We vary the range inside the above range $Q^2\in [0.6,\,2]$~(GeV$^2$);
see Tab.~\ref{tab:VaryFitRangeAdler}.
The central value of $\LMS$ varies inside its error in eq.~\eqref{AdlerFitRes}.
On the other hand, in Tab.~\ref{tab:VaryFitRangeAdler}, 
the values of $\langle G^2\rangle$ marginally overlap with that in
eq.~\eqref{AdlerFitRes}
within the respective errors.
However, the central value
varies outside the error in eq.~\eqref{AdlerFitRes}.
Therefore we need to assign a systematic error of 
(at least) about $0.0015\,{\rm GeV^4}$ to $\langle G^2\rangle$.
Thus, if we take into account this size of systematic error for
$\langle G^2\rangle$,
the OPE is consistent inside the tested range.

In order to estimate the effects of higher order corrections,
we calculate $X^D_{\rm FTRS}$ in the $\rm N^4LL$ approximation 
with the 5-loop perturbative coefficient, which we  
estimate using the large-$\beta_0$ approximation for
$a_4^{\beta_0}(\sqrt{\tau})$
\cite{Beneke:1992ch}
and the 5-loop $\beta$ function \cite{Baikov:2017ayn} for $\hat{H}$. 
Then we perform a fit to estimate $\LMS$ and $A$. 
In the case that 
the $u=-1$ UV renormalon contribution is subtracted from $a_4^{\beta_0}$,
we obtain
\bea
\LMS^{\rm (est)}=0.264(25)\,{\rm GeV},~~~
\langle G^2\rangle^{\rm (est)}=-0.0125(8)\,{\rm  GeV^4}\,. \quad ({\rm N^4LL\,\,estimates})
\eea
The scale dependence becomes smaller than that in the $\rm N^3LL$ approximation.
On the other hand, if $a_4^{\beta_0}$ is used without subtraction 
of the $u=-1$ UV renormalon contribution, 
the fit result is given by 
$\LMS^{\rm (est)}=0.256(30)\,{\rm GeV},\,\langle G^2\rangle^{\rm (est)}=-0.0127(10)\,{\rm  GeV^4}$.
This may indicate
that it is important to deal with the $u=-1$ UV renormalon 
to improve accuracy by going to higher orders. 

Let us discuss the convergence properties of the higher order
corrections.
$X^D_{\rm FTRS}(Q)$ can be separated into two parts as in 
eq.~(\ref{decomp-X-Adler}).
We concentrate on the leading power correction $\sim\LMS^2/Q^2$, 
which seems to limit the accuracy of the fit.
With N$^k$LL prediction the scale dependence cancels up to
${\cal O}(\alfs^{k+1})$ and the residual scale dependence 
can be estimated as
\be
\Delta X^D_{\rm FTRS}(Q)\Big|_{\rm N^kLL}\sim
\frac{1}{\pi i Q^2}\int_{C_{*}}
\!\!d\tau \,b_0^{k+1}\alpha_s(\sqrt{\tau})^{k+2}
\sim \frac{1}{2 b_0}\Bigg(\frac{\LMS}{Q}\Bigg)^{2}\frac{1}{(k+1)!}.
\label{errorFTRS}
\ee
We assume that, if we change the scale by a factor 2, the coefficient of
the ${\cal O}(\alfs^{k+2})$ term varies by order one,
and the LL running coupling is used to make a rough estimate,
\be
b_0\alpha_s(\sqrt{\tau})\simeq \frac{1}{\log(\tau/\LMS^2)}.
\ee
Thus, $\Delta X^D_{\rm FTRS}(Q^2)\big|_{\rm N^kLL}$ 
is expected to be more suppressed at higher orders.
Our results above 
and $\rm N^4LL$ estimation are consistent with this expectation.
However, 
we need to develop a method for resummation of the UV renormalon
in the original series for this argument to be valid at high orders.

Finally, we comment on the relation of our result for the
local gluon condensate with other
previous determinations.
Unfortunately all the other previous determinations in the PV scheme
are performed in the quenched approximation ($n_f=0$), so they
cannot be compared directly to our result.
For instance,
recent determinations by refs.~\cite{Bali:2014sja}, \cite{Lee:2015bci}
and \cite{Ayala:2020pxq}, respectively,
give $\langle G^2 \rangle^{n_f=0}=0.077(87)$,
$0.14$ and $0.076(4)$~GeV$^4$.
All of them use the lattice plaquette action and 
the latter two
subtracted the ${\cal O}(\LQ^4)$ renormalon.

\section{\boldmath Observable II : $B$ semileptonic decay width \\
$\Gamma(B\to X_u l\overline{\nu})$}
\label{Sec.4}

In this section, we apply the FTRS method to the OPE of 
$\Gamma(B\to X_u l\overline{\nu})$, the 
$B$ meson partial
decay width for the charmless semileptonic channel. 
First,  we review the OPE in HQET
and the $u=1/2$ renormalon cancellation by change of mass scheme 
from the pole mass to  the $\msbar$ mass.
Secondly, we explain how to subtract the $u=1$ renormalon by 
the FTRS method.
Finally, we examine the effects of renormalon subtraction.
Through the analysis in Sec.~\ref{Sec.4}, we set $m_c=0$ in loop corrections
for simplicity.

\subsection{\boldmath OPE in HQET and $u=1/2$ renormalon cancellation}
\label{Sec.4.1}

HQET is an effective field theory for describing dynamics of a 
heavy meson $H$, which is a
bound state of a heavy quark $h$ and a light quark $l$.
The mass of $h$ is assumed to be much larger than the
QCD scale, $m_h\gg\LQ$.
In this theory  \cite{Manohar:2000dt}
the heavy quark field is denoted as 
$h_v(x)$, satisfying
\be
h_v(x)=\frac{1+v\!\!\!/}{2}h_v(x)\,.
\ee
$v$ is defined by the four momentum of the hadron $H$,
\EQ{v^\mu=P_H^\mu/m_H\,,}
where $m_H$ is the mass of $H$.
Thus, in the rest frame of $H$, $v^\mu=(1,\vec{0})$, and 
$h_v$ is a two-component quark field.\footnote{
It corresponds to
the upper two component in the Dirac representation of the $\gamma$ matrices.
Namely,
the lower two-component antiquark field has been integrated out from the theory.}

Let $H$ be the $B$ meson,  
and we consider the observable $\Gamma(B\to X_u l\overline{\nu})$.
In HQET,
the OPE of $\Gamma(B\to X_u l\overline{\nu})=\Gamma(m_b)$ is given by $1/m_b$ expansion as
\be
\Gamma(m_b)=\Gamma(m_b)_{\rm LO}\Bigg[X^\Gamma(m_b)\langle{\bar{b}_v}b_v\rangle-X_{\rm kin}^\Gamma(m_b)\frac{\mu^2_\pi}{2m_b^2}+X_{\rm cm}^\Gamma(m_b)\frac{\mu^2_G}{2m_b^2}+\cdots\Bigg],
\label{OPE-GammaB}
\ee
where
$\Gamma(m_b)_{\rm LO}$ denotes the partonic decay width without QCD corrections,
\be
\Gamma(m_b)_{\rm LO}=\frac{G_F^2|V_{ub}|^2}{192\pi^3}m_b^5.
\ee
The Wilson coefficients $X^\Gamma\equiv X_{\bar{b}b}^\Gamma$ 
is known up to ${\cal O}(\alpha_s^2)$ \cite{Pak:2008cp}.
Recently, the ${\cal O}(\alpha_s^3)$ correction has been computed
in expansion in $(1-m_c/m_b)$ \cite{Fael:2020tow}, which is
useful even in the $m_c= 0$ case.
We denote it as
\be
X^\Gamma(m_b)=1+g_0\alpha_s(m_b)
+g_1\alpha_s(m_b)^2+g_2\alpha_s(m_b)^3+{\cal O}(\alpha_s^4).
\ee
$X_{\rm kin}^\Gamma,\,X_{\rm cm}^\Gamma$ are known 
up to ${\cal O}(\alpha_s)$ \cite{Becher:2007tk, Alberti:2013kxa}.
In this analysis, we set $X_{\rm kin}^\Gamma=X_{\rm cm}^\Gamma=1$ for simplicity.
The state normalization is given by
\be
\langle B(p')|B(p)\rangle=2E_p (2 \pi)^3 \delta^3(\vec{p}-\vec{p}\,'),~~
E_p= \sqrt{m_B^2+|\vec{p}\,|^2}.
\ee
In HQET, hadron states $|B(p)\rangle$ are defined with $p=m_B v_r$ and $v_r^\mu=(1,\vec{0})$. 
Then the leading matrix element is given by
\be
\langle\bar{b}_vb_v\rangle\equiv\frac{\langle B(p)|\bar{b}_vb_v|B(p)\rangle}{2m_B}=1,
\ee
because of the $b$-quark number conservation. 

In this OPE, there are no contributions from 
the dimension-four operators, 
$\bar{b}_viv\!\cdot\! Db_v$ and 
the operators of the light sector alone \cite{Neubert:1994wq}. 
Although such operators would give ${\cal O}(\LQ)$ contributions
to $\Gamma(B\to X_u l\overline{\nu})$, 
they are prohibited due to the following reason.
$\bar{b}_viv\!\cdot\! Db_v$ can be eliminated using the equation of
motion, while the insertions of the weak current operator
require $b_v$ and $\bar{b}_v$, hence the operators
of the light sector alone are not allowed.
In this system, the typical energy scale is much larger than $\LQ$ since the
weak decay process $b\to uW\to ul\overline{\nu}$ has a large momentum transfer.
Hence, it is reasonable that low energy gluons which cause an 
${\cal O} (\LQ)$ mass shift cannot appear between the $b$ quark operator insertions
and ${\cal O} (\LQ)$ contributions are absent.

$\mu^2_\pi,\, \mu^2_G$ denote 
the ${\cal O}(\LQ^2)$ non-perturbative matrix elements of the dimension-five operators,
\be
\mu_{\pi}^2=\frac{\langle B(p) |\bar{b}_v D^2_{\perp} b_v |B(p)\rangle}{2 m_B},~~~
\mu_G^2=\frac{\langle B(p) |\bar{b}_v \frac{g}{2} \sigma_{\mu \nu} G^{\mu \nu} b_v |B(p)\rangle}{2 m_B},\label{ME}
\ee
where $D_\perp^\mu=D^\mu-(v\cdot D)v^\mu$.
($D_\perp^2$ is equal to
$-\vec{D}^2$ in the $B$ rest frame.)
Thus, the corresponding terms in eq.~\eqref{OPE-GammaB}
represent the non-relativistic kinetic energy and 
the chromomagnetic energy of the $b$ quark, respectively.

In HQET, $m_h$ represents the pole mass of the heavy quark $h$. 
The pole mass is defined by the pole of the quark propagator, 
which is equivalent to the energy of the quark in its rest frame. 
Its perturbative coefficient at each order is IR finite, 
but the all-order sum is ill-defined.
This is due to the IR renormalons of the quark pole mass.
If we use the pole mass, the perturbative calculation of the Wilson coefficient $X^\Gamma$, is badly affected. 
To overcome this problem, 
we must change the mass scheme from the
pole mass to a short-distance mass $m_b^{\rm short}$. 
Then the OPE of $\Gamma(m_b)$ is given by
\be
\Gamma(m_b)=\Gamma(m_b^{\rm short})_{\rm LO}
\biggl(\frac{m_b}{m_b^{\rm short}}\biggr)^{5}
\Bigg[X^\Gamma(m_b)-\frac{\mu^2_\pi}{2m_b^2}+\frac{\mu^2_G}{2m_b^2}+\cdots\Bigg].\label{GammaOPE}
\ee
The ratio  $m_b/m_b^{\rm short}$ has IR renormalons originating from the pole mass.
We choose the $\msbar$ mass for the
short-distance mass, and the pole-$\msbar$ mass relation is given by
\bea
&&
\frac{m_{b}}{\mbar_b}\equiv
c_m(\mbar_b)=1+\sum_{n=0}^\infty d_n(\mu/\mbar_b)\alpha_s^{n+1}
=1+\sum_{n=0}^\infty d_n(1)\alpha_s(\mbar_b)^{n+1},
\label{poleMS}
\eea
where
\bea
\mbar_b\equiv m^{\msbar}_b(m^{\msbar}_b).
\eea
denotes the $\msbar$ mass renormalized at the $\msbar$ mass scale.
The perturbative series is known up to ${\cal O}(\alpha_s^4)$ \cite{Marquard:2015qpa,Marquard:2016dcn}.
The $u=1/2$ renormalons in eq.~(\ref{poleMS}) should be canceled with that in $X^\Gamma$, 
since there is no non-perturbative term that gives rise to an ${\cal O}(\LQ)$ correction
 in eq.~(\ref{GammaOPE}). 
Thus, 
\be
\Gamma(m_b)=\Gamma(\mbar_b)_{\rm LO}\Bigg[{\bar{X}}^\Gamma(\mbar_b)-\frac{\mu^2_\pi}{2m_b^2}+\frac{\mu^2_G}{2m_b^2}+\cdots\Bigg],
\ee
where the Wilson coefficient is rewritten as
\be
{\bar{X}}^\Gamma(\mbar_b)\equiv
c_m(\mbar_b)^5\,{X}^\Gamma\big(m_b=c_m(\mbar_b)\mbar_b\big)=1+\sum_{n=0}^{\infty}\bar{g}_n\alpha_s(\mbar_b)^{n+1},
\ee
where
\bea
&&
\bar{g}_0=g_0+5d_0,~~~\bar{g}_1=g_1+5g_0d_0+10d_0^2+5d_1,
\\ &&
\bar{g}_2=g_2 + 5 g_1 d_0 + 10 g_0 d_0^2 + 10 d_0^3 + 5 g_0 d_1 + 
 20 d_0 d_1 + 5 d_2 - 2 b_0 d_0 g_0,
~~ \text{etc.}
\eea
${\bar{X}}^\Gamma$ does not have the $u=1/2$ renormalon.
The leading renormalon in ${\bar{X}}^\Gamma$ is expected to be 
at $u=1$, to be canceled by the $\mu^2_\pi,\, \mu^2_G$ terms in the OPE framework. 
In the following, 
we examine the $u=1$ renormalon subtraction by the FTRS method.
We take $Q=\mbar_b$ as a hard scale,
and we investigate the behavior of ${\bar{X}}^\Gamma_{\rm FTRS}(\mbar_b)$
by varying the value  of $\mbar_b$ hypothetically.

\subsection{$u=1$ renormalon subtraction by FTRS method}
\label{Sec.4.2}

To subtract the $u=1$ renormalon from ${\bar{X}}^\Gamma(\mbar_b)$, 
we use the FTRS method with the parameters $(a,u')=(1,-1)$.
In this case the renormalons
are suppressed at $u=1,2,3,\cdots$ in the momentum-space Wilson coefficient
$\tilde{\bar{X}}^\Gamma(\tau)$. 
The explicit form is given as follows.
\be
{\bar{X}}^\Gamma_{\rm FTRS}(\mbar_b)=1+{\bar{X}}^\Gamma_{0}(\mbar_b)+{\bar{X}}^\Gamma_{\rm pow}(\mbar_b),
\ee
\be
{\bar{X}}^\Gamma_{0}(\mbar_b)=\frac{-1}{2\pi^2\mbar_b}\int_0^\infty dt\,t\,e^{-t/\mbar_b}{\rm Im}\,[\tilde{\bar{X}}^\Gamma(\tau=it)],
\ee
\be
{\bar{X}}^\Gamma_{\rm pow}(\mbar_b)=\frac{1}{4\pi^2 i\,\mbar_b}\int_{C_*} d\tau\,\tau\,\tilde{\bar{X}}^\Gamma(\tau).
\ee
After resummation of the artificial UV renormalon at $u=-1/2$,
we obtain the Wilson coefficient in momentum space
in the $\rm N^2LL$ approximation as
\bea
\tilde{\bar{X}}^\Gamma(\tau)\biggr|_{\rm N^2LL}
&=&\frac{4\pi}{\tau}\Biggl[
\biggl[\sin({\pi}u_*)\Gamma(2u_*)-\frac{1}{2u_*+1}\biggr]_{u_*\to\hat{H}}
\sum_{n=0}^2\bar{g}_n\alpha_s(\tau)^{n+1}
\Biggr]_\text{up to ${\cal O}(\alfs({\tau})^3)$}
\nonumber\\
&&~~ +\frac{4\pi}{\tau}\int_0^1dv \,\sum_{n=0}^2\bar{g}_n\alpha_s(\tau /v)^{n+1}\nonumber\\
&\approx&\frac{4\pi}{\tau}\Bigg[0.7728 a_\tau+ 1.717 a_\tau^2+ 3.801 a_\tau^3\non
&&~~+\int_0^1 dv \,\Big(1.354 a_{\tau, v} +2.714 a_{\tau, v}^2+6.110 a_{\tau, v}^3 \Big)\Bigg]\,,
\label{FTRSformula-Bdecay}
\eea
where $a_\tau=\alfs({\tau})$, $a_{\tau, v}=\alfs(\tau /v)$.
Scale variation in the numerical analysis is studied according to the same procedure in Sec 3.2. 

\subsection{\boldmath 
Convergence properties and $u=1$ renormalon}
\label{Sec.4.3}

Since $\mbar_b$ cannot be varied in experiments,
we cannot make a consistency check of the OPE of
$\Gamma(B\to X_u l\overline{\nu})$ in a manner
similar to the Adler function case.
Here, we examine convergence properties of
the leading Wilson coefficient to see effects
of the $u=1$ renormalon, as we vary the hypothetical value
of $\mbar_b$.

\begin{figure}[p]
        \centering
        \includegraphics[width=15.5cm]{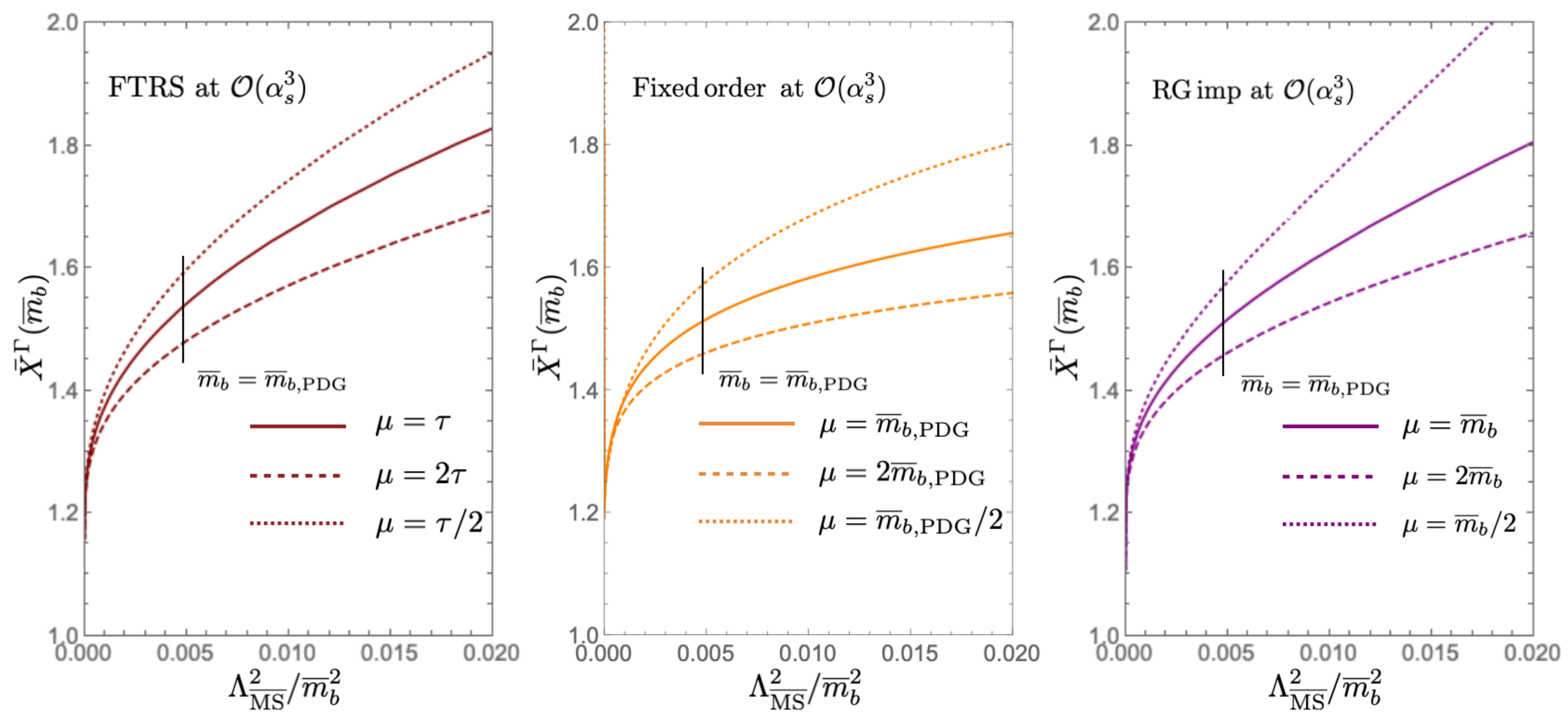}
        \caption
        {\small
        Comparison of ${\bar{X}}^\Gamma$ by FTRS method, and by fixed-order
        and RG-improved calculations
        up to order ${\cal O}(\alpha_s^3)$.
        We take $n_f=4$, $\Lambda_{\overline{\rm MS}}=0.292~\rm GeV$ and $\mbar_c=0$ as inputs.
        We vary $\mu$ within $[\tau/2,2\tau]$, 
        $[\mbar_{b,{\rm PDF}}/2,2\mbar_{b,{\rm PDF}}]$
        and $[\mbar_b/2,2\mbar_b]$, respectively, where $\mbar_{b,{\rm PDF}}=4.18$~GeV.
        }
        \label{BNLL}
        \centering
\vspace*{10mm}
        \includegraphics[width=15.5cm]{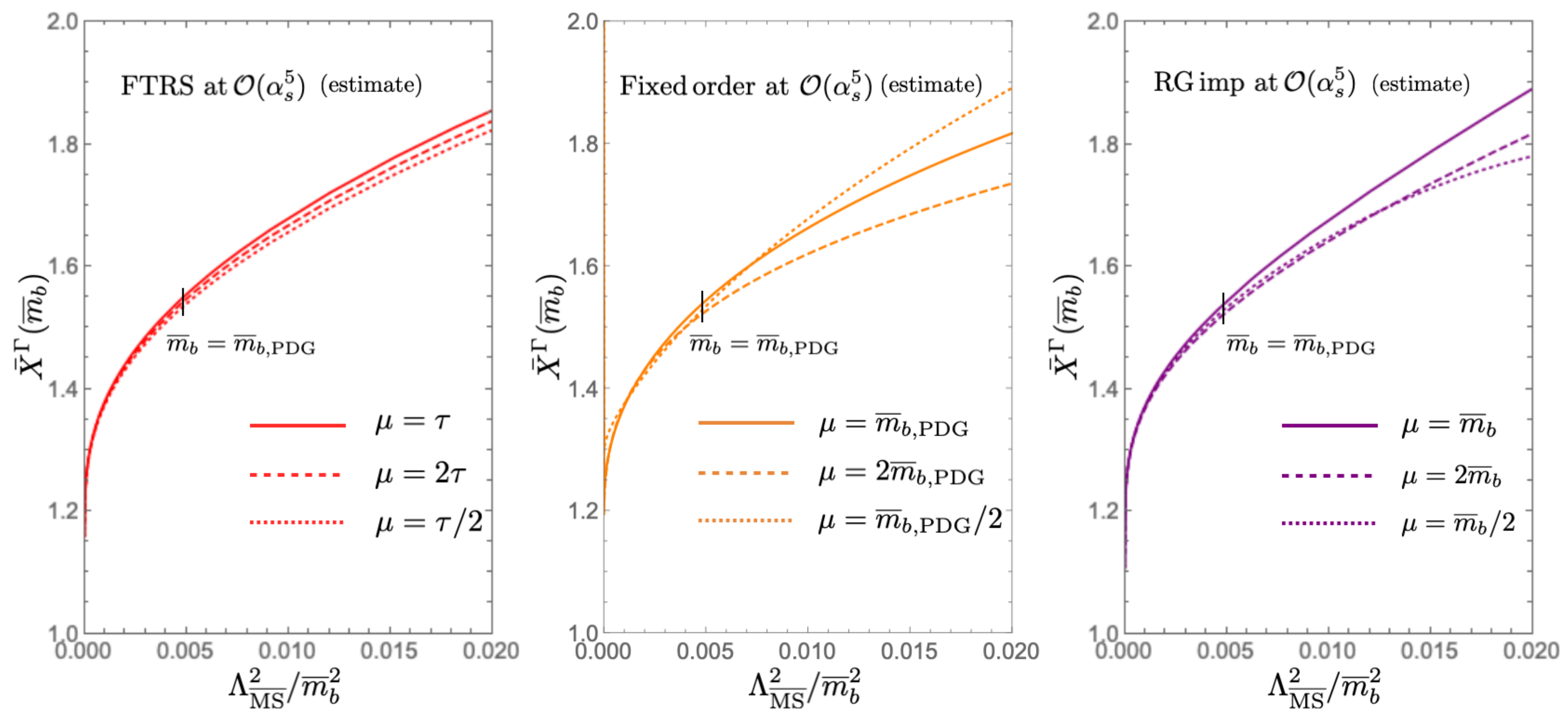}
        \caption
         {\small
        Comparison of ${\bar{X}}^\Gamma$ by FTRS method, and by fixed-order
        and RG-improved calculations
        up to order ${\cal O}(\alpha_s^5)$, 
        which are estimated using 5-loop beta function.
        Other parameters and notations are the same as in Fig.~\ref{BNLL}.
        }
        \label{BNNNNLL}
\end{figure}

We compare ${\bar{X}}^\Gamma$ by the FTRS calculation with 
the fixed-order (FO) 
and RG-improved calculations.
Explicitly, we define
\bea
&&
{\bar{X}}^\Gamma_{\rm FO}(\mbar_b)=\mbar_b\left[ 1+ \sum_{n=0}^2 \bar{g}_n(\mu_0/\mbar_b)\alpha_s(\mu_0)^{n+1} \right],
\\&&
{\bar{X}}^\Gamma_{\rm RG}(\mbar_b)=\mbar_b\left[ 1+ \sum_{n=0}^2 \bar{g}_n(1)\alpha_s(\mbar_b)^{n+1}
\right] .
\eea
Fig.~\ref{BNLL} shows a comparison of ${\bar{X}}^\Gamma_{\rm FTRS}$, 
${\bar{X}}^\Gamma_{\rm FO}$ and ${\bar{X}}^\Gamma_{\rm RG}$.
There are no significant differences in the scale dependence between them.
The scale dependence of ${\bar{X}}^\Gamma_{\rm FTRS}$ at the current accuracy 
is still rather large, where the corrections up to
${\cal O}(\alpha_s^3)$ are known. 
If the size of the $u=1$ renormalon is large in the original
perturbative series, 
we expect that the convergence behavior of ${\bar{X}}^\Gamma_{\rm FTRS}$ is
better than ${\bar{X}}^\Gamma_{\rm FO}$ or
${\bar{X}}^\Gamma_{\rm RG}$, since we have eliminated it
from the former. 

We make an estimate of higher order results by including log dependent terms, 
dictated by the RGE, at the 5-loop level. 
We perform RG improvement of the perturbative series in $\tau$ space 
rather than $1/\mbar_b$ space, 
in accordance with the construction 
that there are no IR renormalons in the $\tau$-space quantity. 
(Note that RG improvements in two spaces are not equivalent 
when we do not know the exact perturbative coefficient of the
non-logarithmic term. 
In this treatment, the $u=1$ renormalon is induced in the perturbative series in $1/\mbar_b$ space.
Then, ambiguities arise in ${\bar{X}}^\Gamma_{\rm FO}$ and
${\bar{X}}^\Gamma_{\rm RG}$.)
Fig.~\ref{BNNNNLL} shows a comparison of these 5-loop estimates. 
If $\mbar_b$ is around the physical mass of the $b$ quark, 
the scale dependence of the FTRS calculation is slightly better than the 
fixed-order and RG-improved calculations. 
Nevertheless, we cannot see significant differences between these three cases.
In particular, the scale dependence of ${\bar{X}}^\Gamma_{\rm FO}$ and
${\bar{X}}^\Gamma_{\rm RG}$
is smaller than that of the FTRS result at the NLL level
(see ref.~\cite{Hayashi:2020ylq}).

The above observations are consistent with
the possibility that the size of the $u=1$ renormalon is small
for this partial decay width.
However, in order to estimate the size of the $u = 1$ renormalon reliably, 
we need more terms of the perturbative series.
On the other hand, in the lower mass range
$\LMS^2/\mbar_b^2\gtrsim 0.01$
($\mbar_b\lesssim$ 3 GeV),
we find a better convergence  by the FTRS method
than the  usual fixed-order or RG-improved calculations.
This can be due 
to circumventing the unphysical singularity of the running coupling
constant in the FTRS 
method.
It may indicate that
we can make use of our method to  lighter flavor observables (e.g. $D$ meson decay width).

In this analysis, although we use the $\overline{\rm MS}$ mass,
it has been known that the use of other short distance masses can lead 
to better convergence of the perturbative series. 
This is because the $\overline{\rm MS}$ mass is far from the ``physical'' mass 
and needs a large perturbative correction to approximate it.
In this study, we choose the $\overline{\rm MS}$ mass in order to use the FTRS method
straightforwardly; if we instead use other short distance masses, 
additional issues arise, e.g., how to treat a factorization scale, which is introduced 
in many short distance masses but is not considered in the FTRS method. 
In addition, the disadvantage of the poor convergence can be compensated once 
the perturbative series of the decay and the mass relation are known to sufficiently high orders.
In this regard, we showed results using estimated higher order perturbative series 
in Fig. 6. 
Thus we use the $\overline{\rm MS}$ mass for the first test of our formulation.
Nevertheless, the use of other short distance masses may improve our result. 
We leave it as our future investigation.

\section{\boldmath Observable III : $B$, $D$ meson masses}

In this section we analyze the masses of the $B$ and $D$ mesons
by the OPE in HQET ($H=B^{(*)},D^{(*)}$).
The renormalons at $u=1/2$ and $u=1$, corresponding to
the ${\cal O}(\LQ)$ and ${\cal O}(\LQ^2)$ renormalons,
are subtracted simultaneously.
We check consistency with theoretical expectations and
extract the non-perturbative parameters $\bar{\Lambda}$, $\mu^2_\pi$
using the $B^{(*)},D^{(*)}$ masses and our current knowledge of
$\mbar_b$, $\mbar_c$, $\alfs(M_Z)$.

\subsection{\boldmath OPE of $H$ meson mass in HQET}

The OPE of the $H$ mass is given by \cite{Falk:1992wt}
\be
M_{H,\,\rm OPE}^{(s)}=m_h+\bar{\Lambda}+\frac{\mu^2_\pi}{2m_h}+w(s)X_{\rm cm}^H(m_h)\frac{\mu^2_G}{2m_h}+{\cal O}\Big(\frac{\LQ^3}{m_h^2}\Big)\,,
\ee
where $s=0,1$ denotes the spin of $H$;
$m_h$ represents the pole mass of the heavy quark ($h=b,c$).
$\bar{\Lambda}$ is an ${\cal O}(\LQ)$ non-perturbative parameter  defined by
\be
\bar{\Lambda}=\lim_{m_h\to\infty}\big[m_H-m_h\big] \,,
\ee
which represents the contribution from light degrees of freedom.
To give an unambiguous definition of $\bar{\Lambda}$ we need to specify
which regularization is applied to the pole mass $m_h$, and we adopt the
PV prescription here.
$\mu^2_\pi$ and $\mu^2_G$ are defined by eq.~(\ref{ME}).
The Wilson coefficient of $\mu^2_\pi$ is 
exactly one according to the reparametrization invariance,
while that of $\mu^2_G$ is known up
to order ${\cal O}(\alpha_s^3)$ \cite{Grozin:2007fh}.
The spin dependent weight is given by
$w(0)=1,\,w(1)=-1/3$. 
Due to the heavy quark symmetry of HQET, 
the non-perturbative parameters $\bar{\Lambda}$, $\mu^2_\pi$, $\mu^2_G$
have common values for $H=B$ and $D$ \cite{Manohar:2000dt}.

We project out $\mu^2_G$ by taking the linear combination,
\bea
\overline{M}_{H,\,\rm OPE}
&\equiv&\frac{M_{H,\,\rm OPE}^{(0)}
+3M_{H,\,\rm OPE}^{(1)}}{4}\nonumber\\
&=&m_h+\bar{\Lambda}+\frac{\mu^2_\pi}{2m_h}
+{\cal O}\Big(\frac{\LQ^3}{m_h^2}\Big).
\eea
The IR renormalons in the pole mass $m_h$ are canceled by 
$\bar{\Lambda}$, $\mu_\pi^2$, $\cdots$ in the OPE framework.
We separate the ${\cal O}(\LQ)$ and ${\cal O}(\LQ^2)$ renormalons 
from $m_h$ by the FTRS method. 
Since the Wilson coefficients of $\bar{\Lambda}$ and $\mu_\pi^2$ 
are exactly one and
the ${\cal O}(\LQ)$ and ${\cal O}(\LQ^2)$ renormalons have 
no logarithmic corrections, these renormalons are completely removed within our
formalism (without using the extended formalism explained in App.~\ref{appB}).

\subsection{\boldmath ${dm_b}/d{\mbar_b}$: 
FTRS, fixed-order and RG-improved calculations}
\label{dmdm}



First, we compare the FTRS method with 
the fixed-order and RG-improved calculations.
We are interested in the effects of the $u=1$
[${\cal O}(\LQ^2)$] renormalon in particular.
Hence, we calculate
$dm_{b}/d\mbar_b$, which is free from
the $u=1/2$ [${\cal O}(\LQ)$] renormalon ambiguity
even in the fixed-order and RG-improved calculations.
In this subsection, we use the perturbative series for the $b$ quark pole mass
in the theory with four massless quarks ($u, d, s, c$).
In the OPE analysis in the following subsections, however, we need to pay 
attention to the treatment of finite (non-zero) mass effects of the $c$ quark,
in order to be compatible with the flavor universality of the non-perturbative 
matrix elements (heavy quark symmetry).
Since we do not need to care about the flavor universality in this subsection,
we use the simple perturbative series for clarity.

We construct ${m_b^{\rm FTRS}}(\mbar_b)$ as follows.
To subtract the ${\cal O}(\LQ)$ and ${\cal O}(\LQ^2)$ renormalons from $m_b$,
we set the parameters $(a,u')=(2,-1/2)$. 
The renormalons at $u=1/2,1,3/2,\cdots$ are suppressed
in momentum space. 
Explicitly, we have
\be
{m_b^{\rm FTRS}}(\mbar_b)=\mbar_b+m_{b,0}(\mbar_b)+m_{b,\rm pow}(\mbar_b),
\label{mbFTRS}
\ee
\be
m_{b,0}(\mbar_b)
=-\frac{\sqrt{\mbar_b}}{2\pi^2}\int_0^\infty dt\,t\,e^{-t/\sqrt{\mbar_b}}
{\rm Im}\,[\tilde{c}_{m}(\tau=it)]
\ee
\be
m_{b,\rm pow}(\mbar_b)=\frac{\sqrt{\mbar_b}}{4\pi^2i}\int_{C_*} d\tau\,\tau\biggl(1-\frac{\tau^2}{2\mbar_b}\biggr)\tilde{c}_{m}(\tau) \,,
\ee
where $c_m(\mbar)$ is defined in eq.~\eqref{poleMS}.
After resummation of the artificial UV renormalons at $u=-1/4$ and $-3/4$,
$\tilde{c}_m(\tau)$ in the $\rm N^3LL$ approximation is calculated
from ${m_{b}}/{\mbar_b}-1=\sum_{n=0}^3 d_n(\mu/\mbar_b)\alpha_s^{n+1}$ 
as
\bea
&&
\tilde{c}_{m}(\tau)\biggr|_{\rm N^3LL}
=\frac{4\pi}{\tau}
\Biggl[
\biggl[\sin(2{\pi}u_*)\Gamma(4u_*)-\Big(\frac{1}{4}\frac{1}{u_*+1/4}-\frac{1}{24}\frac{1}{u_*+3/4}\Big)
\biggr]_{u_*\to\hat{H}}
\nonumber\\ &&
~~~~~~~~~~~
~~~~~~~~~~~
~~~
\times
\sum_{n=0}^\infty d_n\alpha_s(\tau^2)^{n+1}
\Biggr]_\text{up to ${\cal O}(\alfs(\tau^2)^4)$}
\nonumber\\ &&
~~~~~~~~~~~
~~~~~~~~~~~
+\frac{4\pi}{\tau}\int_0^1 \frac{dv}{2v}\, \Big(v^{1/2}-\frac{1}{6}v^{3/2}\Big)\sum_{n=0}^3d_n\alpha_s(\tau^2 /v)^{n+1}\nonumber\\
&&
~~~~~
\approx\frac{4\pi}{\tau}\Bigg[
0.2658a_\tau+0.6730a_\tau^2+1.803
a_\tau^3+3.290 a_\tau^4
\nonumber\\
&&
~~~~~
~~~
+\int_0^1 \frac{dv}{2v}\, \Big(v^{1/2}-\frac{1}{6}v^{3/2}\Big)
\Big(0.4244a_{\tau, v}+0.9401a_{\tau, v}^2+3.039a_{\tau, v}^3+12.65a_{\tau, v}^4\Big)\Bigg],\nonumber\\
\label{alphamb}
\eea
where $a_\tau=\alfs({\tau}^2), a_{\tau, v}=\alfs(\tau^2/v)$. 
In this analysis (in Sec.~\ref{dmdm}), we use $d_n=\tilde{d}_n{(n_l=4)}$ 
of eqs.~\eqref{dtilde1}--\eqref{dtilde3}
without finite mass corrections inside loops $(m_c\to0)$,
i.e., eq.~\eqref{app:formula5.2}.
Scale variation in the numerical analysis is studied according to the same procedure in Sec 3.2. 
\begin{figure}[t]
        \centering
        \includegraphics[width=15cm]{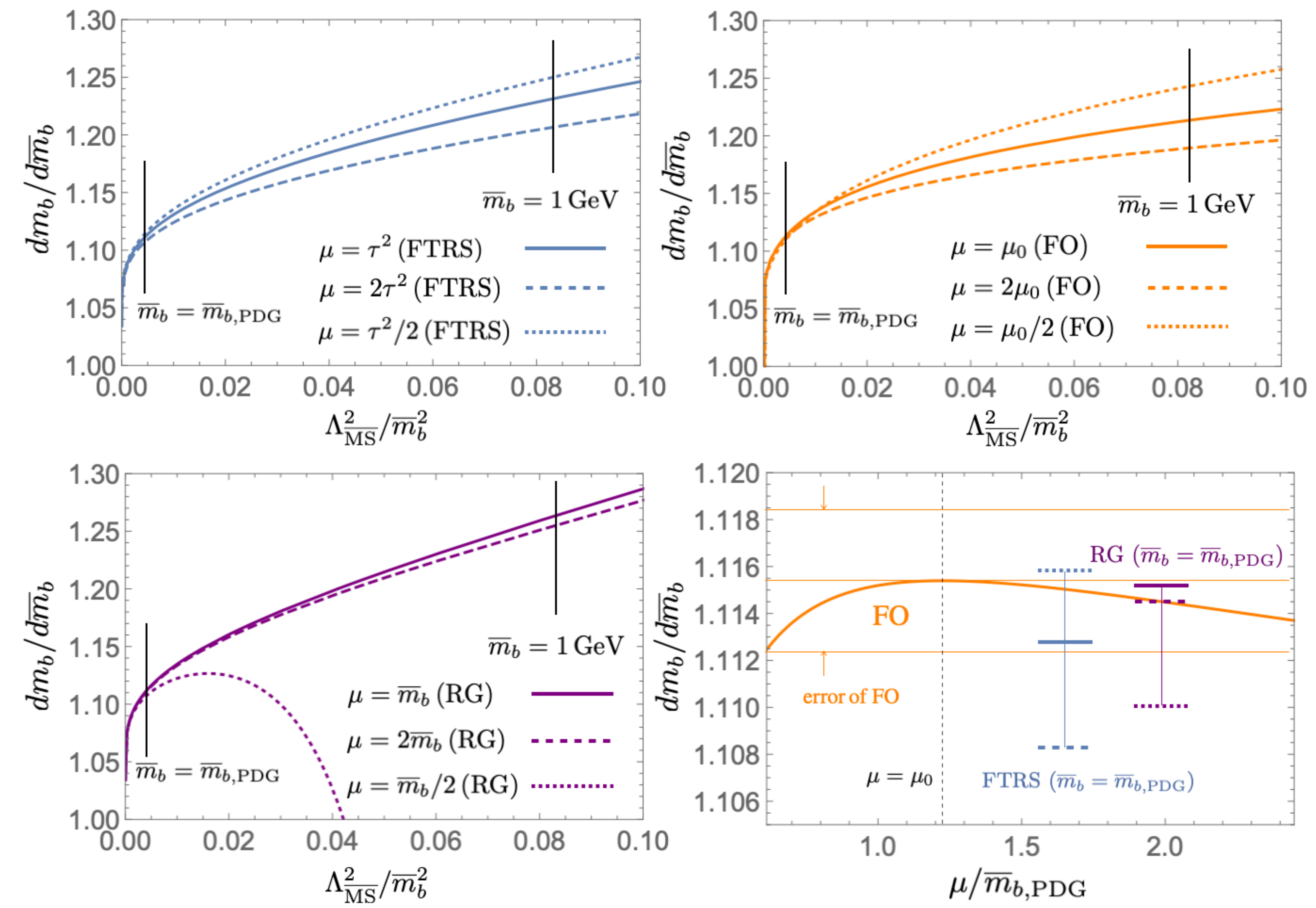}
        \caption
         {\small 
        Comparison of $dm_b/d\mbar_b$ by FTRS method, by fixed-order calculation,
        and by RG improvement
        up to order ${\cal O}(\alpha_s^4)$.
        $\mu$ is varied within $[\tau^2/2,2\tau^2]$,
        $[\mu_0/2,2\mu_0]$,
        and $[\mbar_b/2,2\mbar_b]$, respectively, where
        $\mu_0=17.71\LMS$ (see text).
        The bottom-right figure compares the scale dependence of the three calculations.
        The input is $\LMS^{n_f=4}=0.289\,{\rm GeV}$.
        } \label{polembar}
\end{figure}

We compare $dm_b/d\mbar_b$ by the FTRS calculation with 
the fixed-order (FO) 
and RG-improved calculations.
Explicitly, we define
\bea
&&
m_b^{\rm FO}(\mbar_b)=\mbar_b\left[ 1+ \sum_{n=0}^3 d_n(\mu_0/\mbar_b)\alpha_s(\mu_0)^{n+1} \right],
\\&&
m_b^{\rm RG}(\mbar_b)=\mbar_b\left[ 1+ \sum_{n=0}^3 d_n(1)\alpha_s(\mbar_b)^{n+1}
\right] .
\eea
We take the derivative of $m_b$ with respect to $\mbar_b$.
Then the ${\cal O}(\LQ)$ renormalon ambiguity is absent
in all three quantities, while
only the FTRS calculation is free from the $u=1$ renormalon.
Fig.~\ref{polembar} compares the scale dependence.
For each hypothetical value of $\mbar_b$,
the scale $\mu$ is varied between $[\tau^2/2,2\tau^2]$,
$[\mu_0/2,2\mu_0]$
and $[\mbar_b/2,2\mbar_b]$ in the FTRS, fixed-order and RG-improved
calculations, respectively.
Here, we choose $\mu_0=17.71\LMS$, which coincides with
the minimal sensitivity scale of $dm_b^{\rm FO}/d\mbar_b$
for $\mbar_b=\mbar_{b,\rm PDG}=4.18$~GeV \cite{Zyla:2020zbs}.
The bottom-right figure shows the scale dependence 
of each calculation near $\mbar_b=\mbar_{b,\rm PDG}$.
There is no significant difference between them.
It is consistent with the result of the  previous study 
that the $u=1$ renormalon is small 
and hardly visible from the currently known perturbative
coefficients \cite{Ayala:2019hkn}.

\subsection{\boldmath Internal quark mass effects in pole--$\msbar$ mass relation}
\label{InternalMassEffect}

First, we review a non-trivial aspect
in eliminating the $u=1/2$ and $u=1$ renormalons in 
the pole masses, $m_{b}$ and $m_c$, 
by the universal parameters $\bar{\Lambda}$ and $\mu^2_\pi$.
After that we consider an application to the FTRS method.

Naively one may expect that the IR structure of
the bottom (charm) quark is described well in the theory with $n_l=4$ ($n_l=3$) 
massless quarks. 
($n_l$ denotes the number of massless quarks.)
Hence,
the renormalons of the different quark pole masses 
could originate from the different IR structures. 
It is known, however, that internal massive quarks in loops do not contaminate the IR structure, 
that is, the
internal charm quark with a non-zero (finite)  mass does not 
contribute to the renormalon divergences of the bottom quark pole mass.
Therefore, to be consistent with the universality of the non-perturbative
parameters, we need to 
consider the theory with three massless quarks plus massive charm and bottom quarks
in the OPE analysis.
In this theory 
the $B$ and $D$ mesons share the same light sector and hence
the same renormalon structure,
and we can subtract the renormalons consistently with the universality of 
the non-perturbative matrix elements in HQET.

This feature is demonstrated in the large-$\beta_0$ approximation
with a finite charm quark mass included in loops, as follows.
The bottom quark
pole mass in this approximation is given by the loop momentum integration,
\be
\frac{m_{b,\rm pole}-\mbar_b}{\mbar_b}=\int d^4 k\, F(k,\mbar_b)\,\frac{1}{k^2}\,\frac{\alpha_s^{(4)}(\mu)}{1-\alpha_s^{(4)}(\mu) \Pi(k^2,\mbar_c)}
\label{LBpoleMS},
\ee
where $F(k,\mbar_b)$ is a kinematic function
and $\alpha_s^{(n_f)}$ represents the strong coupling 
constant of the $n_f$-flavor theory.
We have rewritten $\alpha_s^{(5)}$ by $\alpha_s^{(4)}$, which absorbs
the effect of the bottom quark loops.
$\Pi(k^2,\mbar_c)$ is the one-loop 
vacuum polarization which includes a massive charm quark loop,
\be
\Pi(k^2,\mbar_c)=\Pi_{\rm light}(k^2)+\Pi_{c}(k^2, \mbar_c) ,
\ee
where $\Pi_{\rm light}$ represents the contribution from 
the light degrees of freedom (gluons and massless quarks)
while $\Pi_c$ represents the charm quark contribution. 
In the IR region $k^2 \simeq 0$, they are given by
\be
\Pi_{\rm light}(k^2)=b_0^{(3)}\left[\log\left(-\frac{\mu^2}{k^2}\right)+\frac{5}{3}\right],  \label{Pilight}
\ee
\be
\Pi_{c}(k^2,\mbar_c)=-\frac{1}{6\pi}\log\left(\frac{\mu^2}{\mbar_c^2}\right)+{\cal O}\left(\frac{k^2}{\mbar_c^2}\right)\, ,
\label{Pic}
\ee
with $b_0^{(n_l)}=(11-2n_l/3)/(4\pi)$.
$\Pi_{c}$ does not have $\log(k^2)$ behavior in the IR region \cite{Ball:1995ni},
since the charm quark mass works as an IR regulator in the fermion loop integration.
As a result the charm quark contributions do not give renormalons.

We can absorb the $\log(\mu^2/\mbar_c^2)$ term in eq.~\eqref{Pic}
if we use the coupling constant of the 3-flavor theory,
\bea
\frac{\alpha_s^{(4)}(\mu)}{1-\alpha_s^{(4)}(\mu) \Pi(k^2,\mbar_c)}
=\frac{\alpha_s^{(3)}(\mu)}{1-\alpha_s^{(3)}(\mu) \Pi_{\rm light}(k^2)+\mathcal{O}(k^2/\mbar_c^2)} ,
\eea
due to the one loop threshold correction in the $\msbar$ scheme
\be
\frac{1}{\alpha_s^{(4)}(\mu)}=\frac{1}{\alpha_s^{(3)}(\mu)}-\frac{1}{6\pi}\log(\mu^2/\mbar_c^2) .
\ee
After this rewriting, one can clearly see that the renormalon we encounter 
coincides with the one in the $n_f=3$ theory:
\be
\frac{m_{b,\rm pole}-\mbar_b}{\mbar_b} \approx K \sum_{n=0}^{\infty} (2 b_0)^n n! [\alpha_s^{(3)}(\mbar_b)]^{n+1}
\,.
\ee
In fact, it was pointed out that one should use $\alpha_s^{(3)}$ rather than 
$\alpha_s^{(4)}$ as the expansion parameter \cite{Ayala:2014yxa}.

We now show that, 
in order for the FTRS method to work with our parameter choice $(a, u')=(2, -1/2)$, 
it is necessary to express the perturbative series in the 3-flavor coupling
constant.
In fact, renormalons remain in $\tau$ space if we use the 4-flavor coupling constant.
With the 4-flavor coupling constant, from eqs.~\eqref{Pilight} and \eqref{Pic},
the Borel transform close to the first IR renormalon is given by
\be
B^{\text{4-flavor}}(u) \approx  \left(\frac{\mu^2 e^{5/3}}{\mbar_b^2} \right)^{\frac{b_0^{(3)}}{b_0^{(4)}} u} \left(\frac{\mu^2}{\mbar_c^2} \right)^{\frac{u}{6 \pi b_0^{(4)}}} \frac{1}{1-2 \frac{b_0^{(3)}}{b_0^{(4)}} u } ,
\ee
where $u=b_0^{(4)} t$.
The singularity is located at $u=b_0^{(4)}/(2 b_0^{(3)})$.
Then the Borel transform of the $\tau$-space perturbative series is given by
\bea
&&
\tilde{B}^{\text{4-flavor}}(u) \approx \left(\frac{\mu^2 e^{5/3}}{\tau^{2a}} \right)^{\frac{b_0^{(3)}}{b_0^{(4)}} u} \left(\frac{\mu^2}{\mbar_c^2} \right)^{\frac{u}{6 \pi b_0^{(4)}}} \frac{1}{1-2 \frac{b_0^{(3)}}{b_0^{(4)}} u } \sin\left(\pi a \left(\frac{b_0^{(3)}}{b_0^{(4)}} u+u' \right) \right) 
\nonumber\\&&~~~~~~~~~~~~~~~~~~~
\times
\Gamma \left(2 a \left(\frac{b_0^{(3)}}{b_0^{(4)}} u+u' \right)+2 \right) .
\eea
The singularity at $u=b_0^{(4)}/[2 b_0^{(3)}]$ cannot be eliminated in $\tau$ space
with the parameter set $(a, u')=(2, -1/2)$, which is chosen to eliminate the $u=1/2$ renormalon.\footnote{
One can eliminate the renormalons by setting the parameters to $(a, u')=(2, -b_0^{(3)}/(2 b_0^{(4)}))$.}
Using the 3-flavor coupling constant
instead, the Borel transform has the $u=1/2$ renormalon,
\be
B^{\text{3-flavor}}(u) \approx  \left(\frac{\mu^2 e^{5/3}}{\mbar_b^2} \right)^{u} \frac{1}{1-2 u } ,
\ee
hence, we can properly eliminate the $u=1/2$ renormalon in $\tau$ space
with $(a, u')=(2, -1/2)$:
\bea
&&
\tilde{B}^{\text{3-flavor}}(u) \approx  \left(\frac{\mu^2 e^{5/3}}{\mbar_b^2} \right)^{u} \frac{1}{1-2 u }
\sin\left(\pi a \left(u+u' \right) \right) \Gamma \left(2 a \left(u+u' \right)+2 \right)
\nonumber\\&&~~~~~~~~~~~~~
\approx \mathcal{O}\left( \left(u-\frac{1}{2} \right)^0 \right).
\eea

Based on the above consideration, we will use $\alpha_s^{(3)}$ in the following calculations.
In App.~A, we give the explicit perturbative series of the pole-$\msbar$ mass relation for the bottom and charm quarks
[eqs.~\eqref{mb3f} and \eqref{mc3f}] in terms of $\alpha_s^{(3)}$.
Here we use the most precise perturbative coefficients available today:
in the 5-flavor theory, the finite charm mass effects are
included in the bottom pole mass, while
the non-decoupling effects of the bottom mass 
are included in the charm pole mass, respectively, 
up to the $\mathcal{O}(\alpha_s^3)$ order \cite{Fael:2020bgs}.
Then we use the matching relations to rewrite the coupling constant
$\alpha_s^{(5)} \to \alpha_s^{(3)}$ and
the $\overline{\rm MS}$ mass $\mbar_c^{(5)} \to \mbar_c^{(4)}$ (which is needed to
use the input  mass parameter by the Particle Data Group \cite{Zyla:2020zbs}.)
The  perturbative series are numerically
close to each other and
also they are close to the pole-$\overline{\rm MS}$ mass relation for 
a heavy quark with
three-massless quarks only [eq.~\eqref{mh3f}].
These features indicate that these two perturbative series indeed exhibit 
the renormalon divergence of the $n_l=3$ flavor theory, 
as expected from the above argument in the large-$\beta_0$ approximation.
It is especially noteworthy that the renormalon behavior of the
bottom quark pole mass is dominated only by the three light flavors,
whose typical scales are less than $\LQ$, and not affected by 
the charm quark, also beyond the large-$\beta_0$ approximation.
For this reason, we approximate the $\mathcal{O}(\alpha_s^4)$ coefficients
of the  finite mass corrections, which are not known today,
by that of the heavy quark pole mass with three massless quarks only, for 
both bottom and charm quarks,
i.e., $\delta d_3=0$ (when expressed by $\alpha_s^{(3)}$).

\subsection{\boldmath Extracting
$\bar{\Lambda}_{\rm FTRS},\,(\mu^2_\pi)_{\rm FTRS}$}
\label{det-LambdaBar-mupi}

Using $m_{h}^{\rm FTRS}$, we define 
\be
\overline{M}_{H,\,\rm OPE}=m_{h}^{\rm FTRS}+\bar{\Lambda}_{\rm FTRS}+\frac{(\mu^2_\pi)_{\rm FTRS}}{2m_{h}^{\rm FTRS}}+{\cal O}\Big(\frac{\LQ^3}{m_h^2}\Big).
\ee
 $m_{b}^{\rm FTRS}$ ($m_{c}^{\rm FTRS}$) is given by eqs.~(\ref{mbFTRS})--(\ref{alphamb})
with $d_n=\tilde{d}_n({n_l=3})+\delta d_n$ in eq.~(\ref{mb3f}) (eq.~(\ref{mc3f})) up to ${\cal O}(\alpha_s^4)$,
which includes non-zero mass corrections from 
the charm quark
(bottom quark).
Explicitly $\tilde{c}_m$ are given by
\bea
\tilde{c}_{m}(\tau)\biggr|_{m_b,\,{\rm N^3LL}}
\!\!\!\!\!\!\!&&
\approx\frac{4\pi}{\tau}\Bigg[0.2658a_\tau+0.7407a_\tau^2+2.240a_\tau^3+5.096a_\tau^4\nonumber\\
&&
+\int_0^1 \frac{dv}{2v}\, \Big(v^{1/2}-\frac{1}{6}v^{3/2}\Big)
\Big(0.4244a_{\tau, v}+1.037a_{\tau, v}^2+3.744a_{\tau, v}^3+17.44a_{\tau, v}^4\Big)\Bigg],\nonumber\\
\label{FTRSformula-mb}
\eea
\bea
\tilde{c}_{m}(\tau)\biggr|_{m_c,\,{\rm N^3LL}}
\!\!\!\!\!\!\!&&
\approx\frac{4\pi}{\tau}\Bigg[0.2658a_\tau+0.7447a_\tau^2+2.252a_\tau^3+5.080a_\tau^4\nonumber\\
&&
+\int_0^1 \frac{dv}{2v}\, \Big(v^{1/2}-\frac{1}{6}v^{3/2}\Big)
\Big(0.4244a_{\tau, v}+1.044a_{\tau,v}^2+3.757a_{\tau, v}^3+17.44a_{\tau, v}^4\Big)\Bigg],\nonumber\\
\label{FTRSformula-mc}
\eea
where $a_\tau=\alfs({\tau}^2)$, $a_{\tau, v}=\alfs(\tau^2 /v)$. 
Scale variation in the numerical analysis is studied according to the same procedure in Sec 3.2. 

We compare $\overline{M}_{B,\rm OPE}$ and $\overline{M}_{D,\rm OPE}$ with 
the experimental values
\be
\overline{M}_{B,{\rm exp}}=5.313\,{\rm  GeV},\quad\overline{M}_{D,{\rm exp}}=1.971\,{\rm  GeV},
\ee
to determine the values of $\bar{\Lambda}_{\rm FTRS}$ and $(\mu^2_\pi)_{\rm FTRS}$.
The input parameters of $\overline{M}_{H,\,\rm OPE}$ are taken as \cite{Zyla:2020zbs},
\be
\mbar_b^{(5)}=4.18^{+0.03}_{-0.02}\,{\rm GeV},\quad\mbar_c^{(4)}=1.27\pm0.02\,{\rm GeV},
\ee
and 
\be
\LMS^{n_f=3}=0.332\pm{0.015}\,\rm GeV\,,
\ee
corresponding to $\alfs^{n_f=5}(M_Z)=0.1179\pm0.0010$.
The above $\LMS$ is obtained with the four-loop beta function and
three-loop matching relation. We use the above value ${\LMS^{n_f=3}}|_{\text{4-loop}}$ as the input of 
$\LMS^{n_f=3}|_{\text{5-loop}}$; we discard the difference 
between ${\LMS^{n_f=3}}|_{\text{4-loop}}$ and ${\LMS^{n_f=3}}|_{\text{5-loop}}$.
The result reads
\bea
&&
\bar{\Lambda}_{\rm FTRS}=0.495(15)_\mu (49)_{\overline{m}_b} (12)_{\overline{m}_c}
(13)_{\alfs} (0)_{\rm f.m.}\text{GeV} \label{Lambdabarfinal}
\,, 
\\&&
(\mu_\pi^2)_{\rm FTRS}=-0.12(13)_\mu (15)_{\overline{m}_b} (11)_{\overline{m}_c}
(4)_{\alfs} (0)_{\rm f.m.}\text{GeV}^2, \label{myupifinal}
\eea
where the errors denote, respectively,
that from the scale dependence for $\tau^2/2\leq \mu\leq 2\tau^2$,
from the errors of the input $\mbar_b$, $\mbar_c$, $\alfs$, 
and from turning off the $\mbar_c/\mbar_b$ corrections of $d_2$ and use 
$d_2(n_l=3)$.
We note that the removal of the finite mass effects on 
the $\mathcal{O}(\alpha_s^3)$ coefficient
does not change the result at the precision level we work.
Combining the errors in quadrature,
we obtain
\bea
&&
\bar{\Lambda}_{\rm FTRS}=0.495\pm0.053~\text{GeV} 
\,, 
~~~~~
(\mu_\pi^2)_{\rm FTRS}=-0.12\pm 0.23~\text{GeV}^2
\,.
\label{resLambdamu2final2a}
\eea

The error from the scale dependence is a measure of perturbative
ambiguity. 
With our parameter choice, 
the renormalons at $u=1/2$ and $1$ should be absent from our result.
For $\bar{\Lambda}_{\rm FTRS}$, about 3 per cent 
error shows successful subtraction of the $u=1/2$ renormalon. 
On the other hand, the scale dependence of $(\mu_\pi^2)_{\rm FTRS}$
is not smaller than ${\cal O}(\LQ^2)$.
We estimate that this is not because of the contribution from the
$u=1$ renormalon but due to the insufficient number of known terms
of the perturbative series.
A 5-loop examination by using the 5-loop beta function
combined with the estimated 5-loop coefficient in the large-$\beta_0$ approximation
[$d_n$'s are given by eq.~(\ref{mb3f}) and eq.~(\ref{mc3f}) up to ${\cal O}(\alpha_s^5)$]
gives the following result:
\be
\bar{\Lambda}^{\rm (est)}_{\rm FTRS}\Big|_{\rm 5-loop}=0.488(4)_\mu\text{GeV}\,,\,
~~
(\mu_\pi^2)^{\rm (est)}_{\rm FTRS}\Big|_{\rm 5-loop}=-0.09(7)_\mu\text{GeV}^2.
\ee
This indicates that the scale dependence of $\bar{\Lambda}_{\rm FTRS}$ and $(\mu_\pi^2)_{\rm FTRS}$ 
can be straightforwardly reduced at higher orders.

As mentioned above, the inclusion of the finite mass effects 
at $\mathcal{O}(\alpha_s^3)$ gives only tiny effects.
It would be interesting to examine what happens if we 
completely neglect the finite mass effects.
With the perturbative series of a heavy quark with three massless flavors only,
for both bottom and charm quarks [eq.~(\ref{mh3f})],
we obtain 
$
\bar{\Lambda}_{\rm FTRS}\big|_{n_l=3}=0.493\,\text{GeV}\,,\,
(\mu_\pi^2)_{\rm FTRS}\big|_{n_l=3}=-0.11\,\text{GeV}^2,
$
where $\mu=\tau^2$. The values hardly change from the above results [eqs.~\eqref{Lambdabarfinal} and \eqref{myupifinal}],
where we take into account the finite bottom and charm mass effects.

Let us compare our results for $\bar{\Lambda}$ and $\mu_\pi^2$
with other determinations.
Ref.~\cite{Bazavov:2018omf} obtained
\bea
\bar{\Lambda}_{\rm PV}=0.435(31)~{\rm GeV}\,,
~~~
(\mu_\pi^2)_{\rm PV}=0.05(22)~{\rm GeV}^2\,.
\eea
In this fit,
only the ${\cal O}(\LQ)$ renormalon is subtracted, while
the ${\cal O}(1/m_h)$ corrections are included.
They estimate that the error of $\mu_\pi^2$ originates from the
${\cal O}(\LQ^2)$ renormalon.
However, our analysis indicates that it is rather due to the lack of 
known perturbative coefficients, since the size of the ${\cal O}(\LQ^2)$ renormalon
is small. 
(This is estimated by the comparison between the predictions with and without
the ${\cal O}(\LQ^2)$ renormalon subtraction.)
Ref.~\cite{Ayala:2019hkn} obtained 
\bea
\bar{\Lambda}_{\rm PV}=
477(\mu)^{-8}_{+17}(Z_m)^{+11}_{-12}(\alfs)^{-8}_{+9}({\cal O}(1/m_h))^{+46}_{-46}
~{\rm MeV}\,,
\eea
in which only the ${\cal O}(\LQ)$ renormalon is subtracted and
the ${\cal O}(1/m_h)$ corrections are ignored (including the $\mu_\pi^2$ term).
Both of these determinations use the PV scheme and
are consistent with our determination
within the assigned errors.

To further examine the convergence property of the FTRS method,
we 
perform an FTRS calculation of the bottom and charm quark pole masses 
(relevant to the above study) by changing the order of the 
approximation.
The result is as follows.
\bea
{\rm N^2LL}&:~~~~~~  
&m_{b}^{\rm FTRS}=4.811(27)~ \text{GeV},~~~  m_{c}^{\rm FTRS}=1.527(34)~ \text{GeV},
\\
{\rm N^3LL}&:~~~~~~  
&m_{b}^{\rm FTRS}=4.831~\,(8)~ \text{GeV},~~~  m_{c}^{\rm FTRS}=1.516(31)~\text{GeV},
\\
{\rm N^4LL}&{\rm (estimate)}:~  
&m_{b}^{\rm FTRS}=4.835 ~\,(6)~ \text{GeV}, ~~~ m_{c}^{\rm FTRS}=1.513(20)~ \text{GeV}.
\eea
The numbers inside parentheses denote the errors from scale variation 
within $\mu\in [\tau/2,2\tau]$. 
In the N$^4$LL estimate, we use the coefficient of the ${\cal O}(\alfs^5)$ term
in the large-$\beta_0$ 
approximation. 
This result indicates convergence of the perturbative series.

\section{Summary and conclusions}

Towards high precision calculations of QCD effects,
it is important to establish a formulation incorporating
IR renormalon cancellation,
beyond the ${\cal O}(\LQ)$ renormalon of the quark pole  mass.
In this paper, we advocate a new method (FTRS method) within the OPE framework,
which realizes separation and
cancellation of IR renormalons of a general observable.
The method can cancel multiple renormalons simultaneously.
We presented first test analyses of this method applied
to three observables (Adler function, $B$-meson semileptonic decay width,
and $B$, $D$ meson masses).
We confirmed
good consistency with theoretical expectations.
In particular we determined $\bar{\Lambda}$ and $\mu^2_\pi$
by subtracting ${\cal O}(\LQ)$ and ${\cal O}(\LQ^2)$ renormalons 
simultaneously for the first time.

The FTRS method is based on the following idea and construction.
We express the leading Wilson coefficient 
of an observable
by a one-parameter
integral, in which the IR renormalons of the integrand are
suppressed.
The integral is constructed by the Fourier transform.
Due to a property of the Fourier transform, one can adjust parameters
in the transform to suppress  the desired renormalons
of the integrand.
Then one can combine the RG and contour deformation to separate and
subtract the renormalons of the observable.
The result coincides with the renormalon-subtracted Wilson coefficient
in the PV prescription at large orders.
This is a generalization of the method used for the static QCD potential,
for which ample tests have already been carried out.

One might have guessed that, in order to separate and eliminate renormalons of a
Wilson coefficient by organizing it in $1/Q^2$ expansion,
analyses of the
IR structures of the individual observables are mandatory
by investigating the details of loop integrals (e.g., using the 
expansion-by-regions technique).
As it turned out, it is possible to construct
the necessary expansion of the Wilson coefficient
without knowing the deep structures.
The only knowledge of the usual perturbative coefficients is sufficient.
The obtained formula has a concise form.
For instance, we do not need to perform
fits to extract the normalization of individual renormalons.

It is also interesting that
one can derive the same formula in another way, which is more
closely connected to the Wilsonian picture,
by introducing a factorization scale in the artificial ``momentum space.''
It may be useful to develop a physical insight into the FTRS formulation.

%

In the latter part of the paper,
we performed test analyses of the FTRS method.
In the first example, 
we subtracted the ${\cal O}(\LQ^4)$ renormalon from the Adler function 
so that the gluon condensate is also well defined. 
The scale dependence of the FTRS calculation 
is smaller than 
the fixed-order and RG-improved calculations, showing
improvement of convergence by the renormalon subtraction.
From comparison with a phenomenological model, consistency
with the OPE is observed.
Then we estimated the values of the gluon condensate and $\LMS^{n_f=2}$
from this comparison, the latter of which agrees with its lattice determinations
from various observables.
The matching was performed in the range where the non-perturbative
$1/Q^4$ contribution plays a significant role, and
the consistency
with the OPE seems non-trivial.
Our estimate of the 5-loop correction
indicates that 
we may need to take care of the $u=-1$ UV renormalon in the Adler function
to improve accuracy by going to higher orders.

In the second example, we subtracted the ${\cal O}(\LQ^2)$ renormalon from 
the $B\to X_ul\overline{\nu}$ decay width,
after canceling the ${\cal O}(\LQ)$ renormalons by
using the $\msbar$ mass.
Incorporating the recently calculated
${\cal O}(\alpha_s^3)$ corrections,
we scarcely observed differences in the scale dependence
of the FTRS, fixed-order and RG-improved calculations.
We estimated a higher-order correction using the 5-loop beta function.
In this estimate the FTRS calculation showed smaller
scale dependence than the fixed-order and RG-improved calculations.
It indicates that we can improve accuracy at higher orders by the 
${\cal O}(\LQ^2)$ renormalon 
subtraction.

In the last example, 
we subtracted the ${\cal O}(\LQ)$ and ${\cal O}(\LQ^2)$ renormalons 
simultaneously from the
$B$ and $D$ meson masses.
We determined
$\bar{\Lambda}$ and $\mu_\pi^2$, which became well defined
by the renormalon subtraction.
With the perturbative corrections up to ${\cal O}(\alpha_s^4)$, 
we compared the derivative of the $b$ quark pole mass
with respect to the $\msbar$ mass, $dm_b/d\mbar_b$, 
between the FTRS calculation, fixed-order calculation, 
and RG-improved calculation.
The scale dependence is similar (except when we
choose a very small scale).
This is consistent with the previous observation that 
the $u=1$ renormalon in the heavy quark pole mass is small.
Comparing to the experimental values of the $B, \,D$ meson masses,
we determined the universal non-perturbative parameters
$\bar{\Lambda}_{\rm FTRS}$ and $( \mu_\pi^2 )_{\rm FTRS}$
taking the PDG values of $\mbar_b,\,\mbar_c,\,\alpha_s(M_Z)$ as inputs.
We obtained
\bea
&&
\bar{\Lambda}_{\rm FTRS}=0.495\pm0.053~\text{GeV} 
\,, 
~~~~~
(\mu_\pi^2)_{\rm FTRS}=-0.12\pm 0.23~\text{GeV}^2
\,.
\label{resLambdamu2final2}
\eea
To ensure universality (heavy quark symmetry) of 
$\bar{\Lambda}_{\rm FTRS}$ and $( \mu_\pi^2 )_{\rm FTRS}$ 
we need to work in the theory with three massless plus massive 
$c$, $b$ quarks consistently.
Furthermore, expansion in the 3-flavor coupling constant
$\alfs^{(3)}(\tau^2)$ is necessary to fully eliminate the
${\cal O}(\LQ)$ and ${\cal O}(\LQ^2)$ renormalons
in the FTRS method. 
The scale dependence
of $\bar{\Lambda}_{\rm FTRS}$ is small  (0.015~GeV),
whereas
$(\mu_\pi^2)_{\rm FTRS}$ has relatively large scale dependence (0.13~GeV$^2$)
with respect to the ${\cal O}(\LQ^2)$ renormalon subtraction. 
We estimate that the latter feature 
is due to the insufficient number of known perturbative coefficients. 
A 5-loop estimate indicates that the error would reduce
if higher order perturbative calculation of the pole mass is achieved.

In all the above analyses we observed good consistency with
theoretical expectations, such as better convergence and stability,
by subtracting renormalons beyond the ${\cal O}(\LQ)$ renormalon.
Nevertheless, we need to know more terms of the
relevant perturbative series in order to 
make conclusive statements about the effects of
subtracting the renormalons.
Our analyses also show that the 5-loop QCD beta function is
a crucial ingredient in improving accuracies.
Application of the FTRS method to the determination of fundamental
physical parameters is under preparation.
It is desirable that higher order perturbative calculations 
of Wilson coefficients will be
advanced in conjunction.
We thus anticipate that the FTRS method can be
a useful theoretical tool for precision QCD calculations 
in the near future.

In these first analyses we used the FTRS method ignoring
the $\log(Q)$ corrections
and the anomalous dimensions $\gamma(\alfs)$ of the 
subtracted renormalons.
(Some of the renormalons have no corrections exactly.)
In principle, these corrections can be incorporated using the method
discussed in App.~\ref{appB}.
Its application to practical analyses is left to our future investigations.


\section*{Acknowledgements}
Y.H.\ acknowledges support from GP-PU at Tohoku University.
The work of Y.H.\ was also supported in part by Grant-in-Aid for JSPS Fellows (No. 21J10226) from MEXT, Japan.
The works of Y.S.\ and H.T., respectively, were supported in part by Grant-in-Aid for
scientific research (Nos.\  20K03923 and 19K14711) from
MEXT, Japan.

\newpage

\section*{{Appendix}}
\appendix
\vspace{-0.32cm}
\section{List of perturbative coefficients}
\label{App:PertCoeffs}

In this appendix, we collect the perturbative coefficients
necessary for the analyses in this paper.

The QCD $\beta$ function is known up to ${\cal O}(\alpha_s^6)$ (5-loop accuracy)
 \cite{Baikov:2017ayn}.
\be
\beta(\alpha_s)=-\sum_{i=0}^4b_i\alpha_s^{i+2},
\ee
\be
b_0=\frac{1}{4\pi}\Big(11-\frac{2}{3}n_f\Big),\quad
b_1=\frac{1}{(4\pi)^2}\Big(102-\frac{38}{3}n_f\Big),
\ee
\be
b_2=\frac{1}{(4\pi)^3}\Big(\frac{2857}{2}-\frac{5033}{18}n_f+\frac{325}{54}n_f^2\Big),
\ee
\bea
b_3=\frac{1}{(4\pi)^4}\Big[\frac{149753}{6}
&+&3564\zeta_3-\Bigg(\frac{1078361}{162}+\frac{6508}{27}\zeta_3\Big)n_f\nonumber\\
&+&\Big(\frac{50065}{162}+\frac{6472}{81}\zeta_3\Big)n_f^2+\frac{1093}{729}n_f^3\Bigg],
\eea
\bea
b_{4}&=&\frac{1}{(4\pi)^{5}}\Bigg[\frac{8157455}{16}+\frac{621885}{2} \zeta_{3}-\frac{88209}{2} \zeta_{4}-288090 \zeta_{5}\nonumber\\
&+&n_{f}\left(-\frac{336460813}{1944}-\frac{4811164}{81} \zeta_{3}+\frac{33935}{6} \zeta_{4}+\frac{1358995}{27} \zeta_{5}\right)\nonumber\\
&+&n_{f}^{2}\left(\frac{25960913}{1944}+\frac{698531}{81} \zeta_{3}-\frac{10526}{9} \zeta_{4}-\frac{381760}{81} \zeta_{5}\right)\nonumber\\
&+&n_{f}^{3}\left(-\frac{630559}{5832}-\frac{48722}{243} \zeta_{3}+\frac{1618}{27} \zeta_{4}+\frac{460}{9} \zeta_{5}\right)+n_{f}^{4}\left(\frac{1205}{2916}-\frac{152}{81} \zeta_{3}\right)\Bigg].
\eea
$n_f$ is the number of active quark flavors, and $\zeta_n=\zeta(n)=\sum_{k=1}^\infty k^{-n}$
denotes the Riemann zeta function.

The Adler function is known up to ${\cal O}(\alpha_s^4)$ \cite{Baikov:2010je,Baikov:2012zn}.
\be
D(Q^2)_{PT}=3\sum_f Q_f^2+\sum_{n=0}^3 a_n\alpha_s(Q)^{n+1},
\ee
where 
\be
a_n=3\sum_f Q_f^2a_n^{NS}+3\Big(\sum_f Q_f\Big)^2a_n^{SI},
\ee
\be
a_0^{NS}=\frac{1}{\pi},\quad a_0^{SI}=0,
\ee
\be
a_1^{NS}=3\sum_f Q_f^2\frac{1}{\pi ^2}\left[\frac{365}{24}-{11 \zeta_3}+  n_f \left(\frac{2}{3}\zeta_3-\frac{11}{12}\right)\right],~~
a_1^{SI}=0,
\ee
\bea
a_2^{NS}&=&\frac{1}{\pi ^3}\Bigg[ \frac{62645}{288}-\frac{1103 \zeta_3}{4}+\frac{275 \zeta_5}{6}\non
&&~~~~~+  n_f \left(-\frac{7847}{216}+\frac{262 \zeta_3}{9}-\frac{25 \zeta_5}{9}\right)+n_f^2 \left(\frac{151}{162}-\frac{19 \zeta_3}{27}\right)\Bigg],
\eea
\be
a_2^{SI}=\frac{1}{\pi ^3}\Bigg[\frac{55}{216}-\frac{5\zeta_3}{9}\Bigg],
\ee
\bea
a_3^{NS}&=&\frac{1}{\pi^4}\Bigg[\frac{144939499}{20736}+\frac{5445 \zeta _3^2}{8}-\frac{5693495
   \zeta _3}{864}+\frac{65945 \zeta _5}{288}-\frac{7315 \zeta
   _7}{48}\non
   &&~~~+ n_f\left(-\frac{13044007}{10368}-55 \zeta _3^2+\frac{12205 \zeta
   _3}{12}+\frac{29675 \zeta _5}{432}+\frac{665 \zeta
   _7}{72}\right)\non
    &&~~~+n_f^2\left(\frac{1045381}{15552}+\frac{5 \zeta _3^2}{6}-\frac{40655 \zeta _3}{864}-\frac{260 \zeta _5}{27}\right) \non
    &&~~~+n_f^3\left(-\frac{6131}{5832}+\frac{203 \zeta _3}{324}+\frac{5 \zeta _5}{18}\right)\Bigg],
   \eea
\bea
a_3^{SI}&=&\frac{1}{\pi^4}\Bigg[\frac{5795}{576}-\frac{55 \zeta _3^2}{12}-\frac{8245 \zeta
   _3}{432}+\frac{2825 \zeta _5}{216}+\left(-\frac{745}{1296}+\frac{5 \zeta _3^2}{18}+\frac{65 \zeta _3}{72}-\frac{25 \zeta
   _5}{36}\right) n_f\Bigg] \,.
\nonumber\\
\eea
$Q_f$ represents the electric charge of the quark $f$.
($2/3$ for the up-type quark and $-1/3$ for the down-type quark.)

The pole-$\msbar$ mass relation
is known up to ${\cal O}(\alfs^4)$ in the limit that all the masses of
the quarks in internal loops are 
zero \cite{Marquard:2015qpa,Marquard:2016dcn}.
The corrections $\delta d_n$ from the  non-zero mass of 
one of the internal quarks 
(or the non-decoupling effects of an internal heavy quark)
are known up to ${\cal O}(\alfs^3)$ \cite{Fael:2020bgs}.
These are given as follows.
\be
\frac{m_h}{\mbar_h}=c_m(\mbar_h)=1+\sum_{n=0}^3d_n\alpha_s(\mbar_h)^{n+1},
\ee
where $\alpha_s(\mbar_h)=\alpha_s^{(n_l)}(\mbar_h)$ with $n_l=n_f-1$.
(There are $n_l$ massless and one massive internal quarks, besides
the heavy quark $h$.)
\be
d_0=\frac{4}{3\pi},
\ee
and we decompose $d_n$ into two parts as
\be
d_n=\tilde{d}_n+\delta d_n\quad{\rm for}\quad n\geq1.
\ee
\be
\tilde{d}_1=\frac{1}{\pi ^2}\Bigg[\left(-\frac{71}{144}-\frac{\pi ^2}{18}\right) {n_l}-\frac{\zeta_3}{6}+\frac{\pi ^2}{3}+\frac{307}{32}+\frac{1}{9} \pi ^2 \log (2)\Bigg] \,,
\label{dtilde1}
\ee
\bea
\tilde{d}_2&=&\frac{1}{\pi ^3}\Bigg[n_l^2 \left(\frac{7 \zeta_3}{54}+\frac{2353}{23328}+\frac{13 \pi
   ^2}{324}\right)+n_l\Big(\frac{8
   \text{Li}_4\left(\frac{1}{2}\right)}{27}-\frac{241 \zeta_3}{72}-\frac{231847}{23328}-\frac{991 \pi ^2}{648}+\frac{61 \pi
   ^4}{1944}\nonumber\\
   &+&\frac{2}{81} \pi ^2 \log ^2(2)+\frac{\log ^4(2)}{81}-\frac{11}{81} \pi ^2
   \log (2)\Big)-\frac{220 \text{Li}_4\left(\frac{1}{2}\right)}{27}+\frac{1975 \zeta_5}{216}-\frac{1439 \pi ^2 \zeta_3}{432}+\frac{58 \zeta_3}{27}\nonumber\\
   &-&\frac{695 \pi
   ^4}{7776}+\frac{652841 \pi ^2}{38880}+\frac{8462917}{93312}-\frac{55 \log
   ^4(2)}{162}-\frac{22}{81} \pi ^2 \log ^2(2)-\frac{575}{162} \pi ^2 \log (2)\Bigg].
\eea
The complete analytical formula of $\tilde{d}_3$ is still unknown,
and numerically it is given by
\be
\tilde{d}_3\approx -0.006962 n_l^3+0.4455 n_l^2-7.651 n_l+36.57.
\label{dtilde3}
\ee
In Sec.~$\ref{dmdm}$,
we use the series with the coefficient $\tilde{d}_n$ up to $n=3$ with $n_l=4$,
\be
\frac{m_b}{\mbar_b}\approx1+0.424413\alpha_s+ 0.940051\alpha_s^2+3.03854\alpha_s^3+12.6474\alpha_s^4\,.
\label{app:formula5.2}
\ee

The full forms of the corrections $\delta d_n$ are too lengthy to be shown here.
The series we used in Sec.~\ref{det-LambdaBar-mupi}, 
including the non-zero $m_c$ corrections up to ${\cal O}(\alpha_s^3)$, is given by
\be
\frac{m_b}{\mbar_b^{(5)}}
\approx 1 + 0.424413\alpha_s +  1.03744 \alpha_s^2 + 3.74358 \alpha_s^3 + 
17.4376 \alpha_s^4+97.5872\alpha_s^5,
\label{mb3f}
 \ee
where $\alpha_s=\alpha_s^{(3)}(\mbar_b)$ with $\mbar_b=\mbar_b^{(5)}$,
and
\be
\frac{m_c}{\mbar_c^{(4)}}
\approx 1 + 0.424413 \alpha_s +1.04375\alpha_s^2 +  3.75736 \alpha_s^3 + 
17.4376 \alpha_s^4+97.5872\alpha_s^5,
\label{mc3f}
 \ee
where $\alpha_s=\alpha_s^{(3)}(\mbar_c)$ with $\mbar_c=\mbar_c^{(4)}$ and
non-decoupling bottom effects are included.
In obtaining the right-hand sides, we used the inputs $\mbar_c^{(4)}=1.27~{\text{GeV}}$ 
and $\mbar_b^{(5)}=4.18~{\text{GeV}}$.
In each formula, ${\cal O}(\alpha_s^4)$ correction is $\tilde{d}_3(n_l=3)$, 
and ${\cal O}(\alpha_s^5)$ correction is approximated in the large-$\beta_0$ approximation.
In our central analysis (to obtain eqs.~\eqref{Lambdabarfinal} and \eqref{myupifinal})
we use the series up to $\mathcal{O}(\alpha_s^4)$.
Both of ${\cal O}(\alpha_s^2)$ and ${\cal O}(\alpha_s^3)$ corrections are quite similar 
to each other 
and each series can be approximated by $\sum_n \tilde{d}_n\alpha_s^{n+1}$ with $n_l=3$,
given by
\be
\frac{m_h}{\mbar_h}\approx1+0.424413\alpha_s+ 1.04556\alpha_s^2+3.75086\alpha_s^3+17.4376\alpha_s^4.
\label{mh3f}
\ee

The partial decay width $\Gamma(B\to X_ul\overline{\nu})$
is known up to ${\cal O}(\alpha_s^3)$ \cite{Pak:2008cp,Fael:2020tow}, 
where all the quarks inside loops are massless.
It is given by
\be
\Gamma(m_b)_{PT}=\frac{G_F^2|V_{ub}|^2}{192\pi^3}m_b^5X^\Gamma(m_b),
\ee
\be
X^\Gamma(m_b)=1+g_0\alpha_s(m_b)+g_1\alpha_s(m_b)^2+g_2\alpha_s(m_b)^3,
\ee
where $\alpha_s(m_b)=\alpha_s^{(n_l)}(m_b)$.
\bea
g_0=\frac{1}{\pi}\Bigg(\frac{25}{6}-\frac{2\pi^2}{3}\Bigg),
\eea
\bea
g_1&=&\frac{1}{\pi ^2}\Bigg[ \frac{7660327}{93312}
-\frac{6959 \zeta_3}{162}+\frac{289 \pi ^4}{648}-\frac{106775 \pi
   ^2}{11664}-\frac{53}{27} \pi ^2 \log (2)\non
   &&~~+\left(-\frac{1009}{576}+\frac{4 \zeta _3}{3}+\frac{77 \pi
   ^2}{432}\right) n_l\Bigg].
\eea
The order $\alpha_s^3$ correction is calculated with uncertainty of about 10 \%,
using expansion in $1-m_c/m_b$ up to 12th order, which leads to
\be
g_2=\frac{4}{3\pi^3}\big(-202\pm20\big)\,,
\label{g2-Num}
\ee
for $m_c=0$.
(The effect of the error in the above value is negligible in our analysis.)

\section{Including logarithmic corrections to FTRS method}
\label{appB}
In Sec.~\ref{sec:FTRS}, 
we saw the suppression of renormalons in ``momentum space''
by an appropriate choice of the parameters $(a,u')$, see eq.~(\ref{FTdelX}).
It can be extended to a more general case 
with logarithmic (perturbative) corrections of the renormalons
or non-zero values of the anomalous dimensions of the corresponding operators. 
We demonstrate how these can be incorporated into the FTRS method.

For heuristic reasons we present most argument in expansion in
$\log(Q_0^2/Q^2)$ with respect to an arbitrary chosen scale $Q=Q_0$.
In this way we can start from the limit where we know the answer
already (the case without logarithmic corrections).
Nonetheless, the
expansion in $\log(Q_0^2/Q^2)$ can be resummed using RG.
At the end of this appendix we show how to resum $\log(Q_0^2/Q^2)$'s
and obtain $Q_0$ independent expressions.

First, let us consider the case for suppressing only the leading renormalon at $u=u_*$.
The renormalon ambiguity has the form $\delta X_{u_*}\propto
(b_0\alpha_s(Q))^{\gamma_0/b_0}Q^{-2u_*}\sum_{n}s_n(1)\alpha_s(Q)^{n}$
according to eq.~(\ref{renu}).
We expand $\alfs(Q)$ in $\log(Q_0^2/Q^2)$ about $Q=Q_0$ using
\be
\alfs(Q)=\sum_{n=0}^\infty\frac{\log^n(Q_0^2/Q^2)}{n!}\Big[-\beta(\alpha_s(Q_0))\frac{\partial}{\partial \alpha_s(Q_0)}\Big]^n\alpha_s(Q_0),
\ee
where $Q_0$ is an arbitrary expansion point.\footnote{
It is natural to take $Q_0$ within the energy range where we use the OPE,
such that $\alfs(Q_0)\log(Q_0^2/Q^2)$ can be regarded as a small parameter.
}
Then
$\delta X_{u_*}$ of eq.~\eqref{renu}
can be written in the form 
\be
\delta X_{u_*}=A(u_*)\sum_{n=0}^\infty q_n \log^n(Q^{-2})Q^{-2u_*}
=A(u_*)\sum_{n=0}^\infty q_n\Big(\frac{\partial}{\partial u_*}\Big)^nQ^{-2u_*},
\label{def-qn}
\ee
\be
A(u_*)=\frac{\pi}{b_0}\frac{N_{u_*}}{\Gamma(1+\nu_{u_*})}{u_*}^{1+\nu_{u_*}}\LMS^{2u_*},\quad
\nu_{u_*}=u_*\frac{b_1}{b_0^2}-\frac{\gamma_0}{b_0}.
\ee
$q_n$ is given by a combination of 
$s_i$, $\alpha_s(Q_0)$, $\log Q_0^2$, and the coefficients of the beta function 
and anomalous dimension ($b_i$ and $\gamma_i$).

To construct $\tilde{X}(\tau)$, define
the series $\sum_{m=0}^\infty p_m x^m$ by
\be
\sum_{m=0}^\infty p_m x^m=\dfrac{1}{\sum_{n=0}^\infty q_n x^n}\quad
\stackrel{\text{equiv.}}{\Longleftrightarrow}
\quad 
\sum_{m,n=0}^\infty p_mq_n x^{m+n}=1,
\ee
or,
\be
p_0=1/q_0\,,
~~~~~
\sum_{i=0}^Np_iq_{N-i}=0~~{\rm for}~~ N=1,2,3,\cdots.
\ee
Then we can define 
\be
\tilde{X}(\tau)=\sum_{m}p_m\Big(\frac{\partial}{\partial u'}\Big)^m\int d^3\vec{x}\,e^{-i\vec{\tau}\cdot\vec{x}}r^{2au'}X(r^{-a});\quad r=Q^{-1/a},\label{tildeXLO}
\ee
which is a generalized version of eq.~(\ref{tildeX}).
In fact,
\bea
\delta \tilde{X}&=&\sum_{m}p_m\Big(\frac{\partial}{\partial u'}\Big)^m\int d^3\vec{x}\,e^{-i\vec{\tau}\cdot\vec{x}}r^{2au'}A(u_*)\sum_{n=0}^\infty q_n\Big(\frac{\partial}{\partial u_*}\Big)^nr^{2au_*}\nonumber\\
&=&A(u_*)\sum_{m,n}p_mq_n\Big(\frac{\partial}{\partial u'}\Big)^m\Big(\frac{\partial}{\partial u_*}\Big)^nf\big(a(u_*+u');\tau\big)\nonumber\\
&=&A(u_*)\sum_{m,n}p_mq_n\Big(\frac{\partial}{\partial u'}\Big)^{m+n}f\big(a(u_*+u');\tau\big)\nonumber\\
&=&A(u_*)f\big(a(u_*+u');\tau\big),
\eea
where
\be
f(s;\tau)=
\int d^3\vec{x}\,e^{-i\vec{\tau}\cdot\vec{x}}r^{2s}
=-4\pi\,\frac{\sin(\pi s)\Gamma(2s+2)}{\tau^{2s+3}}.
\ee
To suppress the renormalon, we can take the same $(a,u')$ as the ones without
the logarithmic corrections, e.g., $u'=-u_*$ and $a=1$. 
Then $\delta \tilde{X}_{u=u_*}=0$ up to an arbitrary order in the $\log(Q_0^2/Q^2)$ expansion.
This means that we can construct an appropriate $\tilde{X}$ using the sequence $c_n$ 
[cf.\  eq.~\eqref{PTXcn}] and Fourier transform.

The inverse transform can be constructed as 
\bea
X(r^{-a}) =
r^{-2au'}\sum_{n=0}^\infty q_n (2a\log(r))^n\int \frac{d^3\vec{\tau}}{(2\pi)^3}\,e^{i\vec{\tau}\cdot\vec{x}}\tilde{X}(\tau) \,.
\label{invtransf-logcorr}
\eea
In fact, the right-hand side can be written as
\bea
&& \! \! \!
r^{-2au'}\sum_{m,n}p_mq_n
\int d^3\vec{x'}\int \frac{d^3\vec{\tau}}{(2\pi)^3}\,e^{-i\vec{\tau}\cdot(\vec{x'}-\vec{x})}r'^{2au'}
(2a\log(r))^n (2a\log(r'))^m
X(r'^{-a})\nonumber\\
&=&
r^{-2au'}\sum_{m,n}p_mq_n
\int d^3\vec{x'}\delta^3(\vec{x'}-\vec{x})r^{2au'}
(2a\log(r))^{m+n} X(r^{-a})\nonumber\\
&=&
r^{-2au'}\sum_{m,n}p_mq_n
r^{2au'}(2a\log(r))^{m+n}X(r^{-a})\nonumber\\
&=&
X(r^{-a}).
\eea
The formulation in Sec.~\ref{sec:FTRS}  is a special case where $q_1=q_2=\cdots=0$.

Secondly, we suppress the leading and next-to-leading renormalons simultaneously.
For simplicity of calculation, let us assume that they are 
at $u=u_*,\,u_*+1$, so that we take $a=1$.
For $k=0,1$, we write the renormalons as
\be
\delta X_{k}\equiv \delta X_{u=u_*+k}
=A^{(k)}\sum_{n=0}^\infty q^{(k)}_n\Big(\frac{\partial}{\partial u_*}\Big)^nQ^{-2(u_*+k)},
\ee
with
\be
\quad A^{(k)}=\frac{\pi}{b_0}\frac{N_{u_*+k}}{\Gamma(1+\nu_{u_*+k})}{(u_*+k)}^{1+\nu_{u_*+k}}\LMS^{2(u_*+k)}.
\ee

Similarly to the previous case, we define $\tilde{X}(\tau)$ as
(note that $a=1$)
\be
\tilde{X}(\tau)=\sum_{l=0}^1\tau^{2l}\sum_{m}p_m^{(l)}\Big(\frac{\partial}{\partial u'}\Big)^m\int d^3\vec{x}\,e^{-i\vec{\tau}\cdot\vec{x}}r^{2(u'+l)}X(r^{-1}).\label{tildeXNLO}
\ee
To suppress the renormalons at $u=u_*,\,u_*+1$ simultaneously,
the following equations need to be satisfied for $k=0,1$:
\be
\delta\tilde{X}_k(\tau)=0\,\Leftrightarrow\,\sum_{N=0}^\infty\Bigg[\sum_{l=0}^1\sum_{i=0}^N\tau^{2l}p_i^{(l)}q_{N-i}^{(k)}\Big(\frac{\partial}{\partial u'}\Big)^{N}
f(u_*+u'+l+k;\tau)\Bigg]=0.
\ee
As a trial analysis, let us truncate the summation at a fixed $N$ and
see if we can find a solution for $\{p_i^{(l)}\}$. 
The condition reads
\be
\tau^{2s+3}\sum_{i=0}^N\bigg[p_i^{(0)}q_{N-i}^{(k)}\Big(\frac{\partial}{\partial s}\Big)^{N}
f(s;\tau)+p_i^{(1)}q_{N-i}^{(k)}\Big(\frac{\partial}{\partial s}\Big)^{N}
\tau^{2}f(s+1;\tau)\bigg] \, \Biggr|_{s=u_*+u'+k}=0,
\ee
for $k=0,1$.
Let us set $u'=-u_*$. 
Noting that $f(s;\tau), \tau^2 f(s+1;\tau)\propto 1/\tau^{2s+3}$,
the left-hand side is an $(N\!-\!1)$th-order
polynomial of $\log\tau$, and
we write
$\tau^2f(s+1;\tau)=-(2s+3)(2s+2)f(s;\tau)=g(s;\tau)$.
Hence,
\be
\tau^{2s+3}\sum_{i=0}^N\bigg[p_i^{(0)}\frac{\partial^N f(s;\tau)}{\partial s^N}+p_i^{(1)}\frac{\partial^N g(s;\tau)}{\partial s^N}\bigg]q_{N-i}^{(k)}
\Biggr|_{s=k}=0,\label{conN}
\ee
For $N=0$ the condition is satisfied for arbitrary $(p_0^{(0)},p_0^{(1)})$
since $f(k;\tau)=g(k;\tau)=0$ for $k=0,1$.
For general $N$, equating each coefficient of $\log^m(\tau)$ to zero in eq.~(\ref{conN}),
there are $2N$ linear equations for $2(N+1)$ variables 
$\{p_0^{(0)},p_0^{(1)}\},\cdots ,\{p_N^{(0)},p_N^{(1)}\}$, 
which indicates that there is a non-trivial solution for $\{p_i^{(l)}\}$.

We can construct a non-trivial solution
in a more sophisticated way as follows.
We define 
\be
G_{k,l}(r)=\sum_nq_n^{(k)}\Big(\frac{\partial}{\partial u_*}\Big)^nr^{2(u_*+u'+l+k)},
\label{def-matG}
\ee
and $F_l(r)$ satisfying 
\be
\left(
\begin{array}{cc}
G_{0,0} & G_{1,0} \\
G_{0,1} & G_{1,1}\\
\end{array}
\right)
\left(
\begin{array}{c}
F_0(r) \\
F_1(r)\\
\end{array}
\right)
=
\left(
\begin{array}{c}
\alpha_0 \\
\alpha_1r^2\\
\end{array}
\right)
r^{2(u_*+u')},
\ee
where $\alpha_0,\alpha_1$ are $r$-independent arbitrary coefficients.\footnote{
These parameters correspond to $\{p_0^{(0)},p_0^{(1)}\}$.
In the case that the matrix on the left-hand side does not have its inverse, 
we adjust $\alpha_0/\alpha_1$ such that $F_l(r)$ have a non-trivial solution.
}
In particular, $\alpha_k,\,F_l(r)$ are $\tau$-independent.
If we define
\be
\tilde{X}(\tau)=\sum_{l=0}^1\int d^3\vec{x}
\,e^{-i\vec{\tau}\cdot\vec{x}}r^{2(u'+l)}
X(r^{-1})F_l(r) \,,
\ee
we find that
$\delta\tilde{X}_k(\tau)=0$ for $k=0,1$. 
\bea
\because
\delta\tilde{X}_k(\tau)
&=&\sum_{l=0}^1\int d^3\vec{x}
\,e^{-i\vec{\tau}\cdot\vec{x}}r^{2(u'+l)}
\delta X_k(r^{-1})F_l(r)\nonumber\\
&=&A^{(k)}\sum_{l=0}^1\int d^3\vec{x}
\,e^{-i\vec{\tau}\cdot\vec{x}}
\sum_{n=0}^\infty q^{(k)}_n\Big(\frac{\partial}{\partial u_*}\Big)^nr^{2(u_*+u'+l+k)}F_l(r)
\nonumber\\
&=&
A^{(k)}\int d^3\vec{x}
\,e^{-i\vec{\tau}\cdot\vec{x}}
\sum_{l=0}^1G_{k,l}(r)F_l(r)
\nonumber\\
&=&
A^{(k)}\alpha_k\int d^3\vec{x}
\,e^{-i\vec{\tau}\cdot\vec{x}}
r^{2(u_*+u'+k)}
=A^{(k)}\alpha_kf(u_*+u'+k;\tau)=0.
\eea

The inverse transform is given by
\bea
&&
X(r^{-1}) =
\frac{r^{-2u'}}{\alpha_k}\sum_{n}q_n^{(k)}\Big(2\log(r)\Big)^n\int\frac{d^3\vec{\tau}}{(2\pi)^3}e^{i\vec{\tau}\cdot\vec{x}} 
\tilde{X}(\tau) \,.
\label{invtransf-2renorm}
\eea
We can construct it for either $k=0$ or $1$.
(The results are the same.)
In fact the right-hand side can be written as
\bea
&&\!\!\!
\frac{r^{-2u'}}{\alpha_k}\sum_{l,n}q_n^{(k)}\Big(2\log(r)\Big)^n
\int d^3\vec{x'}\int\frac{d^3\vec{\tau}}{(2\pi)^3}e^{-i\vec{\tau}\cdot(\vec{x'}-\vec{x})}
r'^{2(u'+l)}
X(r'^{-1})F_l(r)
\nonumber\\
&=&\frac{r^{-2(u_*+u'+k)}}{\alpha_k}\sum_{l}G_{k,l}(r)F_l(r)X(r^{-1})
=X(r^{-1}).
\eea

Finally we present the formulas corresponding to
eqs.~\eqref{tildeXLO}, \eqref{invtransf-logcorr}, \eqref{def-matG}
and \eqref{invtransf-2renorm},
after resummation of  $\log(Q_0^2/Q^2)$'s.
According to eqs.~\eqref{renu} and \eqref{def-qn}, the relation between
the expansion coefficients in $\log(Q_0^2/Q^2)$ and in $\alfs(Q)$
is given by
\bea
\sum_{n=0}^\infty q_n \log^n(Q^{-2}) = 
(b_0\alfs(Q))^{\gamma_0/b_0}\sum_{m=0}^\infty s_m(1)\,
\alfs(Q)^{m}
\,.
\eea
Then it is readily seen that the following expression is 
equivalent to eq.~\eqref{tildeXLO}:
\be
\tilde{X}(\tau)=
\int d^3\vec{x}\,
\frac{e^{-i\vec{\tau}\cdot\vec{x}}\,r^{2au'}X(r^{-a})}
{(b_0\alfs(r^{-a}))^{\gamma_0/b_0} \sum_m s_m(1)\,
\alfs(r^{-a})^{m}}
\,.
\ee
This is an RG invariant expression.
To obtain an explicit expression 
of $\tilde{X}(\tau)$ up to N$^k$LL, we express the
integrand in expansion in $\alfs\equiv\alfs(\mu)$, 
Fourier transform order by order in $\alfs$
up to the $k$-th order,
and then set $\mu=\tau$.
Eq.~\eqref{invtransf-logcorr} can be written as
\bea
X(r^{-a}) =
r^{-2au'}\,
(b_0\alfs(r^{-a}))^{\gamma_0/b_0}\,
\sum_m s_m(1)\,
\alfs(r^{-a})^{m}\,
\int \frac{d^3\vec{\tau}}{(2\pi)^3}\,e^{i\vec{\tau}\cdot\vec{x}}\tilde{X}(\tau) \,.
\eea
Similarly the relation between the expansion coefficients in the case
with two
renormalons reads
\bea
\sum_{n=0}^\infty q_n^{(k)} \log^n(Q^{-2}) = 
(b_0\alfs(Q))^{\delta_k}\sum_{m=0}^\infty s_m^{(k)}(1)\,
\alfs(Q)^{m}\,,
\eea
where $\delta_k=\gamma_0^{(k)}/b_0$.
The following formulas achieve
resummation of logarithms in eqs.~\eqref{def-matG}
and \eqref{invtransf-2renorm}:
\bea
&&
G_{k,l}(r)=(b_0\alfs(r^{-1}))^{\delta_k}
 \sum_m s_m^{(k)}(1)\,
\alfs(r^{-1})^{m}\,
r^{2(u_*+u'+l+k)},
\\
&&
X(r^{-1}) =
\frac{r^{-2u'}}{\alpha_k}\,(b_0\alfs(r^{-1}))^{\delta_k}
\sum_m s_m^{(k)}(1)\,\alfs(r^{-1})^{m}\,
\int\frac{d^3\vec{\tau}}{(2\pi)^3}e^{i\vec{\tau}\cdot\vec{x}} 
\tilde{X}(\tau) \,.
\eea

We anticipate that the formulation presented in this appendix can be extended to the case 
$a\neq1$ and $k=2,3,4,\cdots$.

\section{Relation between contour integrals of Borel transform and Fourier transform
}
\label{AppC}

In this appendix, we derive a relation between a regularized inverse Borel
transform and a regularized inverse Fourier transform.
The relation is used to show equivalence
of the renormalon-subtracted 
Wilson coefficient in the usual PV prescription and
that in the FTRS method.

We consider\footnote{
The reason why we limit $k$ to this range is that
we do not know the QCD beta function beyond five loops and
hence in the argument below the singularity structure in the complex
$p$ plane cannot be made definite.
} 
 the observable in the $\tau$ space
$\tilde{X}(\tau)=\tilde{X}^{(k)}(\tau)$  in the N$^k$LL approximation
for $0\le k\leq 4$, as given by eq.~\eqref{FTXpert}.
From it, the perturbative
coefficients of 
$X(Q)=\sum_{n}{c}_n(\mu/Q)\alfs^{n+1}$
can be computed up to arbitrarily high orders,
by inverse
Fourier transformation at each order of expansion in $\alfs$.
We assume that 
the Borel transform of $X(Q)$, 
\bea
B_{X}(u)=\sum_{n=0}^\infty\frac{c_n}{n!}\left(\frac{u}{b_0}\right)^n \,,
\label{app:BXu}
\eea
does not have singularities in the right half $u$-plane, ${\rm Re}\,u>0$,
except on the positive real $u$-axis.\footnote{
For $k=0$ (LL) the assumption indeed holds, see below.
Although we believe that this assumption can be
checked for $1\le k\le 4$, up to now we do not know the proof or disproof.
}
If $|u|$ is smaller than the distance to the renormalon closest to the origin,
the series converges and
$B_X(u)$ is single-valued.
At larger $|u|$, $B_X(u)$ is defined by analytic continuation.
For $k=0$ the closest renormalon is a pole, while for $1\le k\le 4$ it is a
branch point, where the branch cut extends along the real axis to the right
\cite{Beneke:1998ui}.

\begin{figure}[t]
        \centering
        \includegraphics[width=7cm]{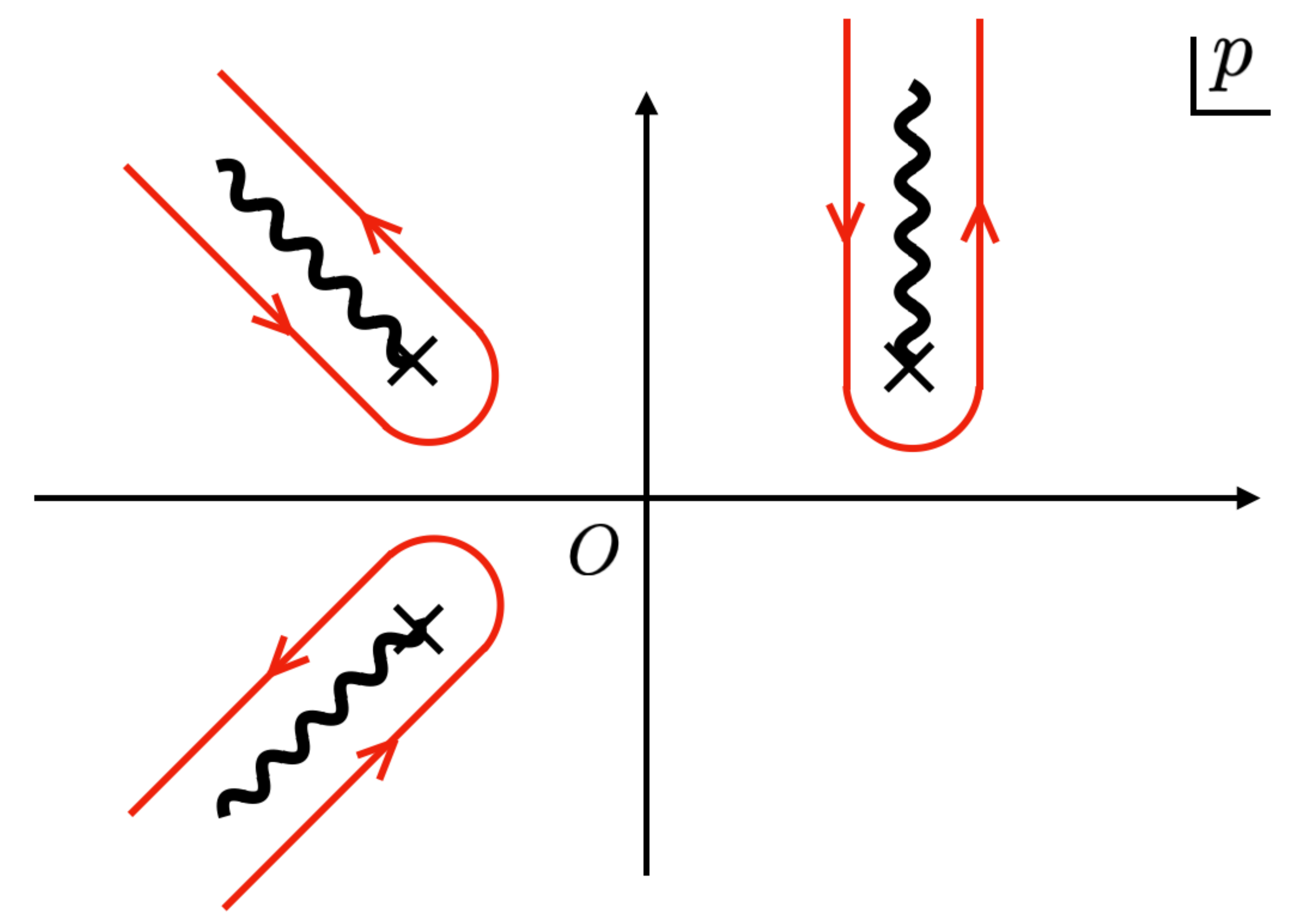}~~~~~~
        \caption
        {\small
		Schematic diagram of the contour of $p$ integration in 
eq.~\eqref{app:defBX+}.
Each branch cut extends to the direction in which the integral wrapping it
converges.}
        \label{fig:psing-contour}
\end{figure}

Let us define
\bea
&&
X_+(Q)
= \frac{r^{-2au'-1}}{2\pi^2}\int_{C_-} d\tau\,\tau\,\sin\left(\tau r\right)\tilde{X}(\tau)~~~;~~~r=Q^{-1/a},
\label{app:defX+}
\\&&
B_{X+}(u)=\oint\frac{dp}{2\pi i}\, e^{up/b_0}\big[X_+(Q)\big]_{\alfs\to1/p}\,.
\label{app:defBX+}
\eea
$\tilde{X}$ has a singularity corresponding to the (Landau) singularity
of $ [\alfs(\tau^a)]_{\rm N^kLL}$ at $\tau>0$.
Along 
the $\tau$ integration contour $C_-(\tau)$ the singularity is circumvented
to the lower half plane.
Thus, $X_+$ is well defined.
In the second equation, $X_+$ is regarded as a function of
$\alfs\equiv\alfs(\mu)$ and is rewritten in terms of $p=1/\alfs$.
The integral contour of $p$ is taken as a closed path surrounding all the 
singularities of $\big[{X}_+(Q)\big]_{\alfs\to 1/p}$ counterclockwise;
see Fig.~\ref{fig:psing-contour}.
This contour is obtained by a continuous deformation of
the closed contour surrounding the origin $p=0$ in a
small $|p|$ region.

In the case that we apply eqs.~\eqref{app:defX+} and \eqref{app:defBX+}  to 
the series expansion in $\alfs$ (up to arbitrary order), the expansion of
$\tilde{X}(\tau)$ has no singularity at $\tau\in {\bf R}_{> 0}$.
Then, 
$C_-(\tau)$ can be deformed to the positive
real $\tau$ axis, and the expansion of
$X_+(Q)$ coincides with that of $X(Q)$;
the singularities of the expansion of
$\big[{X}(Q)\big]_{\alfs\to 1/p}$ are multiple 
poles at
the origin $p=0$, and 
the expansion of $B_{X+}$ reduces to
eq.~\eqref{app:BXu}, which
follows readily by the residue theorem.
Hence, $B_{X+}$ coincides with $B_{X}$  for small $|u|$,
or in other words, $B_{X+}$ is defined as an analytic continuation of $B_{X}$.

We define the regularized Borel summation representation of
 $X(Q)=\sum_{n}{c}_n(\mu/Q)\alfs^{n+1}$ as
\bea
{X}(Q)_{\rm BI,+}=\frac{1}{b_0}\int_{C_+} du\,e^{-\frac{u}{b_0\alpha_s}}B_{X+}(u)
=i\int_0^\infty ds\,e^{-\frac{is}{\alpha_s}}B_{{X+}}(ib_0s)\,,
\eea
where
the integration contour of $u$ is rotated to the positive imaginary axis $(u=ib_0s)$\footnote{
This equality is invalid if $B_X(u)$ contains singularities in the first quadrant of complex Borel $u$-plane. 
We assume that the contribution from such singularities is negligible in the high energy region,
while it would be a restriction in the low energy region.
For the Adler function, 
this kind of restriction has been discussed recently~\cite{Hoang:2020mkw}.
}. 

The reason why we choose the contour $C_-(\tau)$ in eq.~\eqref{app:defX+}
rather than $C_+(\tau)$ is as follows.
Let us explain in the case that
$\tilde{X}(\tau)$ is given by the LL approximation, 
$\tilde{X}\propto [\alfs(\tau^a)]_{\rm LL}$.
(We explain the NLL approximation and beyond later.)
Thus,
\bea
{X}(r^{-a})_{\rm BI,+}
&=&\frac{r^{-2au'-1}}{4\pi^3}
\oint dp
\int_0^\infty \!\! ds\,e^{is(p-1/\alfs)}
\int_{C_-} d\tau\,\tau\,\sin\left(\tau r\right)
\big[\tilde{X}(\tau)\big]_{\alfs\to1/p}
 \,,
\label{Bl+}
\eea
where
the integration contour of $p$ surrounds the pole
of $\tilde{X}$ at\footnote{
This is the only singularity of $\tilde{X}$ in the LL approximation.
}
$p=b_0\log(\mu^2/\tau^{2a})$, which originates from
$[\alfs(\tau^a)]_{\rm LL}=1/[p-b_0\log(\mu^2/\tau^{2a})]$.
We would like to integrate over $s$ first.
To ensure convergence at $s\to\infty$, the imaginary part of $p$ should be 
non-negative.
(Note $\alfs>0$.)
By taking the path of $\tau$
(slightly) in the lower-half plane, i.e., along $C_-(\tau)$, the pole 
$p=b_0\log(\mu^2/\tau^{2a})$ lies
in the upper-half $p$ plane, and the entire contour of $p$ can be taken
in the upper-half plane.
After integration over $s$, we have
\bea
{X}(Q)_{\rm BI,+}
&=&\frac{ir^{-2au'-1}}{4\pi^3}\oint dp\frac{1}{p-1/\alfs}
\int_{C_-} d\tau\,\tau\,\sin\left(\tau r\right)
\big[\tilde{X}(\tau)\big]_{\alfs\to1/p}
 \,.
\label{Bl+2}
\eea
This argument also shows that $B_{X+}(u)$ is a natural integral
representation of $B_X(u)$ for $u$ in the upper-half plane.

\begin{figure}[t]
        \centering
        \includegraphics[width=7cm]{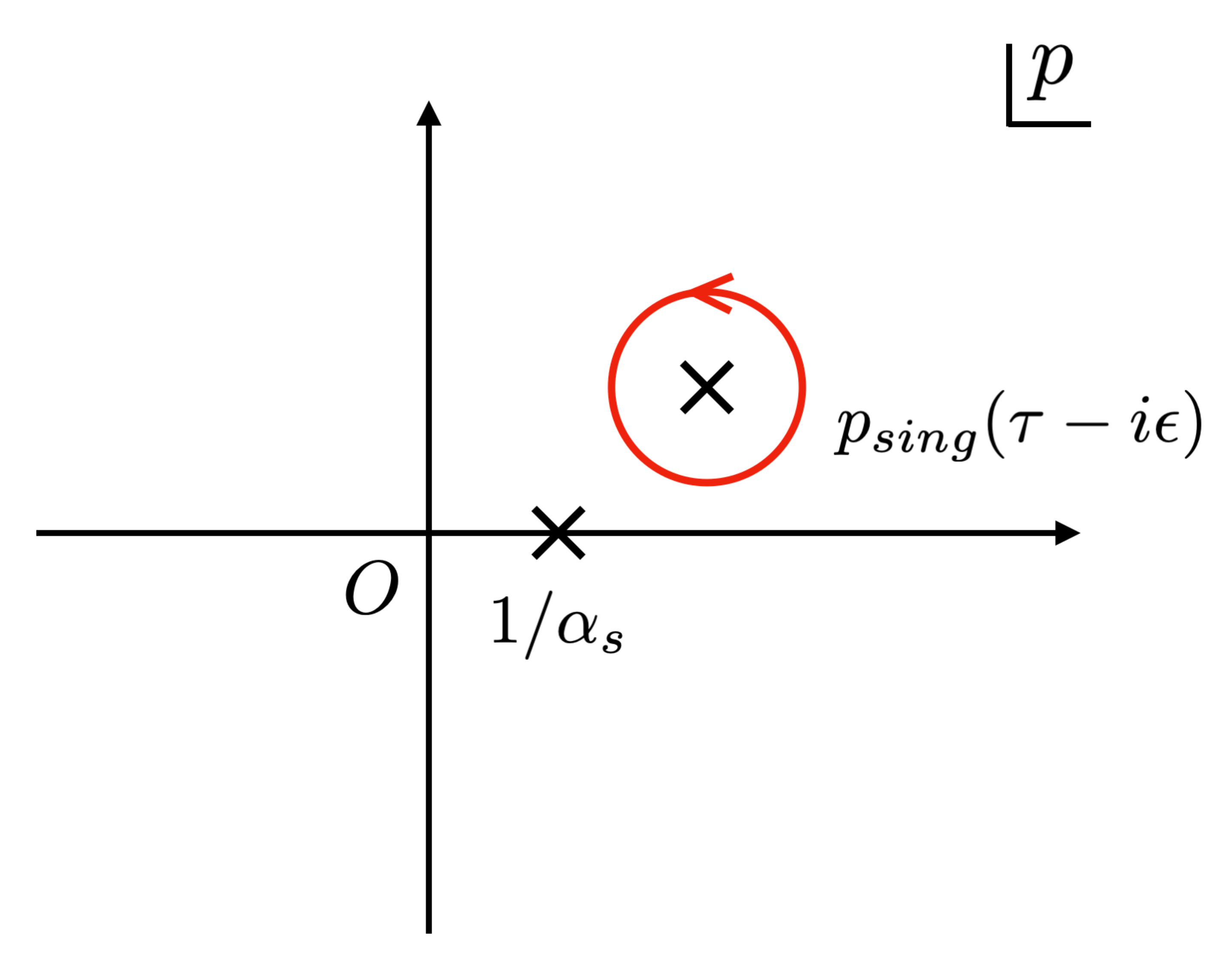}~~~~~~
        \caption
        {\small
		Singularities and integration contour in the complex $p$ plane,
		in the LL approximation for $\tilde{X}(\tau)$.}
        \label{fig:psingLL}
\end{figure}

Next we integrate over $p$.
There are two poles at $p=1/\alfs$ and $p=b_0\log(\mu^2/\tau^{2a})$.
The two poles are well separated as $\tau$ moves along $C_-(\tau)$\,,
see Fig.~\ref{fig:psingLL}.
In fact these two poles coincide only if $\tau^a=\LMS^{\rm LL}$ but
this is circumvented on $C_-(\tau)$.
Hence, the closed contour of $p$ can always be taken to surround only
the pole at $p=b_0\log(\mu^2/\tau^{2a})$.
The integrand reduces to zero
sufficiently rapidly $\sim 1/|p|^2$ as $|p|\to \infty$.
This means that we can take the residue at $p=1/\alfs$ and obtain
\bea
{X}(Q)_{\rm BI,+}
=\frac{1}{b_0}\int_{C_+} du\,e^{-\frac{u}{b_0\alpha_s}}B_{X}(u)
=\frac{r^{-2au'-1}}{2\pi^2}
\int_{C_-} d\tau\,\tau\,\sin\left(\tau r\right)
\tilde{X}(\tau)
 \,.
\label{Bl+3}
\eea

This is the relation between the regularized inverse Borel transform
and the regularized inverse Fourier transform, which
we set out to derive.\footnote{
In the case that we expand in $\alfs$, regularizations are unnecessary.
The difference is that the singularities of $\tilde{X}$ in the $p$ plane
stay fixed at the origin, so that we do not need to rotate the integration contour
of $u$ but it can be kept on the positive real axis.
}
Similarly,
$X(Q)_{\rm BI,-}$ is given by changing $C_\pm \to C_\mp$
in eq.~(\ref{Bl+3}). 
Taking the average of these relations, 
the equivalence of
$[X(Q)]_{\rm PV}$ in eq.~(\ref{PV-Bl}) and
$[X(Q)]_{\rm FTRS}$ in eq.~(\ref{XFTRS}) is shown. 
We emphasize again that it is crucial that
$\tilde{X}(\tau)$ is free of IR renormalons and well-defined
(while $X(Q)$ is not).

In the case that 
$\tilde{X}(\tau)$ is given by the NLL approximation or beyond,
the above argument needs to be modified as follows.

We set $\tilde{X}=\tilde{X}^{(k)}
\propto
\sum_{n=0}^k\tilde{c}_n(0)\alfs(\tau^a)^{n+1}$ for a given $k \in \{ 1,2,3,4 \}$.
According to our current knowledge of RGE at N$^k$LL, 
$\alfs(\tau^a)$ diverges at $\tau=\tau_*\in \mathbf{ R}$ if the running starts from $\mu>\tau^a_*$ with the initial condition $\alfs(\mu)=1/p$.
This causes a singularity on the positive real $p$-axis
for given values of $\tau$ and $\mu$.
The relation between $p$ and $\alfs(\tau^a)$ is determined implicitly by
\bea
&&
\log\biggl(\frac{\mu^2}{\tau^{2a}}\biggr)
=-\int_{1/p}^{\alfs(\tau^a)}\!\!\!\!dx\,\frac{1}{\beta(x)}\,,
~~~~~~~
\beta(\alfs)=-\sum_{n=0}^k b_n\,\alfs^{n+2}\,.
\label{rel-alftau-p}
\eea
One can analyze the positions of the singularities 
of $\big[\tilde{X}(\tau)\big]_{\alfs\to1/p}$ in the complex $p$ plane
and find the following feature.
If $\tau\in {\bf R}_{>0}$, we can choose $\exists p_{\rm ref}\in {\bf R}_{>0}$
independent of $\tau$ such that
all the singularities except one
(let us call it $p_{*,1}$) are located to the left of $p_{\rm ref}$.
We can choose $\alfs$ such that $1/\alfs>p_{\rm ref}$.
In the region $\tau^a<\mu$, $p_{*,1}$ is real positive
and collides with $1/\alfs$ at $\tau =\tau_*$.
In this region, if $\tau$ is shifted slightly to the lower half plane,
$p_{*,1}$ is shifted slightly to the upper half plane.
Thus, $p_{*,1}$ plays the role of the only singularity in the LL case
if $\tau^a<\mu$, although $p_{*,1}$ is a branch point rather than 
a pole.
In the region $\tau^a>\mu$, $p_{*,1}$ is also located to the left of $p_{\rm ref}$.

\begin{figure}[t]
        \centering
        \includegraphics[width=8cm]{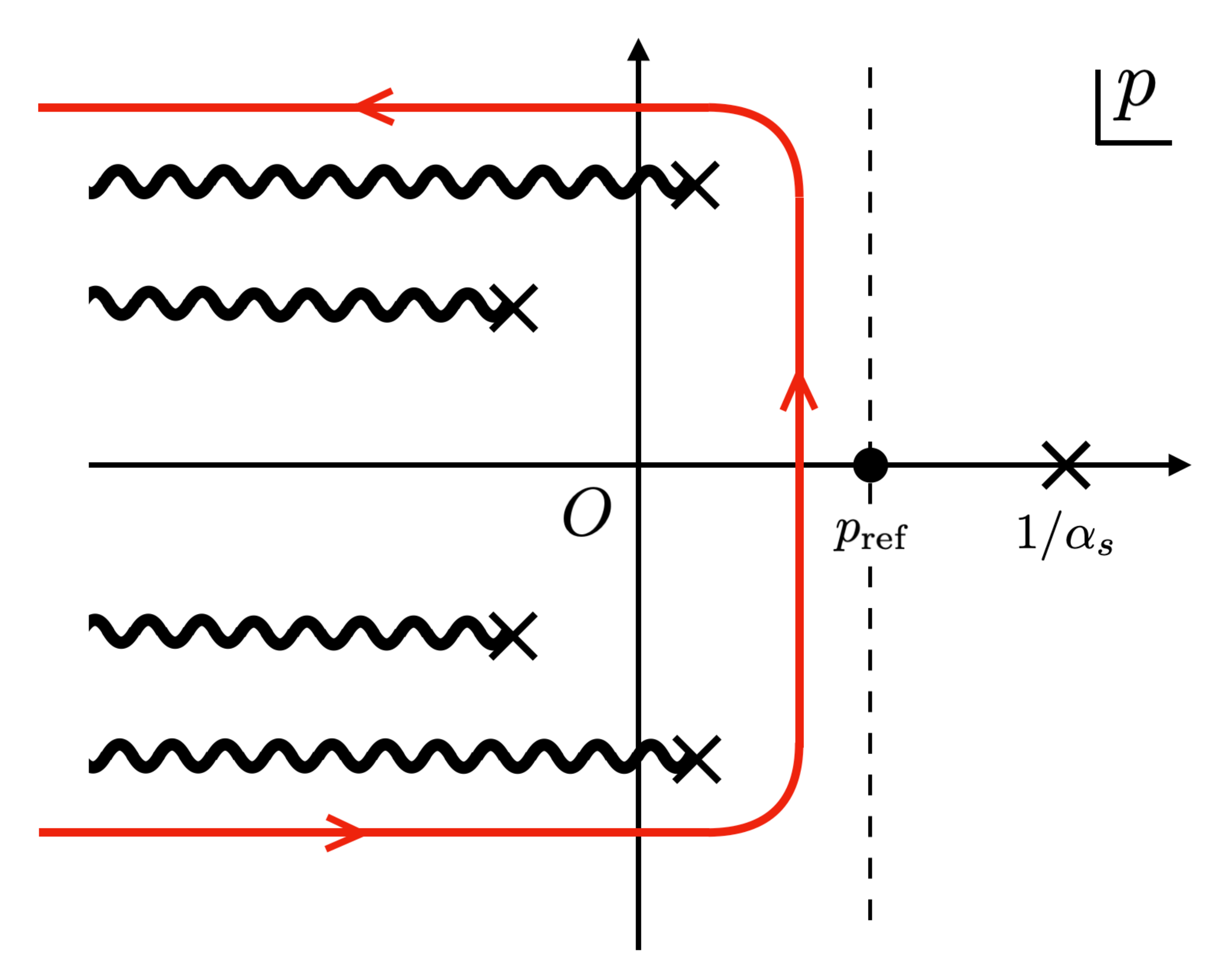}~~~~~~
        \caption
        {\small
Contour of $p$ integration surrounding singularities in the $p$ plane.
These singularities correspond to all the singularities
except $p_{*,1}$
in the region $\tau^a<\mu$, and to all the singularities
in the region $\tau^a>\mu$.
}
        \label{fig:pint_path}
\end{figure}

We separate the integral along $C_-(\tau)$ of eq.~\eqref{Bl+} into the regions
${\rm Re}\,\tau^a>\mu$ and 
${\rm Re}\,\tau^a<\mu$.
In the latter integral, we further separate the integral corresponding to
the contour of $p$
wrapping the branch cut of $p_{*,1}$ from the rest.
The branch cut  of $p_{*,1}$ is taken to extend to $+i\infty$.
For this particular integral, we treat it similarly to the LL case
and integrate over $s$.
For the rest of the integrals, rather than integrating over $s$ from 0 to $\infty$,
we integrate over $u$ from 0 to $\infty$.
(Namely, we transform back from $s$ to $u$, and instead of integrating
along the positive imaginary $u$-axis, we integrate along the positive real $u$-axis.)
The integral over $u$ converges, since ${\rm Re}\,p-1/\alfs<0$ 
for any $p$ along 
the contour of $p$ which is closed to the left; see Fig.~\ref{fig:pint_path}.
After collecting all the integrals, we obtain eq.~\eqref{Bl+2} again.
The essence of the above procedure is that we can find a path
to eq.~\eqref{Bl+2} by an analytic continuation.
The remaining procedure is the same as the LL case.

Below we present an analysis of the singularities
 of $\big[\tilde{X}(\tau)\big]_{\alfs\to1/p}$ in the complex $p$ plane.
We consider the case $\tau\in {\bf R}_{>0}$.
The integral in eq.~\eqref{rel-alftau-p} can be evaluated as
\bea
&&
-\int_{1/p}^{\alfs(\tau^a)}\!\!\!\!dx\,\frac{1}{\beta(x)}
= \int_{1/\alfs(\tau^a)}^{p} \!\!\!\! dq\,\, \frac{q^{k}}{b_0q^{k}+b_1q^{k-1}+\cdots+b_k}
\nonumber\\&&
=\frac{1}{b_0}\left(p-\frac{1}{\alfs(\tau^a)}\right)+\sum_{j=1}^k R_j 
\log\left(\frac{p_j-p}{p_j-1/\alfs(\tau^a)}\right)
=\log\biggl(\frac{\mu^2}{\tau^{2a}}\biggr)
\,.
\eea
Here, the complex roots of 
$-p^{k+2}\beta(1/p)=b_0p^{k}+b_1p^{k-1}+\cdots+b_k=0$
are denoted as ${p}_{j}$ ($1\le j\le k$).
$R_j$ is the residue of $p^k/[-p^{k+2}\beta(1/p)]$ at $p=p_j$.
The roots $p_j$ are logarithmic branch points of $\alfs(\tau^a)$, hence,
they are singularities of $\big[\tilde{X}(\tau)\big]_{\alfs\to1/p}$.
They are branch points independent of $\tau$.
The other class of singularities originate from
divergence of $\alfs(\tau^a)$, and the positions of those  singularities ($p=p_{*,i}$)
are determined by
\bea
&&
\log\biggl(\frac{\mu^2}{\tau^{2a}}\biggr)
=\int_0^{p_{*,i}} dq\,\, \frac{q^{k}}{b_0q^{k}+b_1q^{k-1}+\cdots+b_k}
\nonumber\\&&
~~~~~~~~~~~~~
=\frac{p_{*,i}}{b_0}+\sum_{j=1}^k R_j \log(1-p_{*,i}/p_j)
\,.
\eea
They are branch points, $\alfs(\tau^a)\sim(p-p_{*,i})^{-1/(k+1)}$.

\begin{figure}[t]
        \centering
        \includegraphics[width=10cm]{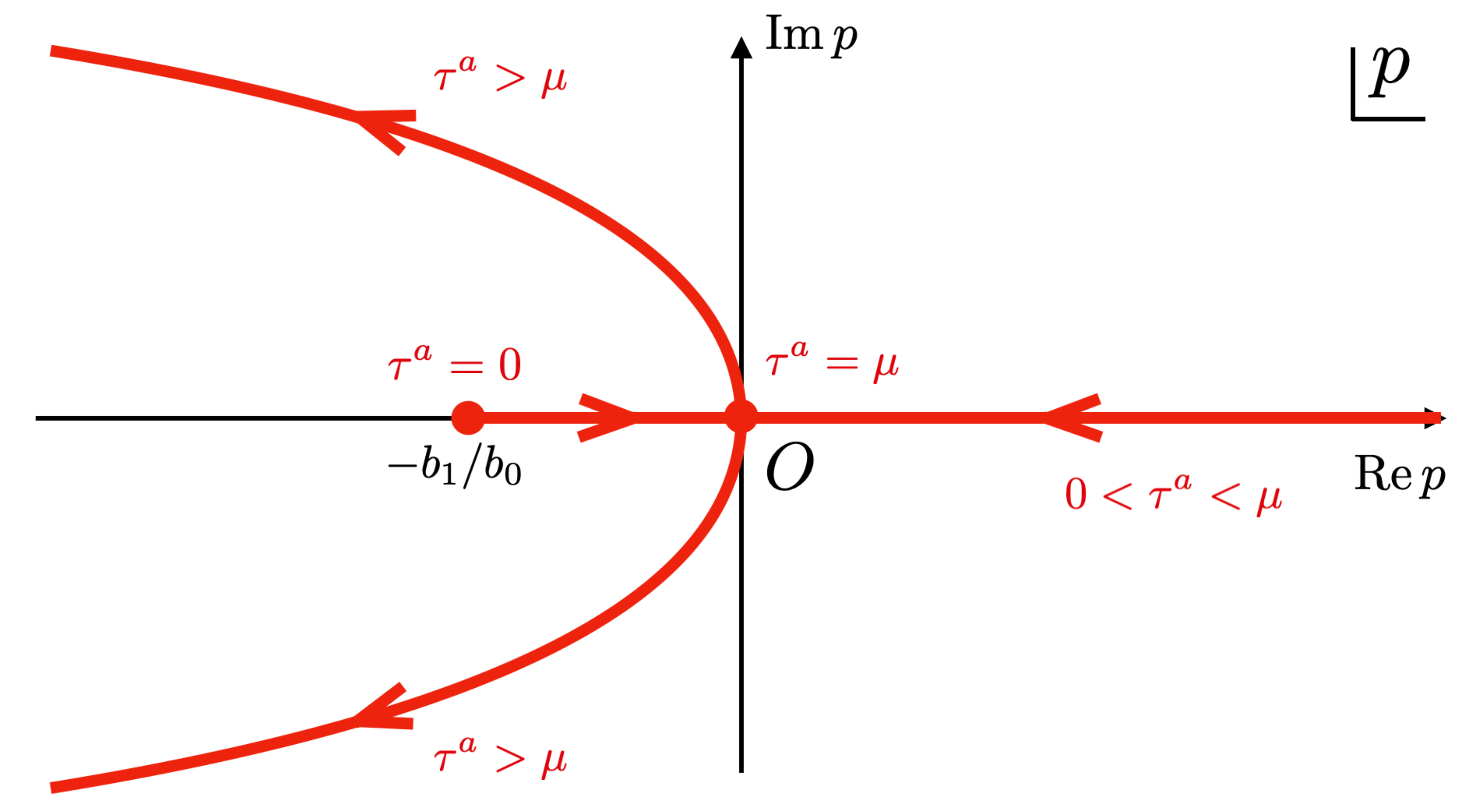}~~~~~~
        \caption
        {\small
Trajectories of 	$p_{*,1}$ and 
$p_{*,2}$ in the NLL approximation of $\tilde{X}$
($k=1$), as $\tau\in{\bf R}$ is raised
from 0 to $\infty$. At $\tau=0$, 	
$p_{*,1}=+\infty$ and $p_{*,2}=p_1=-b_1/b_0$.
}
        \label{fig:psing-trajectory-NLL}
\end{figure}

When $\tau^a/\mu \simeq 1$, $p_{*,i}$ 
($1\le i \le k+1$) are located close
to the origin and determined
approximately by $(k+1)b_k\log(\mu^2/\tau^{2a})= (p_{*,i})^{k+1}$.
In the region $0<\tau^a<\mu$, we take $p_{*,1}$ as the one
corresponding to the Landau singularity $\tau_*$.
That is, $p_{*,1}$ is real positive and
moves from $+\infty$ to 0 as $\tau^a$ is raised from 0 to $\mu$.
The behavior of other $p_{*,i}$ as $\tau \to 0$ belongs to either of
the following categories:
\begin{itemize}
\item[(a)]
 $p_{*,i}$ goes towards left, i.e., ${\rm Re}\, p_{*,i}\to -\infty$ while
${\rm Im}\, p_{*,i}/{\rm Re}\, p_{*,i}\to 0$.
\item[(b)]
 $p_{*,i}\in {\bf R}$ and converges towards one of $p_j$'s which is also real.
\item[(c)]
 $p_{*,i}$ approaches one of $p_j$'s as $p_{*,i}$ rotates around this fixed point
infinitely many times (hence it enters the different sheets).
\item[(d)]
 $p_{*,i}$ rotates around one of $p_j$'s 
infinitely many times (hence it enters the different sheets) as the
distance to this fixed point increases.
\end{itemize}
In the region $\tau^a>\mu$, as $\tau\to\infty$
the behavior of every $p_{*,i}$ (including $p_{*,1}$) belongs to either of the above
categories.
As an example we show the trajectories of $p_{*,i}$ for $k=1$ (NLL case)
in Fig.~\ref{fig:psing-trajectory-NLL}.



\section{FTRS formula in a Wilsonian picture}
\label{AppD}

We derive the FTRS formula eq.~(\ref{XFTRS-decomp})
by introducing a factorization scale ${\mu}_f$ to separate
UV and IR regions in the ``momentum space.''
This has a closer connection to the original idea of Wilson,
who developed the OPE framework to separate UV and IR contributions.

Let us introduce the
factorization scale ${\mu}_f$ in a Wilsonian OPE framework as
\bea
\LQ \ll {\mu}_f \ll Q =r^{-a}.
\eea
In this framework, the Wilson coefficient, which corresponds to the UV
contribution to
$X(Q)$,
can be computed using perturbative QCD.
We introduce 
an IR cut off in the momentum-space integral and resum logarithms
by RG:
\bea
&&
X_{\rm UV}(Q=r^{-a};\tilde{\mu}_f)\equiv 
r^{-2au'}
{\hbox to 18pt{
\hbox to -1pt{$\displaystyle \int$} 
\raise-15pt\hbox{$\scriptstyle \tau>\tilde{\mu}_f$} 
}}~
\frac{d^3\vec{\tau}}{(2\pi)^3} 
\,
e^{i\vec{\tau}\cdot\vec{x}}\,\tilde{X}(\tau)
\nonumber\\&&
=\frac{r^{-2 a u'-1}}{2 \pi^2} \int_{\tilde{\mu}_f}^{\infty} d \tau \, \tau \sin(\tau r) \tilde{X}(\tau) 
=\frac{r^{-2 a u'-1}}{2 \pi^2}\, {\rm Im}\int_{\tilde{\mu}_f}^{\infty} d \tau \, \tau e^{i\tau r} \tilde{X}(\tau) 
\, ,
\label{V_UV}
\eea
where $\tilde{\mu}_f=\mu_f^{1/a}$, and
$\tilde{X}(\tau)=\tilde{X}^{(k)}(\tau)$
of eq.~(\ref{FTXpert}), i.e., that in the N$^k$LL approximation.
The integral is well defined, since the singularity of
the running coupling constant is not included in the integral region.

\begin{figure}[t]
\begin{center}
\includegraphics[width=7cm]{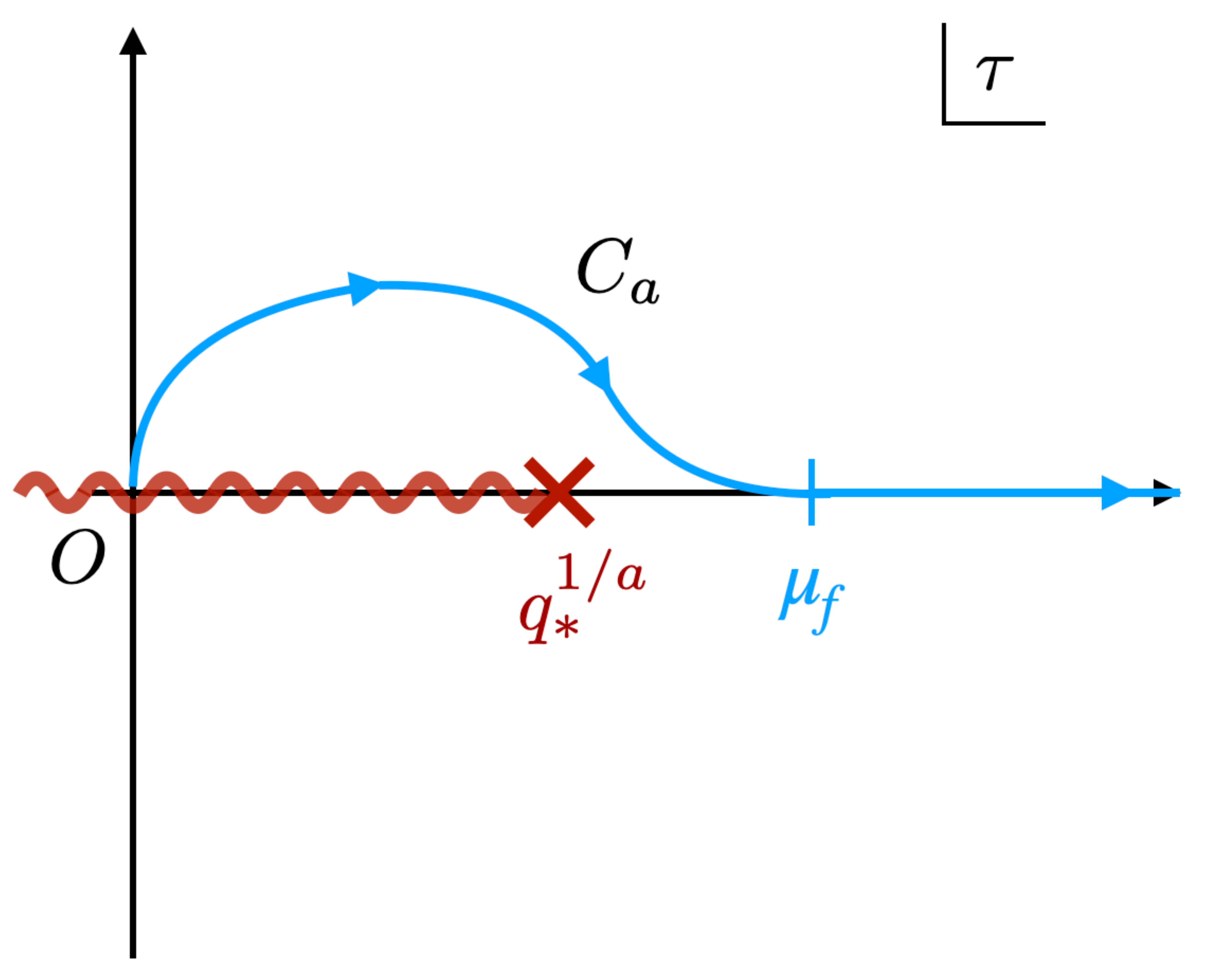}
~
\includegraphics[width=7cm]{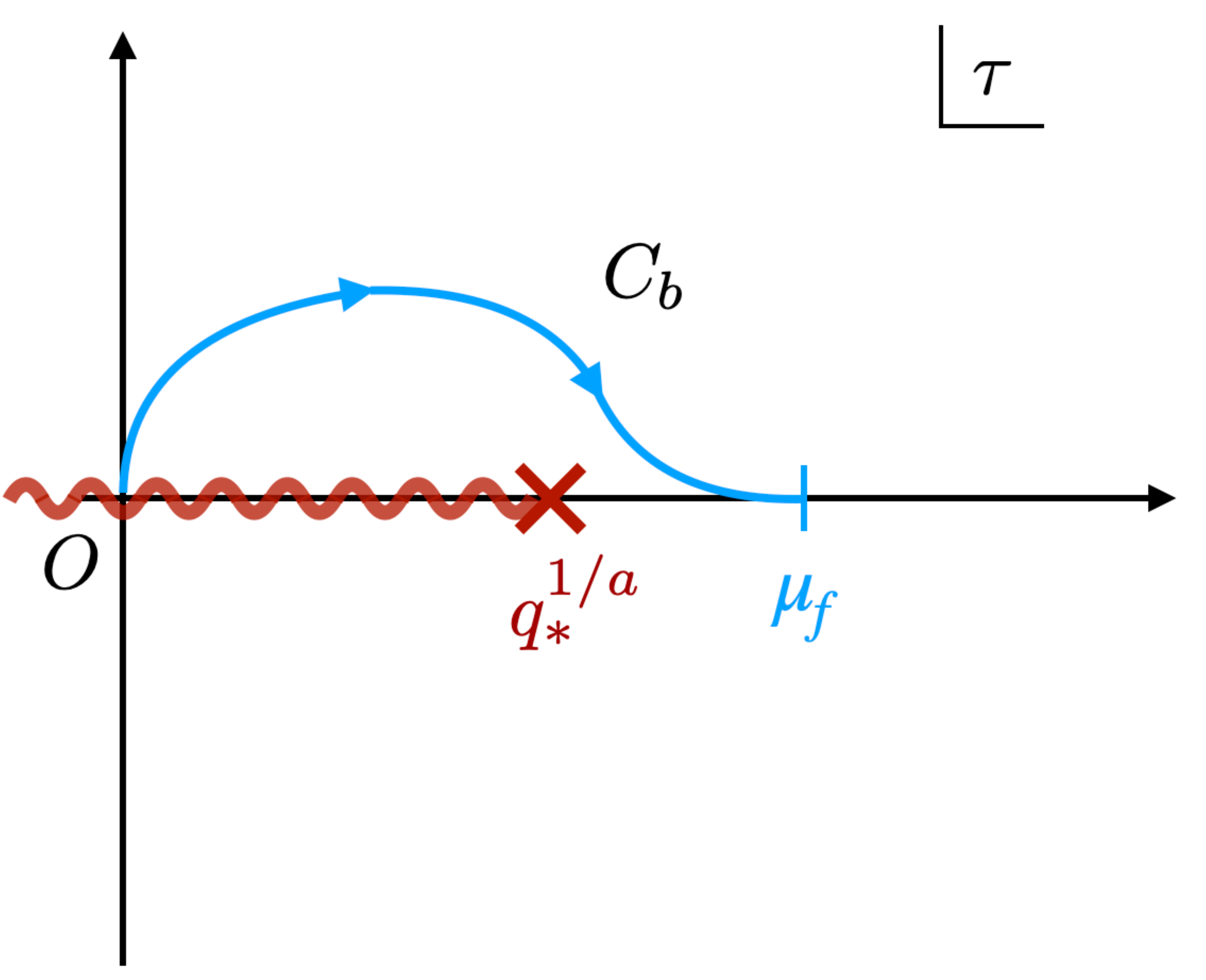}
\end{center}
\caption{\footnotesize
Integral contours in the complex $\tau$-plane shown by
blue lines.
The branch cut of the running coupling constant is also shown.
\label{Contours}}
\end{figure}

We can obtain a short-distance expansion of $X_{\rm UV}$
as follows.
We separate the integral contour
into the difference of two contours $C_a-C_b$
in the complex $\tau$-plane:
\bea
&&
X_{\rm UV}
=\frac{r^{-2 a u'-1}}{2 \pi^2} \,
{\rm Im}\,\int_{C_a-C_b} d \tau \, \tau e^{i\tau r} \tilde{X}(\tau) 
\,,
\eea
see Fig.~\ref{Contours}.

The integral path $C_a$ can be deformed to the positive imaginary axis as
\bea
&&
\frac{r^{-2 a u'-1}}{2 \pi^2}\, {\rm Im}\,\int_{C_a} d \tau \, \tau e^{i\tau r} \tilde{X}(\tau) 
=\frac{-r^{-2 a u'-1}}{2\pi^2} \int_0^{\infty} dt \, t\, e^{-tr}\,{\rm Im}\,[\tilde{X}(it)]
\,,
\eea
where $\tau=it$.
This contribution coincides with $\bar{X}_0(Q)$ defined in eq.~(\ref{X0}).

Since $\tilde{\mu}_f r\ll 1$, along the contour $C_b$ it
is justified to expand the Fourier factor as
\bea
e^{i\tau r}=\sum_{m=0}^\infty \frac{(i\tau r)^m}{m!}
=1+i\tau r+\frac{1}{2}(i\tau r)^2 +\cdots .
\label{FourierFac}
\eea
Then the integral along $C_b$ can be written as
\bea
\frac{r^{-2 a u'-1}}{2 \pi^2} \,{\rm Im}\int_{C_b} d \tau \, \tau e^{i\tau r} \tilde{X}(\tau) 
=\sum_{m=0}^\infty
\frac{Q^{2u'-(m-1)/a}}{2 \pi^2 m!} \,
{\rm Im} \,i^m\int_{C_b} d \tau \, \tau^{m+1}\tilde{X}(\tau) \, .
\eea
For even $m$'s
the integral contour can be deformed to $C_*$ (Fig.~\ref{fig:contour}), which surrounds
the branch cut of the running coupling constant ($m=2n$):
\bea
&&
{\rm Im}\,i^{2n}\int_{C_b} d \tau \, \tau^{2n+1}\tilde{X}(\tau) 
=\frac{(-1)^n}{2i} \int_{C_b-C_b^*} d \tau \, \tau^{2n+1}\tilde{X}(\tau) 
\nonumber\\&&
~~~~~~~~~~~~
~~~~~~~~~~~~
~~~~~~~
=\frac{(-1)^{n+1}}{2i} \int_{C_*} d \tau \, \tau^{2n+1}\tilde{X}(\tau) 
\,,
\eea
where we used 
$[\alpha_s(q)]^*=\alpha_s(q^*)$.
Although originally these integrals appear
to be dependent on $\tilde{\mu}_f$, in fact they reveal to be
independent of $\tilde{\mu}_f$.
Each integral along the closed contour $C_*$ is independent
of the intermediate point $\tilde{\mu}_f$, 
since the integral contour is continuously deformable
with the end points kept fixed.
Thus, this part coincides with
$\bar{X}_{\rm pow}(Q)$ defined in eq.~(\ref{Xpow}), which
is in fact evaluated in expansion in $\tau r$.

In contrast, integrals corresponding to
odd $m$'s cannot be expressed
as integrals along a closed contour 
and are generally dependent on $\tilde{\mu}_f$ ($m=2n+1$):\footnote{
The difference originates from the fact that the integrals can be written only
as ones along $C_b+C_b^*$ rather than
$C_b-C_b^*$, since $i$ remains in eq.~(\ref{FourierFac}).
}
\bea
&&
{\rm Im}\,i^{2n+1}\int_{C_b} d \tau \, \tau^{2n+2}\tilde{X}(\tau) 
={(-1)^n}\,{\rm Re}\int_{C_b} d \tau \, \tau^{2(n+1)}\tilde{X}(\tau) 
\,.
\eea

Combining these, we obtain
\bea
&&
X_{\rm UV}(r;\tilde{\mu}_f)=
\bar{X}_{0}(Q)+\bar{X}_{\rm pow}(Q) + [\delta X(Q;\tilde{\mu}_f)]_{\rm Wilsonian}
\nonumber\\&&
~~~~~~~~~~~~~~
= [X(Q)]_{\rm FTRS} + [\delta X(Q;\tilde{\mu}_f)]_{\rm Wilsonian}
\,,
\label{expXUV}
\eea
where
\bea
&&
[\delta X(Q;\tilde{\mu}_f)]_\text{Wilsonian}=- \sum_{n=0}^\infty
\frac{(-1)^n}{(2n+1)! \pi^2} 
\frac{1}{Q^{2(-u'+n/a)}} \,{\rm Re}\int_{C_b} d \tau \, \tau^{2(n+1)}\tilde{X}(\tau) \, .
\eea
Thus, we have separated $X_{\rm UV}(r;\tilde{\mu}_f)$ into the
part insensitive to the factorization scale (IR cut off) $\tilde{\mu}_f$ and the part
which is sensitive to it.
The former can be regarded as a genuinely UV part of $X_{\rm UV}$
and in fact, 
coincides with the renormalon-free Wilson coefficient $ [X(Q)]_{\rm FTRS}$.
The latter has a form similar to
$[ \delta X ]_{\rm FTRS}$ defined in
eq.~(\ref{delXFTRS})
and can be regarded as corresponding to the contributions from the IR renormalons
in the Wilsonian picture.
Note that, since $\tilde{X}$ is (largely) free from renormalons, 
the IR sensitivity manifests itself only through dependence on $\tilde{\mu}_f$
and we do not need to look into the details of the IR structure of $\tilde{X}$.
This is the reason why we can separate (the dominant part of) the
IR renormalons in a one-parameter
integral form in the Wilsonian picture.

It should be noted, however, that the prescription for subtracting
IR contributions are different between the Wilsonian and FTRS methods,
as the latter subtracts only the $\tilde{\mu}_f$ independent part.
The $\tilde{\mu}_f$ dependent part generally generates power dependences on
${\mu}_f/Q$, which can give large effects and destabilize phenomenological
analyses.
On the other hand, in the end such $\tilde{\mu}_f$ dependences cancel 
against those in the non-perturbative matrix elements in
a physical observable, since an observable is independent of $\tilde{\mu}_f$.
The $\tilde{\mu}_f$ independent subtraction scheme of the FTRS (or PV)
prescription is favorable in the sense that it avoids possible large
power corrections in ${\mu}_f/Q$
at intermediate stages.
(Note that ${\mu}_f\gg\LQ$.)

\section{Resummation of artificial UV renormalons}
\label{AppE}

We derive the formula for resumming the artificial UV renormalons 
at $u=-u'-(2k+3)/(2a)$ ($k=0,1,2,\cdots$)  generated by
the Fourier transform, 
cf.\  eq.~\eqref{FTdelX}.
We use this formula to calculate the momentum-space series
in eqs.~\eqref{FTRSformula-AdlerFn},~\eqref{FTRSformula-Bdecay}, \eqref{alphamb}, \eqref{FTRSformula-mb} and \eqref{FTRSformula-mc}.

As an example,
we consider a Wilson coefficient $X(Q)$ 
which includes IR renormalons at $u=1,2$, whose
perturbative expansion up to ${\cal O}(\alpha_s^{k+1})$ 
is given by
\be
X(Q)=\sum_{n=0}^k c_n\alpha_s(Q)^{n+1}.
\ee
To suppress the IR renormlaons, we define the Fourier transform $\tilde{X}(\tau)$
by choosing $(a,u')=(1,-1)$ as
\bea
\tilde{X}(\tau)&=&
\frac{4\pi}{\tau}\sum_{n=0}^k\tilde{c}_n\alpha_s(\tau)^{n+1}\,,
\eea
where $\tilde{c}_n$'s are determined from
\bea
&&
\Big[\sin(\pi u_*)\Gamma(2u_*)\Big]_{u_*\to\hat{H}}\sum_{n=0}^\infty c_n\alpha_s^{n+1}(\tau)=
\sum_{n=0}^\infty \tilde{c}_n\alpha_s(\tau)^{n+1}\,,
\label{ex-det-tildecn}
\\&&
\hat{H}=-\beta\big(\alpha_s(\tau)\big)\frac{\partial}{\partial \alpha_s(\tau)}
=\sum_{i=0}^\infty b_i\alpha_s(\tau)^{i+2}
\frac{\partial}{\partial \alpha_s(\tau)}\,,
\eea
by equating both sides
of eq.~\eqref{ex-det-tildecn} at each order of series expansion in $\alfs(\tau)$.
More explicitly, the left-hand side of eq.~\eqref{ex-det-tildecn} is given by
\bea
\sum_{m=0}^\infty\frac{1}{m!}\frac{d^m}{du_*^m}\sin(\pi u_*)\Gamma(2u_*)\Bigg|_{u_*=0}\hat{H}^m\sum_{n=0}^\infty c_n\alpha_s(\tau)^{n+1}\,.
\label{FTX}
\eea
According to eq.~\eqref{FTdelX},
$\sin(\pi u_*)\Gamma(2u_*)$ generates UV renormalons at $u_*=-1/2,-3/2,\cdots$.
Hence,
the $\alfs$ series expansion includes the behavior $\tilde{c}_n\sim(-1)^nn!$.
They cause ambiguities in the $\tau$ integral for the truncated calculation,
while the UV renormalon effect can be resummed for an all-order series. 

Let us resum the contributions from
the first two UV renormalons at  $u=-1/2$ and $-3/2$.
We decompose $\sin(\pi u_*)\Gamma(2u_*)$ into two parts as
\bea
\sin(\pi u_*)\Gamma(2u_*)
&=&\Bigg[\sin(\pi u_*)\Gamma(2u_*)-\frac{1}{2u_*+1}+\frac{1}{6}\frac{1}{2u_*+3}\Bigg]\nonumber\\
&&+\Bigg[\frac{1}{2u_*+1}-\frac{1}{6}\,\frac{1}{2u_*+3}\Bigg]
.
\eea
The second term is the sum of 
the UV renormalon poles at $u=-1/2$ and $-3/2$.
These poles are removed in the first square brackets.
Thus, the series expansion after subtracting the UV renormalons is determined
from
\bea
&&
\Biggl[\sin(\pi u_*)\Gamma(2u_*)-\frac{1}{2u_*+1}
+\frac{1}{6}\frac{1}{2u_*+3}
\Biggr]_{u_*\to\hat{H}}\sum_{n=0}^\infty c_n\alpha_s^{n+1}(\tau)=
\sum_{n=0}^\infty \tilde{c}^{\rm subt}_n\alpha_s(\tau)^{n+1}\,.
\nonumber\\
\label{formula-subt}
\eea
On the other hand, the resummed contribution can be calculated
as follows.
Using the formula
\be
\frac{1}{A}=\int_0^\infty ds\,e^{-sA},
\ee
we can resum the pole part as
\bea
&&
\Delta S^{\rm resum}=
\Bigg[\frac{1}{2\hat{H}+1}-\frac{1}{6}\frac{1}{2\hat{H}+3}\Bigg]\sum_{n=0}^k c_n\alpha_s(\tau)^{n+1}\nonumber\\
&&~~~~~
=\int_0^\infty ds\,
\Big(e^{-s(2\hat{H}+1)}-\frac{1}{6}e^{-s(2\hat{H}+3)}\Big)\sum_{n=0}^k c_n\alpha_s(\tau)^{n+1}\nonumber\\
&&~~~~~
=\int_0^\infty ds\,
\Big(e^{-s}-\frac{1}{6}e^{-3s}\Big)\sum_{n=0}^k c_n\alpha_s(\tau e^{s})^{n+1}\non
&&~~~~~
=\int_0^1 \frac{dv}{v}\,
\Big(v-\frac{1}{6}v^3\Big)\sum_{n=0}^k c_n\alpha_s(\tau /v)^{n+1}
,
\label{resumFM}
\eea
since for an arbitrary function $h(\alpha_s(\mu))$,
\be
e^{-2s\hat{H}}h(\alpha_s(\mu))=h(\alpha_s(\mu e^{s})).
\ee

Thus, our formula is given by
\bea
\tilde{X}(\tau)&=&
\frac{4\pi}{\tau}
\left[
\sum_{n=0}^k\tilde{c}^{\rm subt}_n\alpha_s(\tau)^{n+1}
+ \Delta S^{\rm resum}
\right]
\,.
\eea
The point is that we evaluate $\dfrac{1}{2\hat{H}+1}$ and $\dfrac{1}{2\hat{H}+3}$ 
not by Taylor expansion 
but as a running effect of the coupling constant $\alpha_s$. 
Contributions from
UV renormalons further from the origin can be resummed in a similar manner.

We perform the $\tau$-integral of the UV renormalon resummed part, 
$4\pi \Delta S^{\rm resum}/\tau$,
and give the corresponding $Q$-space quantity by the sum of
$\bar{X}_0^{\rm resum} $ and $\bar{X}_{\rm pow}^{\rm resum}$ 
[cf.\  eqs.~\eqref{X0} and \eqref{Xpow}] in the following way.
We can calculate $\bar{X}_0^{\rm resum} $ as
\begin{align}
\bar{X}_0^{\rm resum}&(Q)
=\frac{2}{ \pi Q} \int_0^1\frac{dv}{v}\,\Big(v-\frac{1}{6}v^3\Big)
\int_0^{\infty} dt \, e^{-t/Q} {\rm Re} \, \Big[\sum_{n=0}^k c_n \alpha_s(i t /v)^{n+1}\Big] \non
&=\frac{2}{ \pi Q}\int_0^1\frac{dv}{v}\,\Big(v-\frac{1}{6}v^3\Big)
\int_0^{\infty} dx \, v\, e^{-x v/Q}   {\rm Re} \, \Big[\sum_{n=0}^k c_n \alpha_s(i x)^{n+1}\Big]\non
&=\frac{2}{\pi Q} \int_0^{\infty} dx \, {\rm Re} \,  \Big[\sum_{n=0}^k c_n \alpha_s(i x)^{n+1}\Big]
\int_0^1dv\,\Big(v-\frac{1}{6}v^3\Big)  e^{-x v/Q} .
\label{resum0}
\end{align}
In the second equality we changed the integration variable as $x=t v$,
and in the last equality we interchanged the order of the two integrals.
One can perform the last $s$-integral analytically.
Hence it is sufficient to perform numerical integration only once (for the $x$-integral)
even for $\bar{X}_0^{\rm resum}(Q)$.

$\bar{X}_{\rm pow}^{\rm resum}(Q)$ can be calculated as
\begin{align}
&\!\!\!\bar{X}_{\rm pow}^{\rm resum}(Q)
=\frac{1}{\pi i Q} \int_0^1\frac{dv}{v}\,\Big(v-\frac{1}{6}v^3\Big)
\int_{C_*} d \tau \, \cos(\tau/Q) \sum_{n=0}^k c_n \alpha_s(\tau/v)^{n+1} \non
&~~~~
=\frac{1}{\pi i Q} \int_0^1\frac{dv}{v}\,\Big(v-\frac{1}{6}v^3\Big)
\int_{C_*} d z \,v  \cos(z v/Q) \sum_{n=0}^k c_n \alpha_s(z)^{n+1} \non
&~~~~
=\sum_{m=0}^{\infty} \frac{(-1)^m}{(2m)!}
\int_0^1dv\,\Big(v-\frac{1}{6}v^3\Big) v^{2m} 
\int_{C_*} \frac{d z}{\pi i} \, \frac{z^{2m}}{Q^{2m+1}}  \sum_{n=0}^k c_n \alpha_s(z)^{n+1}\non
&~~~~
=\sum_{m=0}^{\infty} \frac{(-1)^m}{(2m)!}\frac{5m+11}{12(m+1)(m+2)}
\int_{C_*} \frac{d z}{\pi i}  \,  \frac{z^{2m}}{Q^{2m+1}} \sum_{n=0}^k c_n \alpha_s(z)^{n+1} .
\label{resumpow}
\end{align}
In the second equality we changed the integration variable as $z=\tau/ v$
and used the fact that the integration contour can be taken the same after the change of the variable
due to the analyticity of the integrand. In the third equality we expanded $\cos(zv/Q)$. 
In the last equality, we performed the $v$-integral analytically.
A numerical integration is required only for the $z$-integral.

In sec.~\ref{Sec2.3},
we use the modified formula for the resummation such that 
all the artificial UV renormalons are resummed.
We decompose $\sin(\pi u_*)\Gamma(2u_*)$ into two parts as 
\bea
\Big[\sin(\pi u_*)\Gamma(2u_*)-(1-u_*)(2-u_*)f(u_*)\Big]+(1-u_*)(2-u_*)f(u_*).
\label{sep-UVren}
\eea
The function $f(u_*)$ is the sum of the UV renormalons of
\be
\frac{\sin\big(\pi u_*)\Gamma\big(2u_*\big)}{(1-u_*)(2-u_*)}
\ee
at $u_*=-(2m+1)/2$ ($m=0,1,2,\cdots$),
given by
\bea
f(u_*)&=&\sum_{m=0}^\infty\frac{(-1)^m}{(2m+1)!(2m+3)(2m+5)}\frac{1}{2u_*+2m+1}\non
&=&\int_0^\infty ds\,e^{-2u_*s}\sum_{m=0}^\infty\frac{(-1)^m}{(2m+1)!(2m+3)(2m+5)}e^{-s(2m+1)}\non
&=&\int_0^\infty ds\,e^{-2u_*s}\Big( 3 e^{4 s} \sin\big(e^{-s}\big)-3 e^{3s} \cos\big(e^{-s}\big) -e^{2 s}\sin\big(e^{-s}\big) \Big)\non
&=&\int_0^1 dv\,v^{2u_*}\frac{3\sin\big(v\big)-3v\cos\big(v\big)-v^2\sin\big(v\big)}{v^5}.
\label{fu_st}
\eea
The first term of eq.~\eqref{sep-UVren} is free from all the UV renormalon poles,
and it has the structure that the IR renormalon poles at $u_*=1$ and $u_*=2$ are suppressed.
Thus, we can obtain the following series expansion,
\bea
&&
\Biggl[\sin(\pi u_*)\Gamma(2u_*)-(1-u_*)(2-u_*)f(u_*)
\Biggr]_{u_*\to\hat{H}}\sum_{n=0}^\infty c_n\alpha_s^{n+1}(\tau)=
\sum_{n=0}^\infty \tilde{c}^{\rm subt}_n\alpha_s(\tau)^{n+1}\,,
\nonumber\\
\label{formula-subt-full}
\eea
which is free of all the artificial UV renormalons,
and does not have IR renormalons at $u=1$ and $u=2$
which is included in the original series $X(Q)$.
This formula gives eq.~\eqref{tildeXtoy-subt}, which shows a good convergence.

On the other hand, 
for the second term of eq.~\eqref{sep-UVren},
we must resum all the artificial UV renormalon poles
according to the same procedure as in eq.~\eqref{resumFM}.
The formula is given by
\bea
\Delta S^{\rm resum}
&=&(1-\hat{H})(2-\hat{H})f(\hat{H})\sum_{n=0}^kc_n\alfs(\tau)^{n+1}\non
&=&\int_0^1 dv\frac{6\sin\big(v\big)-6v\cos\big(v\big)-2v^2\sin\big(v\big)}{v^5}\non
&&~~~~~~~~~~~~~
\times(1-\hat{H})(2-\hat{H})\sum_{n=0}^kc_n\alfs(\tau/v)^{n+1}\bigg|_{{\rm up\,to\,}{\cal O}(\alpha_s(\tau/v)^{k+1})}.
\label{gen-resumFM}
\eea
Since eqs.~\eqref{FTX} and~\eqref{formula-subt-full} do not contain IR renormalons,
the perturbative series in eq.~\eqref{gen-resumFM} also does not have IR renormalons.
In the case that we use the one-loop beta function (then $\hat{H}=b_0\alpha_s^2(\partial/\partial\alpha_s)$),
the perturbative series in eq.~\eqref{gen-resumFM} is given by
\bea
&&(1-\hat{H})(2-\hat{H})\sum_{n=0}^kc_n\alfs(\tau/v)^{n+1}\bigg|_{{\rm up\,to\,}{\cal O}(\alpha_s(\tau/v)^{k+1})}\non
&&~~~~=c_0\alfs(\tau/v)+\big(c_1-3c_0\big)\alfs(\tau/v)^2\non
&&~~~~~~~+\sum_{n=2}^k\big(c_n-3nc_{n-1}+2n(n-1)c_{n-2}\big)\alfs(\tau/v)^{n+1}\bigg|_{{\rm up\,to\,}{\cal O}(\alpha_s(\tau/v)^{k+1})}\,,
\label{elim-IRren}
\eea
where we can see that 
the IR renormalons at $u=1$ and $u=2$ are eliminated explicitly.

$\bar{X}_0^{\rm resum}(Q)$ and $\bar{X}_{\rm pow}^{\rm resum}(Q)$ can be calculated
in the same way as in eqs.~\eqref{resum0} and \eqref{resumpow}.

\section{\boldmath Phenomenological model of $R$-ratio}
\label{App:PhenoModel}

Ref.~\cite{Bernecker:2011gh} proposed a
model function for the $R$-ratio (for $n_f=3$), which is given by 
\bea
R(s)&=&\theta\left(\sqrt{s}-2 m_{\pi^{\pm}}\right)
\theta\left(4.4 m_{\pi^{\pm}}-\sqrt{s}\right)\frac{1}{4}\left[1-\frac{4 m_{\pi \pm}^{2}}{s}\right]^{3 / 2}\left(0.6473+f_{0}(\sqrt{s})\right) \nonumber\\
&+&\theta\left(\sqrt{s}-4.4 m_{\pi^{\pm}}\right) 
\theta\left(M_{3}-\sqrt{s}\right)\left(\sum_{i=1}^{2} f_{i}(\sqrt{s})\right) \nonumber\\ 
&+&f_{3}(\sqrt{s})+3\left(\sum_{f=u,d,s}Q_f^2\right)\theta\left(\sqrt{s}-M_{3}\right),
\eea
where
\be
f_i(\sqrt{s})=\frac{C_i \Gamma_i^{2}}{4(\sqrt{s}-M_i)^{2}+\Gamma_i^{2}}
\,,
\ee
and the values of the parameters are given as in the table below.
\be
\begin{array}{|c|c|c|c|} 
\hline \quad i \quad & \quad C_{i} \quad & \quad M_{i} / \mathrm{GeV} \quad & \quad \Gamma_{i} / \mathrm{GeV} \quad \\ \hline 0 & 655.5 & 0.7819 & 0.0358 \\ 1 & 8.5 & 0.7650 & 0.130 \\ 2 & 11.5 & 0.7820 & 0.00829 \\ 3 & 50.0 & 1.0195 & 0.00426 \\ \hline \end{array}
\ee
$f_0,f_1$ and $f_2$ represent the peaks of the $\rho$ and $\omega$ mesons;
$f_3$ corresponds to that of the $\phi$ meson which is $s$-flavor.
The last term represents the lowest-order perturbative
contribution $\propto \sum_fQ_f^2$.

For application to the $n_f=2$ case, we replace 
\be
f_3\to0,
\ee
\be
\sum_{f=u,d,s}Q_f^2\to\sum_{f=u,d}Q_f^2.
\ee

\section{\boldmath Effects of unsuppressed singularities
on $[X(Q)]_{\rm FTRS}$}
\label{App:cond(iii)}

We analyze effects of unsuppressed singularities in the Borel plane
on the result of the FTRS method 
$[X(Q)]_{\rm FTRS}$.
Namely, we analyze the case where the assumption (iii) in Sec.~\ref{sec:FTRS}
[below eq.~\eqref{Xpow}]
is violated.

Suppose that an observable in the $\tau$ space (momentum space) 
$\tilde{Y}(\tau)$ has
two unsuppressed singularities in the 
right-half Borel plane at $u=Re^{i\theta}$ and
$u=Re^{-i\theta}$, which are complex conjugate to each other
to guarantee the observable to be real ($R>0$, $0\leq \theta < \pi/2$).
For analyzing the
leading-order behavior, it is sufficient to set $b_1=b_2=\cdots=0$.
Explicitly, we consider\footnote{
This corresponds to the Borel transform (in the momentum space)
$$B_{\tilde{Y}}(u)=\frac{1}{2}\frac{e^{i\theta_0}}{u-Re^{i\theta}}+(\text{c.c.})\,.$$
}
\be
\tilde{Y}(\tau)=\frac{4\pi}{\tau}\sum_{n=0}^\infty \frac{n!}{R^{n+1}}
\cos (n\theta+\theta_0)\,
b_0^n\alpha_s(\tau)^{n+1}
\,,
\ee
with, for definiteness, the parameter choice $(a,u')=(1,-1)$.
(Since the singularities remain unsuppressed, $Re^{i\theta}\notin \mathbf{Z}$.)
The series of the FTRS method truncated at
${\cal O}(\alfs^{k+1})$ is given by
\be
\big[Y(Q)\big]_{\rm FTRS}^{(k)}=
\sum_{n=0}^k\big({\rm Re} [I_n]+J_n\big)
\,,
\ee
where
\bea
I_n
&&=\frac{2}{\pi Q}\int_0^\infty dt\,e^{-t/Q}
\frac{n!}{R^{n+1}}\cos (n\theta+\theta_0)\,
b_0^n\alpha_s(it)^{n+1}\non
&&=\frac{2}{\pi b_0Q}\int_0^\infty dt\,e^{-t/Q}\frac{n!}{R^{n+1}}\cos (n\theta+\theta_0)\,\frac{1}{\big[\log(t^2/\LQ^2)+i\pi\big]^{n+1}}\non
&&=\frac{2}{\pi b_0}\frac{n!}{R^{n+1}}\cos (n\theta+\theta_0)\,\int_0^\infty ds\,e^{-s}
\frac{1}{\big[\log(s^2)+\log(Q^2/\LQ^2)+i\pi\big]^{n+1}}
\label{I_n}
\eea
denotes each term of $\bar{Y}_0(Q)$ before taking the real part.
On the other hand, each term of the power series part $\bar{Y}_{\rm pow}(Q)$
is evaluated by the Cauchy theorem to be
\bea
J_n
&&=\frac{1}{\pi iQ}\int_{C_*}d\tau\,\cos(\tau/Q)\frac{n!}{R^{n+1}}
\cos (n\theta+\theta_0)\,b_0^n\alpha_s(\tau)^{n+1}\non
&&=\frac{1}{\pi ib_0Q}\int_{C_*}d\tau\,\sum_{m=0}^\infty\frac{(-1)^{2m}}{(2m)!}(\tau/Q)^{2m}\frac{n!}{R^{n+1}}\cos (n\theta+\theta_0)\,\frac{1}{\log^{n+1}(\tau^2/\LQ^2)}\non
&&=\frac{1}{b_0R}\cos (n\theta+\theta_0)\,
\sum_{m=0}^\infty\frac{(-1)^{2m}}{(2m)!}\Bigg(\frac{
2m+1}{2R}\Bigg)^n\bigg(\frac{\LQ}{Q}\bigg)^{2m+1}.
\eea

Since $|\cos (n\theta+\theta_0)|\leq 1$, we can set upper bounds as
$|{\rm Re}[I_n]|\leq|I_n|\leq |I_n|_{\cos (n\theta+\theta_0)\to 1}$ and 
$|J_n|\leq |J_n|_{\cos (n\theta+\theta_0)\to 1}$.
Hence, we ignore the cosine factor in the order estimate below.
The cosine factor simply gives an oscillatory factor with modulus one
as $n$ is varied.

We examine $|J_n|$ for each power of $\LQ/Q$, i.e., for each $m$.
For $m<R-\frac{1}{2}$, $|J_n^{(m)}|$ reduces to zero as $n$ increases,
and the sum $\sum_n J_n^{(m)}$ is convergent.
For $m=R-\frac{1}{2}$, (if such $m$ exists), $|J_n^{(m)}|$ is independent
of $n$ and therefore the sum $\sum_n J_n^{(m)}$ diverges.
For $m>R-\frac{1}{2}$, the sum $\sum_n J_n^{(m)}$ is more rapidly divergent.
For a small $n$, $|J_n^{(m)}|$ becomes smaller for larger $m$ by
higher power suppression of $\LQ/Q\ll 1$.
As $n$ is increased, the higher power terms of $\LQ/Q$ (higher $m$ terms)
become relatively more important, and at
$n \approx n_*(R,Q)$, the first divergent term $J_n^{(m_R)}$
and the last convergent term $J_n^{(m_R-1)}$ becomes comparable in 
magnitude,
where $m_R$ is the smallest integer satisfying $m_R \ge R-\frac{1}{2}$,
and
\bea
&&
n_*(R,Q) \approx 
\frac{\log (Q^2/\LQ^2)+\log[(2R-2)(2R-1)]}{-\log\Bigl(1-\frac{1}{R}\Bigr)}
\label{accurate-n*}
\\
&&
~~~~~~~~~~~~
\to \frac{R}{b_0\alfs(Q)} ~~\text{for}~~R\gg 1~\text{and}~
\log (Q^2/\LQ^2)\gg \log(R^2) \,.
\eea
$J_{n_*}^{(m)}$ for higher $m$ are smaller 
in magnitude than $J_{n_*}^{(m_R,m_R-1)}$.
For $n \simgt n_*(R,Q)$, $|J_{n}^{(m_R)}|$ exceeds $|J_{n}^{(m_R-1)}|$
and also
$J_{n}^{(m)}$ for higher $m$ increases rapidly with $n$.
The magnitude of the
minimal term $J_{n_*}^{(m_R)}$ can be estimated to be order $(\LQ^2/Q^2)^R$.
{\Large [}\,$\sum_{m\ge m_R} J_{n_*}^{(m)}$ is also order $(\LQ^2/Q^2)^R$.{\Large ]}

We turn to an estimate of $I_n$.
For $\log (Q^2/\LQ^2) \gg n$, $I_n$ can be approximated by
expanding the integral in eq.~\eqref{I_n} by $\log(s^2)$.
Thus, the leading asymptotic form is given by
\be
I_n \sim I_n^{\rm (est)} \equiv \frac{2}{\pi b_0}\,\frac{n!}{[R\, \{ \log (Q^2/\LQ^2)+i\pi \} ]^{n+1}}\,.
\ee
As long as $n<R \log (Q^2/\LQ^2) =R/[b_0 \alfs(Q)]$, $|I_n^{\rm (est)} |$ decreases with increasing $n$.
At $n \approx R/[b_0 \alfs(Q)]$, $|I_n^{\rm (est)} |$ becomes smallest and order
$(\LQ^2/Q^2)^R$.
However, the estimate $I_n \sim I_n^{\rm (est)}$ is not valid
at $n \approx R/[b_0 \alfs(Q)] $.
In fact, $|I_n|$ is much smaller than $|I_n^{\rm (est)}|$ as $n$ becomes large,
due to a highly oscillatory behavior of the integrand of eq.~\eqref{I_n}
for large $n$.
We were unable to find a good approximate formula for $I_n$ around $n \sim n_*(R,Q)$.
Instead we show the numerical evaluation of the ratio $|I_n^{\rm (est)} |/|I_n |$
(which is independent of $R$)
for a wide range of $n$ and $\log (Q^2/\LQ^2)$ in Tab.~\ref{tab:ratio-Inest-In}.
It 
should suffice to ensure that $|I_{n}|\ll |J_{n}|$ around $n\sim n_*$ for
any practical purposes.

\begin{table}[t]
\begin{center}
\includegraphics[width=16cm]{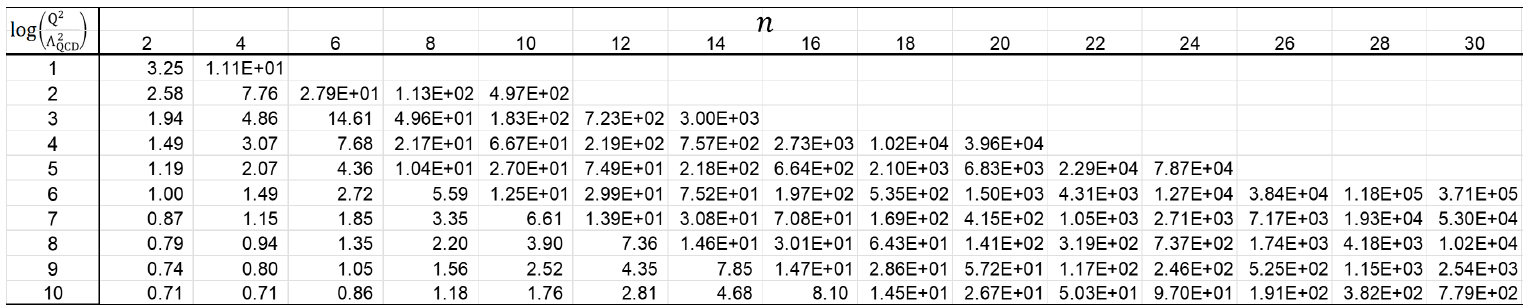}
\caption{\small
Numerical evaluation of the ratio $|I_n^{\rm (est)} |/|I_n |$ for various values of $n$ and 
$\log (Q^2/\LQ^2)$.
This ratio is independent of $R$ and increases rapidly from order one as $n$ increases.
}\label{tab:ratio-Inest-In}
\end{center}
\end{table}

Thus, the major diverging behavior emerges from the power
series part $J_n$ and its uncertainty can be estimated as order $(\LQ^2/Q^2)^R$
in parallel with the conventional estimate of renormalon.

As a last comment, it may make sense to expand the power series part
$Y_{\rm pow}(Q)$ in $\LQ^2/Q^2$ and truncate it at $m = m_R-1$, if
the position of the unsuppressed singularities can be estimated.
This is because, at $n \geq n_*(R,Q)$, $J_n^{(m)}$ grows rapidly for higher
$m$ and tends to deteriorate the behavior of $[Y(Q)]_{\rm FTRS}^{(k)}$
in the small $Q$ region for a fixed $k$, see Fig.~\ref{fig:Xtoymis}.
This is not the case if unsuppressed singularities are absent or negligible, as we
assume in the main body of our analysis, where it makes hardly
any differences to our analysis results whether
we truncate
$X_{\rm pow}(Q)$ by expanding it in $\LQ^2/Q^2$ or not.

\end{document}